\documentclass{aa}  
\usepackage{mathtools}
\usepackage{multirow}
\usepackage{siunitx}
\usepackage{comment}
\usepackage{booktabs}
\usepackage{graphicx}
\usepackage{menukeys}
\usepackage{algorithm}
\usepackage{ulem}
\usepackage[noend]{algpseudocode}
\graphicspath{{Fig/}}
\include{definitions}
\usepackage[colorlinks=true, allcolors=blue]{hyperref}
\usepackage{txfonts}

\newcommand{\arhp}{ArH$^{+}$ }
\newcommand{\ohp}{OH$^{+}$ }
\newcommand{\ohtop}{o-H$_{2}$O$^{+}$ }
\newcommand{\phtop}{p-H$_{2}$O$^{+}$ } 
\newenvironment{rightcases}
  {\left.\begin{aligned}}
  {\end{aligned}\right\rbrace}
\usepackage{cuted}
\makeatletter
\renewcommand*\aa@pageof{, page \thepage{} of \pageref*{LastPage}}
\makeatother
\begin{document} 

   \title{Extending the view of ArH$^{+}$ chemistry in diffuse clouds\thanks{The reduced ArH$^+$ and p-H$_{2}$O$^+$ spectra are only available at the CDS via anonymous ftp to cdsarc.u-strasbg.fr (\url{130.79.128.5}) or via \url{http://cdsweb.u-strasbg.fr/cgi-bin/qcat?J/A+A/}}}

   \author{Arshia M. Jacob\inst{1}\thanks{Member of the International Max Planck Research School (IMPRS) for Astronomy and Astrophysics at the Universities of Bonn and Cologne}
          \and
          Karl M. Menten\inst{1}
          \and
          Friedrich Wyrowski\inst{1}
          \and 
          Benjamin Winkel\inst{1}
          \and 
          David A. Neufeld\inst{2}
          }

   \institute{Max-Planck-Institut f\"{u}r Radioastronomie, Auf dem H\"{u}gel 69, 53121 Bonn, Germany
   \and
   Department of Physics and Astronomy, Johns Hopkins University, 3400 North Charles Street, Baltimore, MD 21218, USA \\
   \email{ajacob@mpifr-bonn.mpg.de}}

   \date{Received August 17, 2020; accepted October 1, 2020}
  \titlerunning{Extending the view of ArH$^{+}$ chemistry in diffuse clouds}
   \authorrunning{A. Jacob et al.}
% \abstract{}{}{}{}{} 
% 5 {} token are mandatory
 
  \abstract
{One of the surprises of the Herschel mission was the detection of ArH$^{+}$ towards the Crab Nebula in emission and in absorption
towards strong Galactic background sources. Although these detections were limited to the first quadrant of the Galaxy, the existing data suggest that \arhp ubiquitously and exclusively probes the diffuse atomic regions of the interstellar medium.}
{In this study, we extend the coverage of \arhp to other parts of the Galaxy with new observations of its $J = 1-0$ transition along seven Galactic sight lines towards bright sub-millimetre continuum sources. We aim to benchmark 
its efficiency as a tracer of purely atomic gas by evaluating its correlation (or lack of correlation as suggested by chemical models) with other well-known
atomic gas tracers such as \ohp and H$_{2}$O$^{+}$ and the molecular gas tracer CH.}
{The observations of the $J = 1-0$ line of \arhp near 617.5~GHz were made feasible with the new, sensitive SEPIA660 
receiver on the APEX 12~m telescope. Furthermore, the two sidebands %capabilities 
of this receiver allowed us to observe the $N_{K_{a}K_{c}} = 1_{1,0}-1_{0,1}$ transitions of para-H$_{2}$O$^{+}$ at 
607.227~GHz simultaneously with the ArH$^+$ line.}
{We modelled the optically thin absorption spectra %under the assumption of local thermodynamic equilibrium and estimated column densities, assuming excitation by the cosmological background radiation only. The \arhp column densities that we derive range from $4\times10^{11}$ to $10^{13}~$cm$^{-2}$ corresponding to 
of the different species and subsequently derived their column densities. By analysing the steady state chemistry of OH$^+$ and o-H$_{2}$O$^+$, we derive on average a cosmic-ray ionisation rate, $\zeta_\text{p}(\text{H})$, of ($2.3\pm0.3$)$\times10^{-16}$~s$^{-1}$ towards the sight lines studied in this work. Using the derived values of $\zeta_\text{p}(\text{H})$ and the observed \arhp abundances we constrain the molecular fraction of the gas traced by \arhp to lie below 2$\times10^{-2}$ with a median value of $8.8\times10^{-4}$. Combined, our observations of ArH$^+$, OH$^+$, H$_{2}$O$^+$, and CH probe different regimes
%phases of the ISM is usually reserved for WIM, WMN, CNM ...
of the interstellar medium, from  diffuse atomic to diffuse and translucent molecular clouds. Over Galactic scales, we see that the distribution of $N(\text{ArH}^+)$ is associated with that of $N(\text{H})$, particularly in the inner Galaxy %($R_{\text{GAL}} > 7~$kpc), w
(within 7~kpc of the Galactic centre) with potentially even contributions from the warm neutral medium phase of atomic gas at larger galactocentric distances. We derive an average ortho-to-para ratio for H$_{2}$O$^+$ of 2.1$\pm$1.0, which corresponds to a nuclear spin temperature of 41~K, consistent with the typical gas temperatures of diffuse clouds. }{}
 \keywords{ISM: molecules -- ISM: abundances -- ISM: clouds -- astrochemistry: cosmic rays}

   \maketitle
   
%
%________________________________________________________________

\section{Introduction} \label{sec:intro}
 Chemists have been fascinated by the possible existence and formation of noble gas compounds since the synthesis of the first noble gas-bearing molecule, XePtF$_{6}$, \citep{bartlett1962} and the molecular ion, HeH$^{+}$, \citep{Hogness1925} in the laboratory.
 In an astronomical setting, the helium hydride cation HeH$^+$ had long been predicted to be the first molecule to form in the Universe \citep{Galli2013} and to be detectable in planetary nebulae (PNe) \citep{Black1978}, many years before it was actually discovered in the PN NGC 7027 \citep{gusten2019astrophysical}. A few years ago, interest in the role of noble gas-bearing molecules in astrochemistry received attention after the detection of argonium, ArH$^{+}$, the first %naturally occurring 
 noble gas compound detected in the interstellar medium (ISM). This serendipitous discovery was made towards the Crab Nebula supernova remnant by \citet{barlow2013detection}, who observed the ${J = 1 - 0}$ and ${J = 2 - 1}$ rotational transitions of \arhp in emission using the Fourier Transform
Spectrometer (FTS) of the Herschel Spectral and Photometric Imaging REceiver (SPIRE) \citep{griffin2010herschel, swinyard2010flight}. This led to the subsequent identification of the ${J= 1 - 0}$ transitions of both $^{36}$\arhp and $^{38}$\arhp in the ISM by \citet{schilke2014ubiquitous}, which could be attributed to spectral features that had remained unidentified for some time in spectra taken with the Heterodyne Instrument for the Far Infrared (HIFI), also on Herschel \citep{de2010herschel}. These authors detected these two \arhp isotopologues in absorption, against the strong sub-millimetre %/far-infrared 
wavelength continuum of the high-mass star-forming regions (SFRs) Sgr~B2(M) and (N), G34.26+0.15, {W31~C}, W49(N), and W51e under the framework of the %Heterodyne Instrument for the Far Infrared \citep[HIFI,][]{de2010herschel} 
HIFI key guaranteed time programmes 
	Herschel observations of EXtra-Ordinary Sources (HEXOS) \citep{bergin2010herschel} and PRobing InterStellar Molecules with Absorption line Studies (PRISMAS) \citep{gerin2010interstellar}. Following this, \citet{Mueller2015} reported $^{36}$\arhp and $^{38}$\arhp absorption along two sight lines through the redshift $z = 0.8858$ foreground galaxy absorbing the continuum of the gravitational lens-magnified blazar, PKS 1830$-$211, using the Atacama Large Millimetre/sub-millimetre Array \citep[ALMA,][]{wootten2009atacama}.

Quantum chemical considerations have shown that \arhp can form via gas phase reactions between Ar$^{+}$ and molecular hydrogen \citep{roach1970potential}. Additionally, \citet{theis2015arh2+} have hypothesised that the formation of \arhp can also proceed via the dissociation of a semi-stable intermediate product, ArH$_{2}^{+}$:
\begin{equation*}
  \text{Ar}  \xrightarrow{\text{CR}} \text{Ar}^{+} \xrightarrow{\text{H}_{2}}
    \begin{cases}
    & \text{ArH}^{+} + \text{H} \hspace{2.25cm} (\Delta E = 1.436~\text{eV}) \\
      & \text{ArH}_{2}^{+}  \rightarrow \text{ArH}^{+} + \text{H} \hspace{0.95cm} (\Delta E =  {1.436~\text{eV}}) \, .
    \end{cases}       
\end{equation*}
 \arhp is formed in diffuse interstellar clouds with predominantly atomic gas and a small molecular hydrogen content
 %$f_{\rm H}_2$, $< 10^{-2}$ 
 and is readily destroyed %in the ISM 
 primarily via proton transfer reactions with neutral species, in particular with H$_{2}$ \citep{schilke2014ubiquitous}, while photo-dissociation has been shown to be less important \citep{roueff2014photodissociation}. %With \molh playing a pivotal role in %both the formation as well as 
 %the destruction of this cation, it is conceivable that \arhp is formed in predominantly %atomic gas. 
 Chemical models by \citet{schilke2014ubiquitous} and \cite{neufeld2016chemistry} suggest that \arhp must reside in low-density gas with very small molecular fractions, ${f_{\text{H}_{2}}=2n(\text{H}_{2})/\left[n(\text{H{\tiny I}})+2n(\text{H}_{2})\right] = 10^{-4} - 10^{-2}}$, and high cosmic-ray ionisation rates, $\zeta_\text{p}(\text{H}) = 4-8\times 10^{-16}$~s$^{-1}$, thereby establishing the unique capabilities of this cation as a tracer of purely atomic gas. Remarkably, to quote  \citet{schilke2014ubiquitous}, ``Paradoxically, the \arhp molecule is a better tracer of almost purely atomic hydrogen gas than H{\small I} itself.''
 
 The chemical significance of \arhp in the ISM has triggered both laboratory and theoretical studies of this molecule \citep{bizzocchi2016first, coxon2016accurate}. Recently, \citet{priestley2017} have discussed ArH$^{+}$ in the extreme environment of the Crab Nebula, while \citet{bialy2019chemical} have studied the role of super-sonic turbulence in determining the chemical abundances of diffuse gas tracers, including ArH$^{+}$. \\
 
 As to observations, after the end of the Herschel mission in mid-2013, to observe the lines of ArH$^{+}$ and many other light hydrides in the sub-millimetre/far-infrared wavelength regime, we must rely on ground based observatories and the air-borne %infrared 
 Stratospheric Observatory for Infrared Astronomy (SOFIA). A large part of this range cannot be accessed from the ground because of absorption in the Earth's atmosphere. Still, observations in certain wavelength intervals are possible, even from the ground, in the so-called sub-millimetre windows. The frequencies of the ArH$^{+}$ transitions, lie at the border of such a window and can be observed from high mountain sites under exceptional conditions. In this paper we present observations of the $J = 1 - 0$ transition of $^{36}$\arhp near 617~GHz along the line-of-sight (LOS) towards a sample of seven sub-millimetre and far-infrared bright continuum sources in the Galaxy, made with the new SEPIA660 receiver on the Atacama Pathfinder Experiment 12~m sub-millimetre telescope (APEX). The observational setup used, along with the data reduction is described in Sect.~\ref{sec:obs}. 
 
 In addition to ArH$^+$, our observing setup %made it
 allowed simultaneous observations of the ${N_{K_{a}K_{c}} = 1_{10}-1_{01}, J=1/2 - 3/2}$ and ${J=3/2 - 3/2}$ transitions of para-H$_{2}$O$^+$ at 604 and 607~GHz, respectively. The H$_{2}$O$^+$ molecular ion plays an important role in elementary processes in the ISM and exists in two symmetric states of opposite parities, ortho- and para-H$_{2}$O$^+$, corresponding to the different spin configurations of the hydrogen atoms. By combining our \phtop data with previously obtained data of o-H$_{2}$O$^+$, which is available for 
 %a subsample of our
 our same sample of targets, %\ohtop, which %towards the same sample of sources 
 we are able to investigate the ortho-to-para ratio of H$_{2}$O$^+$. Studying the relative abundance of the two states can give us significant insight into the efficiency of conversion between them, the formation pathway of H$_{2}$O$^+$ and the thermodynamic properties of the gas that contains it. The acquisition of the data for para-H$_{2}$O$^+$, as well as the retrieval of archival data is also described in Sect.~\ref{sec:obs}. 
 
 Sect.~\ref{sec:results}, presents the observed spectra and the derived physical parameters, which is followed by our analysis and discussion of these results in Sect.~\ref{sec:discussion}. Finally, our conclusions are given in Sect.~\ref{sec:conclusions}.
 %Limited by the instrumental resolution of the SPIRE-FTS on board Herschel, the relatively narrow line width (in comparison to the observed spectral resolution) of the $^{36}$\arhp line was previously not resolved towards extra-galactic sources \citep{rosenberg2014radiative}. Therefore, by utilising the high sensitivity and wide spectral window offered by the SEPIA660 receiver, we extend this analysis towards three extra-galactic sources, as well. 

 \section{Observations}\label{sec:obs} 
 The $J = 1-0$ transition of $^{36}$\arhp (hereafter ArH$^{+}$) was observed in 2019 July--August (Project Id: M-0103.F-9519C-2019) using the Swedish-ESO PI (SEPIA660) receiver \citep{belitsky2018sepia, hesper2018deployable} on the APEX 12~m sub-millimetre telescope. It is a sideband separating (2SB), dual polarisation receiver, covering a bandwidth of 8~GHz, per sideband, with a sideband rejection level \textgreater15~dB. The observations were carried out in wobbler switching mode, using a throw of the wobbling secondary of 120$^{\prime\prime}$ in azimuth at a rate of 1.5~Hz, fast enough to reliably recover the continuum emission of the background sources which represent sub-millimetre bright massive clumps within SFRs in the first and fourth quadrants of the Galaxy, selected from the APEX Telescope Large Area Survey of the GALaxy at 870 $\mu$m (ATLASGAL) \citep{Schuller2009, csengeri2014atlasgal}. Information on our source sample is given in Table~\ref{tab:source_properties}. 
 The receiver was tuned so that the lower sideband (LSB) was centred at a frequency of 606.5~GHz, covering both the ${N_{K_{a}K_{c}} = 1_{10}-1_{01}, J = 3/2 - 1/2}$ and ${J = 3/2 - 3/2}$ fine-structure transitions of \phtop at 604.678 and 607.227~GHz, respectively, while the upper sideband (USB) was centred at a frequency of 618.5~GHz that covers the \arhp $J = 1-0$ transition at 617.525~GHz (and also includes an atmospheric absorption feature at 620.7~GHz). The spectroscopic parameters of the different transitions that are studied in this work are discussed in Tab.~\ref{tab:spectroscopic_properties}.
 
\begin{figure}
    \includegraphics[width=0.48\textwidth]{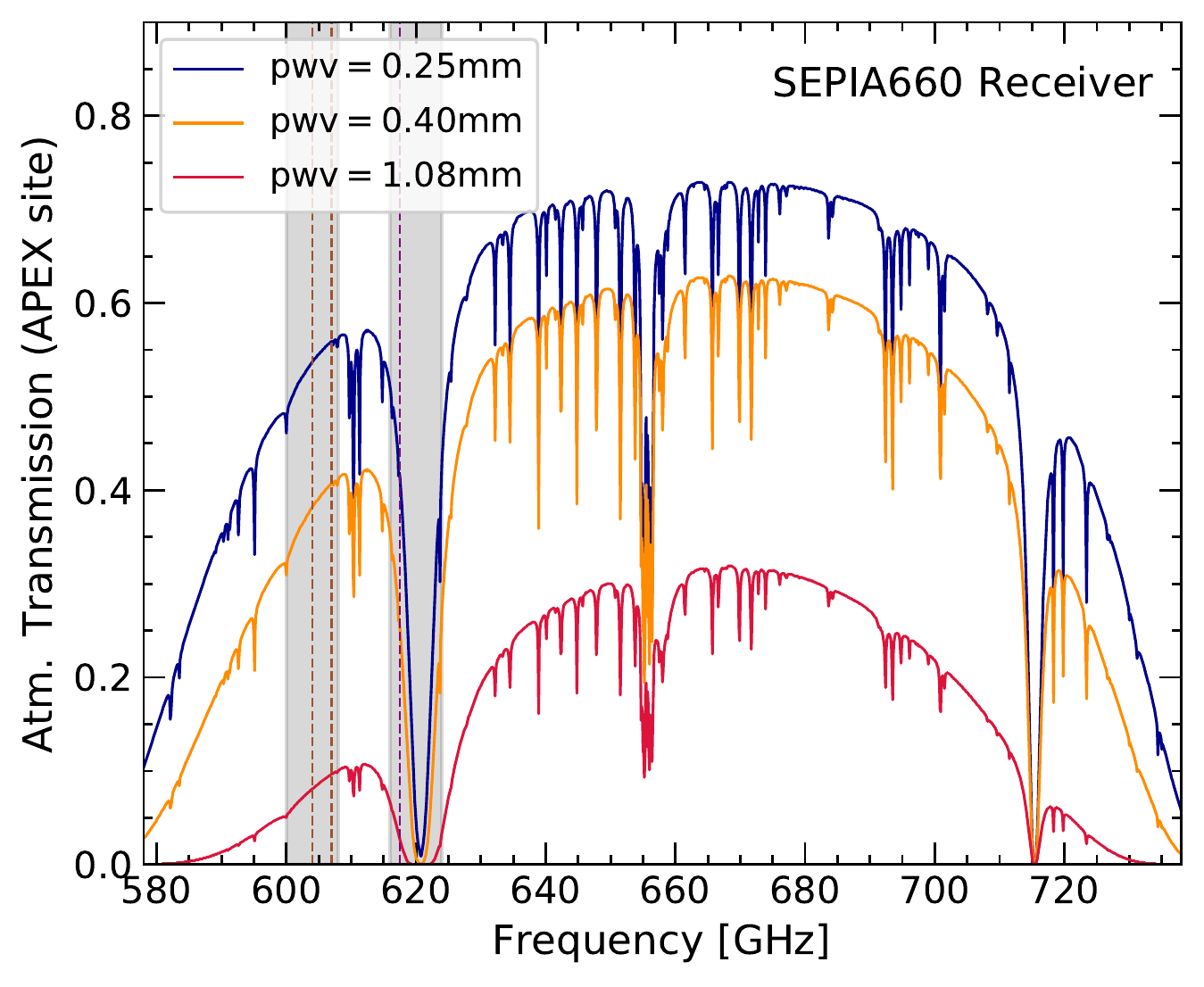}
    \caption{Calculated atmospheric zenith transmission of the entire 450 $\mu$m submillimeter window for the site of the APEX telescope for three values of the precipitable water vapour column: 0.25~mm (dark blue), 0.40~mm (orange), and 1.08~mm (red). The grey shaded regions mark the 8~GHz frequency windows covered by each of the two sidebands (separated by 8~GHz) of the SEPIA660 receiver. The dashed brown and pink lines mark frequencies of the p-H$_{2}$O$^+$ and ArH$^+$ lines discussed in this work, respectively.}
    \label{fig:atm_transmission}
\end{figure}

 Our observations were carried out under excellent weather conditions, with precipitable water vapour (PWV) levels between 0.25 and 0.41~mm, corresponding to an atmospheric transmission at the zenith better than or comparable to 0.5 in both sidebands and a mean system temperature of 880~K, at 617~GHz. In Fig.~\ref{fig:atm_transmission} we display the corresponding atmospheric zenith transmission from the APEX site and, for comparison, we also present the atmospheric transmission for PWV = 1.08~mm, which corresponds to the  50\% percentile\footnote{Meaning that the PWV content was 1.08~mm or lower for half of the covered time period.} of the PWV values measured over a span of more than 12.8 years on the Llano de Chajnantor \citep{Otarolaweather2019}. The atmospheric transmission curves presented in Fig.~\ref{fig:atm_transmission} are computed based on the \textit{am} transmission model\footnote{See \url{https://www.cfa.harvard.edu/~spaine/am/} for more information on the $am$ atmospheric model.} developed by the Smithsonian Receiver Lab at the Smithsonian Astrophysical Observatory. On average we spent a total (on+off) observing time of 2.3~hr towards each Galactic source. The half power beam-width (HPBW) is $10\rlap{.}^{\prime\prime}$3 at 617~GHz.
 The spectra were converted into main-beam brightness temperature units using a forward efficiency of 0.95 and a main-beam efficiency of ${\sim}$0.39 (determined by observing Mars). The calibrated spectra were subsequently processed using the GILDAS/CLASS software\footnote{Software package developed by IRAM, see \url{https://www.iram.fr/IRAMFR/GILDAS/} for more information regarding GILDAS packages.}. The spectra obtained towards the different sources were smoothed to velocity bins of ${\sim} 1.1$~km~s$^{-1}$ %,%, and up to 8.6~hr towards the extra-Galactic sources.  and ${\sim} 4.5$~km~s$^{-1}$, respectively
 and a first order polynomial baseline was removed.

 In addition to the results of the \arhp and \phtop observations described above, we use complementary APEX data of OH$^{+}$, some of which has already been published by \citet{wiesemeyer2016far}. The ancillary \ohtop data used here, was procured as a part of the Water in Star-forming regions with Herschel (WISH) survey \citep{van2011water} using the HIFI instrument on the Herschel Space Observatory and retrieved using the Herschel Science Archive\footnote{http://archives.esac.esa.int/hsa/whsa/}. The spectroscopic parameters of these lines are also summarised in Tab.~\ref{tab:spectroscopic_properties}. 
 
 Furthermore, to be able to constrain the theoretical predictions of \arhp abundances, we obtained archival data of the H{\small I} 21~cm line. We use archival %interferometric 
 data of H{\small I} absorption and emission from the THOR survey\footnote{THOR is The H{\small I}/OH/Recombination line survey of the inner Milky Way observed using the Very Large Array (VLA) in C-configuration \citep{beuther2016hi}.} for AGAL019.609$-$00.234, data presented in \citet{winkel2017hydrogen} for AGAL031.412+0.31 and data from the SGPS database \citep{McClure2005}\footnote{SGPS is the Southern Galactic Plane Survey, which  combines data from the Australia Telescope Compact Array and the Parkes Radio Telescope.}  for the remaining sources. %We corrected for optical depth effects 
 By combining the absorption profiles with emission line data, we were able to determine H{\small I} column densities as described in \citet{winkel2017hydrogen}. The results of the H{\small I} analysis, namely the optical depth, spin temperatures, and H{\small I} column densities along with the corresponding H{\small I} emission and absorption spectra are given in Appendix~\ref{appendix:hi_analysis}. We also compare the \arhp line profiles with those of molecular hydrogen derived from its diffuse gas proxy, CH, using data presented in \citet{jacob2019fingerprinting} for AG330.954$-$00.182, AG332.836$-$00.549 and AG351.581$-$00.352. We also present CH 2~THz data that was previously not published, towards AG10.472+00.027 observed using the upGREAT receiver on board SOFIA. The observational setup used, is akin to that detailed in \citet{jacob2019fingerprinting}, for the other sources.

\begin{table*}
    \centering 
     \caption{Properties of the sources analysed in this work.}
    \begin{tabular}{rrrrrrcrr}
    \hline \hline
        \multicolumn{1}{c}{Source} & \multicolumn{4}{c}{Coordinates (J2000)}  & $d$\tablefootmark{a} & Ref & $\upsilon_{\text{LSR}}$ & $T_{\text{c}}$\tablefootmark{b}\\
        ATLASGAL Name &   $\alpha$~[hh:mm:ss] & $\delta$~[dd:mm:ss] & $l [^{\circ}]$  & $b [^{\circ}]$ & [kpc] & & [km~s$^{-1}$] &  [K] \\
         \hline
        
         AG10.472+00.027 & 18:08:38.20 & $-$19:51:49.60 & 10.472 & +0.028 &  8.6 & [1] & $+$67.6 &  3.50\\
         AG19.609$-$00.234  & 18:27:37.98 & $-$11:56:36.60 & 19.608 & $-$0.233 & 12.7 & [2] & $+$40.8 &  1.85\\
         AG31.412+00.307  & 18:47:34.30 & $-$01:12:46.00 & 31.411 & +0.308 &  5.3 & [3] & $+$98.2 & 1.62 \\
         AG330.954$-$00.182  & 16:09:53.01 &  $-$51:54:54.80 & 330.952 & $-$0.181 & 5.7 & [4] & $-$91.2 & 4.54 \\
         AG332.826$-$00.549  & 16:20:10.65 & $-$50:53:17.60 &  332.824 &$-$0.549 & 3.6 & [3] & $-$57.1 & 2.55 \\
         AG337.704$-$00.054  & 16:38:29.42 & $-$47:00:38.80 & 337.704 & $-$0.053& 12.3 & [5] & $-$47.4 & 1.30\\
         AG351.581$-$00.352 & 17:25:25.03 & $-$36:12:45.30 & 351.581 & $-$0.352 & 6.8 & [5] & $-$95.9 & 2.51 \\
         \hline
         %\multicolumn{9}{c}{Extra-Galactic sources}\\
         %\hline
         %Arp~220 & 15:34:57.20 & $+$23:30:11.00 & 36.627 & +53.028 & 7.2$\times10^{4}$ & [6] & 5434.0 &  0.05 \\
         %NGC~253 &  00:47:32.98 & $-$25:17:15.90 & 97.363 & $-$87.964 & 3.0$\times10^{3}$& [6] & 240.0 &0.15\\
         %NGC~4945 & 13:05:27.48 & $-$49:28:05.60 & 305.271 &  +13.340 & 3.8$\times10^{3}$&  [6] & 563.0 & 0.39 \\
         
         \hline
    \end{tabular}
    \tablefoot{ \tablefoottext{a}{Heliocentric distance to the source.} \tablefoottext{b}{Main-beam brightness temperature of the continuum at 617~GHz.}}
    \tablebib{For the heliocentric distances: [1]~\citet{sanna2014trigonometric}; [2]~\citet{urquhart2014rms}; [3]~\citet{moises2011spectrophotometric}; %(4)~\citet{zhang2009trigonometric};
    [4]~\citet{wienen2015atlasgal}; %5]~\citet{moises2011spectrophotometric}; 
    [5]~\citet{green2011distances};
    %%[6]~\href{http://ned.ipac.caltech.edu/}{Nasa Extragalactic Database (NED)}
    %(8)~\citet{karachentsev2003distances};
    %(9)~\citet{karachentsev2006hubble};
    }
    \label{tab:source_properties}
\end{table*}

\begin{table*}
    \centering
     \caption{Spectroscopic properties of the studied species and transitions.}
    \begin{tabular}{lcccccl}
    \hline \hline
    Species & \multicolumn{2}{c}{Transition} & Frequency & $A_{\text{E}}$ & $E_{\text{u}}$ & Receiver/Telescope  \\
    &  $J^{\prime} - J^{\prime\prime}$ & $F^{\prime} - F^{\prime\prime}$ & [GHz] & [s$^{-1}$] & [K] \\
    \hline 
    
    \arhp  & $1 - 0$  & --- & 617.5252(2) & 0.0045 & 29.63 & SEPIA660/APEX\\
    %\hline
    \phtop & $3/2 - 1/2$& --- & 604.6841(8)& 0.0013 & 59.20 & SEPIA660/APEX \\ 
    ~~$N_{K_a, K_c} = 1_{1,0} - 1_{0,1}$     & $3/2 - 3/2$&  --- & 607.2258(2) & 0.0062 & 59.20 & SEPIA660/APEX\\  %\hline
    %\ohp & $2-1$ & $5/2 - 3/2$ & 971.803\tablefootmark{*} & 0.0182 & 46.64 & HIFI/Herschel\\ 
    %(1$-$0)& & $3/2 - 1/2$ & 971.805 & 0.0152 & \\
    %     & & $3/2 - 3/2$ & 971.919 & 0.0030 & \\ %\hline
    \ohp & $1 - 1$ & $1/2 - 1/2$ & 1032.9985(7) & 0.0141  & 49.58 & 1.05THz Rx./APEX\\
    ~~$N =1-0$&  & $3/2 - 1/2$ & 1033.0040(10) & 0.0035  &  \\
        & & $1/2 - 3/2$ & 1033.1129(7) & 0.0070  & \\
        & & $3/2 - 3/2$ & 1033.1186(10)\tablefootmark{*} & 0.0176  &  \\ %\hline
    \ohtop & $3/2 - 1/2$ & $3/2 - 1/2$ & 1115.1560(8) &  0.0171 & 53.52 & HIFI/Herschel\\
    ~~$N_{K_a, K_c} = 1_{1,1} - 0_{0,0}$ & & $1/2 - 1/2$ & 1115.1914(7) & 0.0274& \\
         & & $5/2 - 3/2$ & 1115.2093(7)\tablefootmark{*} & 0.0309& \\
         & & $3/2 - 3/2$ & 1115.2681(7) & 0.0138& \\
         & & $1/2 - 3/2$ & 1115.3035(8) & 0.0034& \\% \hline
     CH &  $3/2-1/2$ & $1-1$ & 2006.74886(6) &  0.0111& 96.31 & upGREAT/SOFIA\\
      ~~$^2\Pi_{3/2}, N = 2-1$ & & $1-0$ & 2006.76258(6) & 0.0223 & \\
            & & $2-1$ & 2006.79906(6)\tablefootmark{*} & 0.0335 &  \\ \hline    
     \end{tabular}
     \tablefoot{ The spectroscopic data are taken from the Cologne Database for Molecular Spectroscopy \citep[CDMS,][]{muller2005cologne}. The H$_2$O$^+$ frequencies were actually refined considering astronomical observations \citep[see Appendix A of][]{Mueller2016}. For the rest frequencies, the numbers in parentheses give the uncertainty in the last listed digit. \tablefoottext{*}{Indicates the strongest hyperfine structure transition, which was used to set the velocity scale in the analysis.}}
 \label{tab:spectroscopic_properties}
\end{table*}

It is essential to account for calibration uncertainties in the absolute continuum level as the line-to-continuum ratio forms the crux of the analysis that we present in the following sections. In order to attest for the reliability of the quoted continuum brightness temperatures, we measured the fluctuations in the continuum level across scans and also compared the ArH$^+$ continuum fluxes with ancillary continuum data at 870~$\mu$m observed using the LABOCA bolometer at the APEX telescope as part of the ATLASGAL survey. The scatter in the continuum levels across scans is $<12\%$, on average and the ArH$^+$ continuum fluxes, correlate well with that of the 870~$\mu$m continuum emission, with a relative scatter of 5\%. We present a more detailed 
%candid
discussion on the same, in Appendix~\ref{appendix:continuum_calib}.

\section{Results}\label{sec:results}
\begin{figure}
    
     \includegraphics[width=0.485\textwidth]{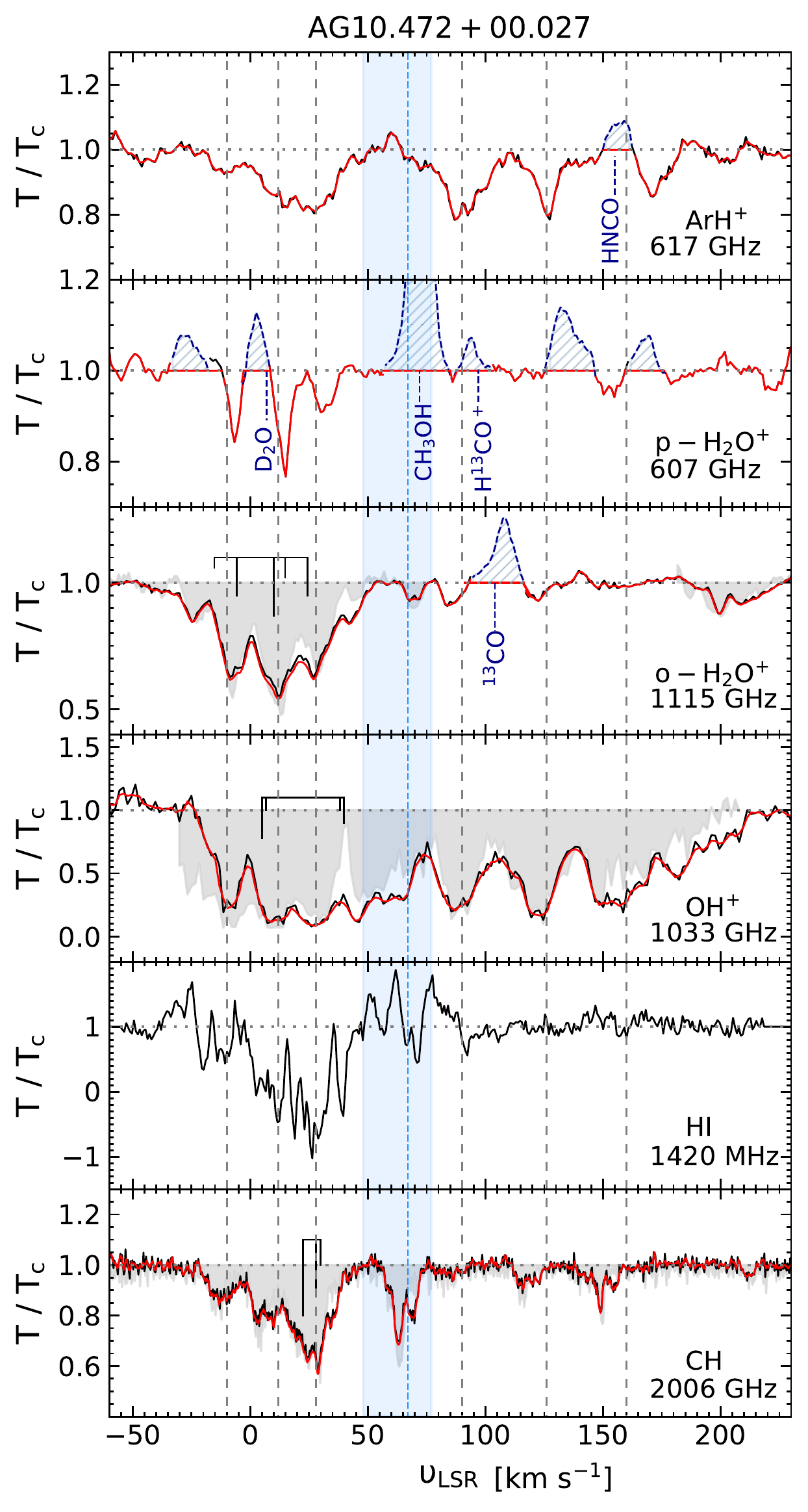} 
    \caption{From top to bottom: Normalised spectra showing transitions of ArH$^{+}$, p-H${_{2}}$O$^{+}$, o-H${_{2}}$O$^{+}$, OH$^{+}$, H{\tiny I}, and CH, towards AG10.472+00.027, in black. The vertical dashed blue line and blue shaded regions mark the systemic velocity and the typical velocity dispersion of the source. The vertical dashed grey lines indicate the main absorption dips. Red curves display the Wiener filter fit to the observed spectra except for the H{\tiny I} spectra. %With very little continuum the signal-to-noise ratio in the H{\small I} spectrum is poor, therefore the negative values seen here result from the large noise levels present in this spectrum.
    The relative intensities of the hyperfine structure (hfs) components of the o-H$_{2}$O$^+$ and the OH$^+$ and CH transitions are shown in black above their respective spectra and the grey shaded regions display their hfs deconvolved spectra. Contaminating emission features are marked by the blue hatched lines and are excluded from the modelled fit and analysis.}
    \label{fig:spectra_G10P47}
\end{figure}

\begin{figure}
   \includegraphics[width=0.485\textwidth]{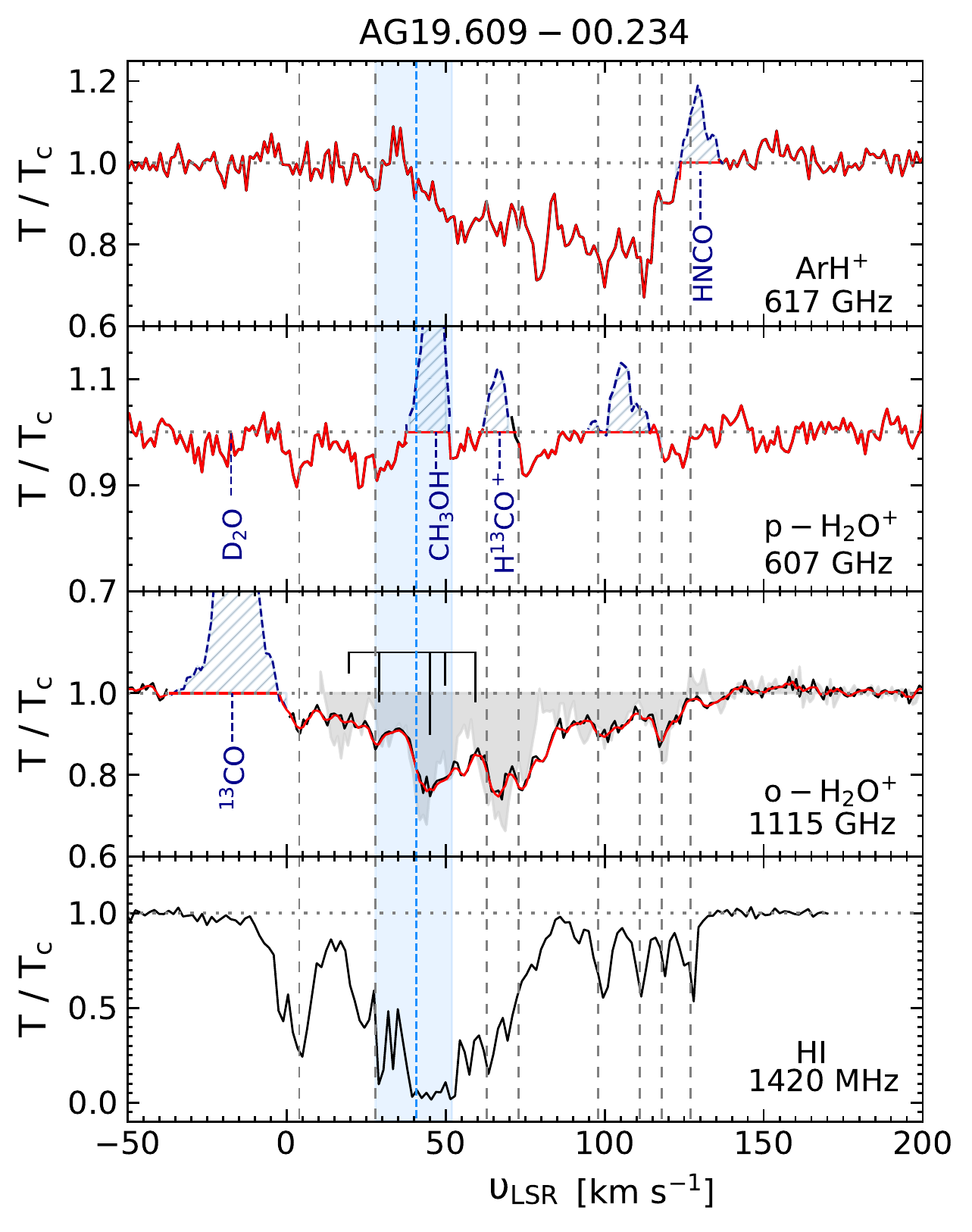}
   \caption{Same as Figure.~\ref{fig:spectra_G10P47} but towards AG19.609$-$00.234. There is no \ohp and CH spectra of the transitions studied here, available for this source.}
   \label{fig:spectra_G19P61}
\end{figure}

\begin{figure}
    \includegraphics[width=0.485\textwidth]{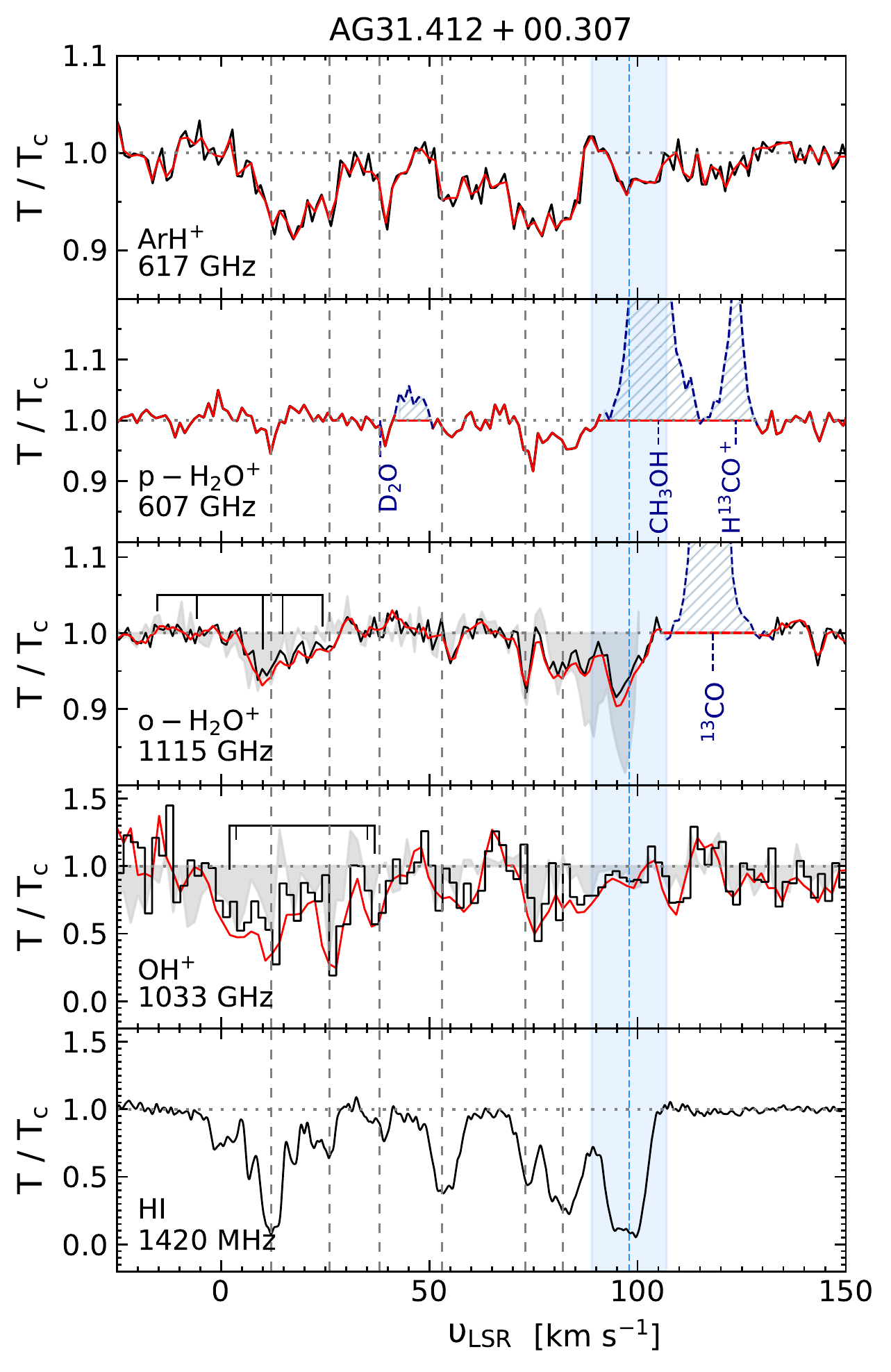}
    \caption{Same as Figure.~\ref{fig:spectra_G10P47} but towards AG31.412+00.307. There is no CH spectrum of the transitions studied here, available for this source.}
    \label{fig:spectra_G31P41}
\end{figure}

\begin{figure}
       \includegraphics[width=0.485\textwidth]{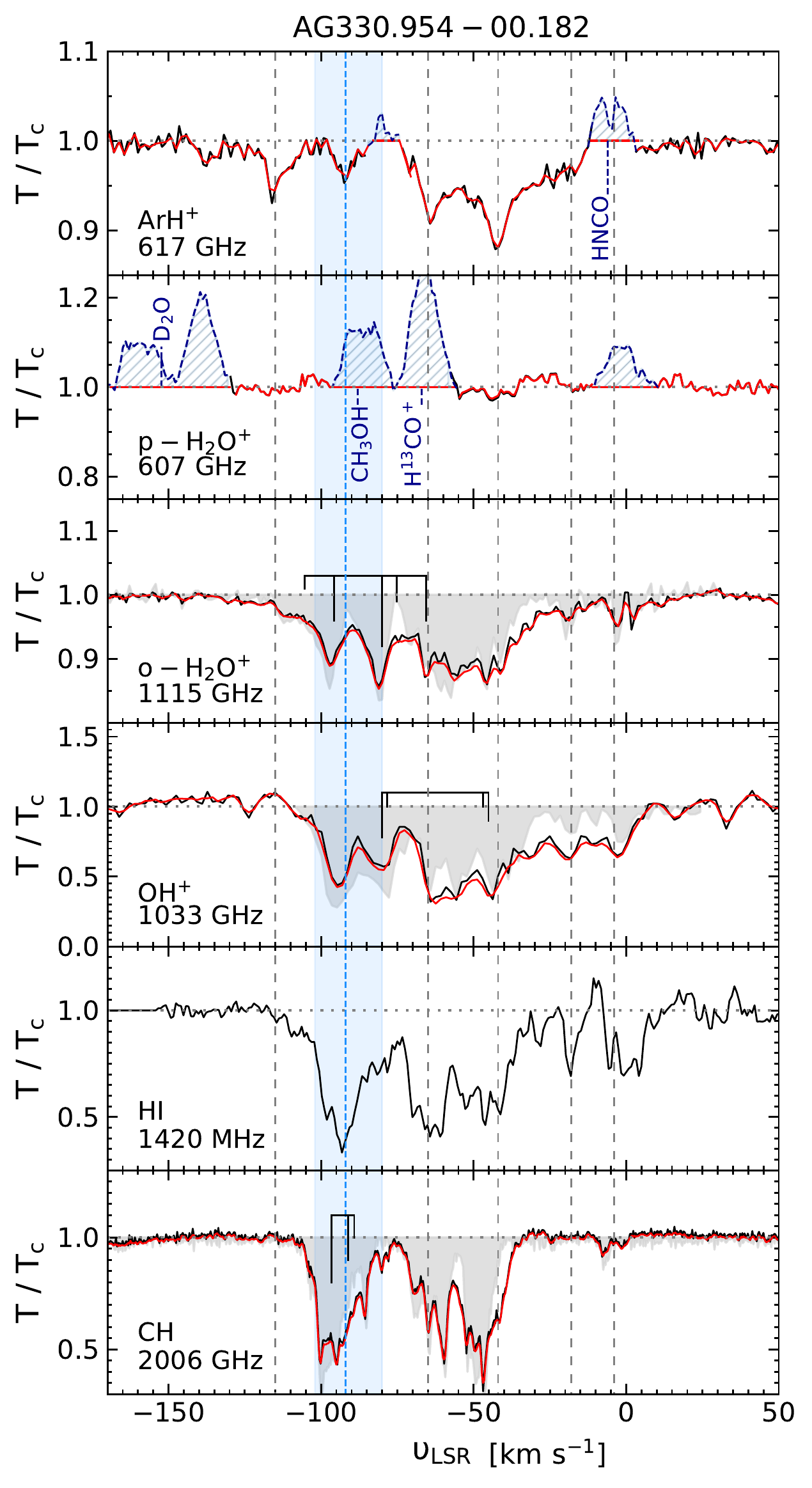} 
       \caption{Same as Figure.~\ref{fig:spectra_G10P47} but towards AG330.954$-$00.182.   }
       \label{fig:spectra_G330P95}
\end{figure}

\begin{figure}
   \includegraphics[width=0.485\textwidth]{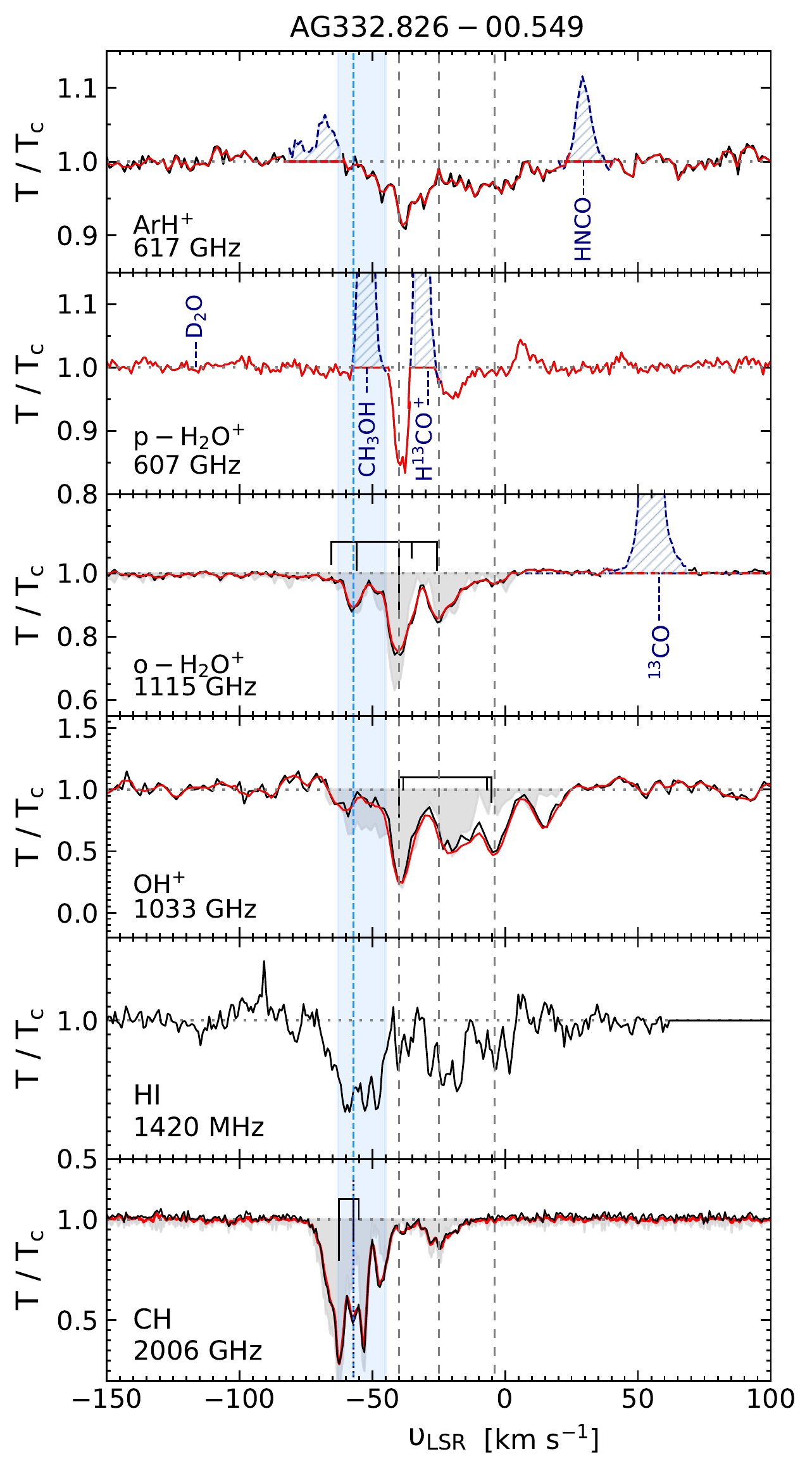}
          \caption{Same as Figure.~\ref{fig:spectra_G10P47} but towards AG332.826$-$00.549.}
          \label{fig:spectra_G332P83}
\end{figure}

\begin{figure}
    \includegraphics[width=0.485\textwidth]{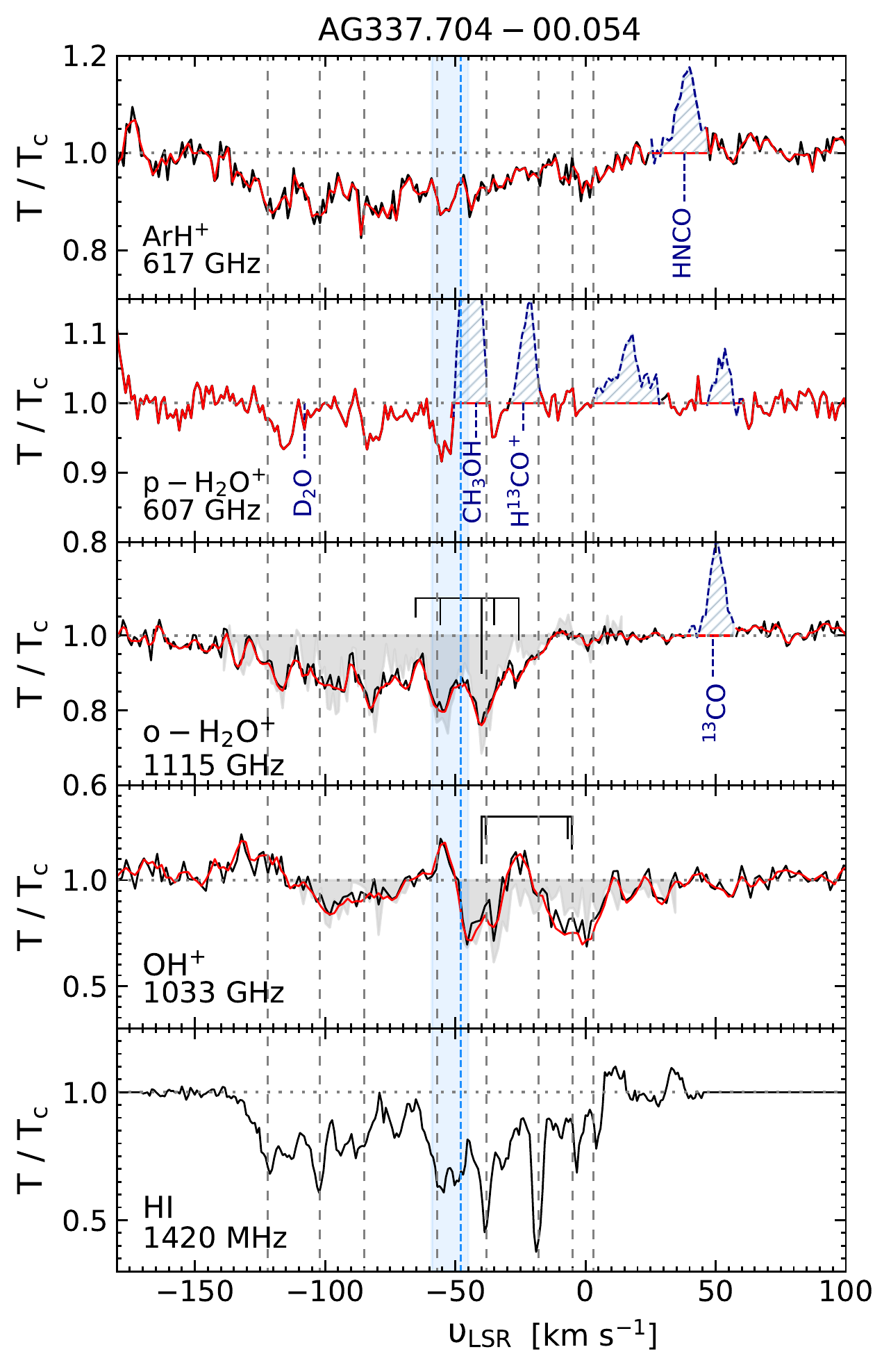}
           \caption{Same as Figure.~\ref{fig:spectra_G10P47} but towards AG337.704$-$00.054. There is no CH spectrum of the transitions studied here, available for this source.}
           \label{fig:spectra_G337P70}
\end{figure}

\begin{figure}
    \includegraphics[width=0.485\textwidth]{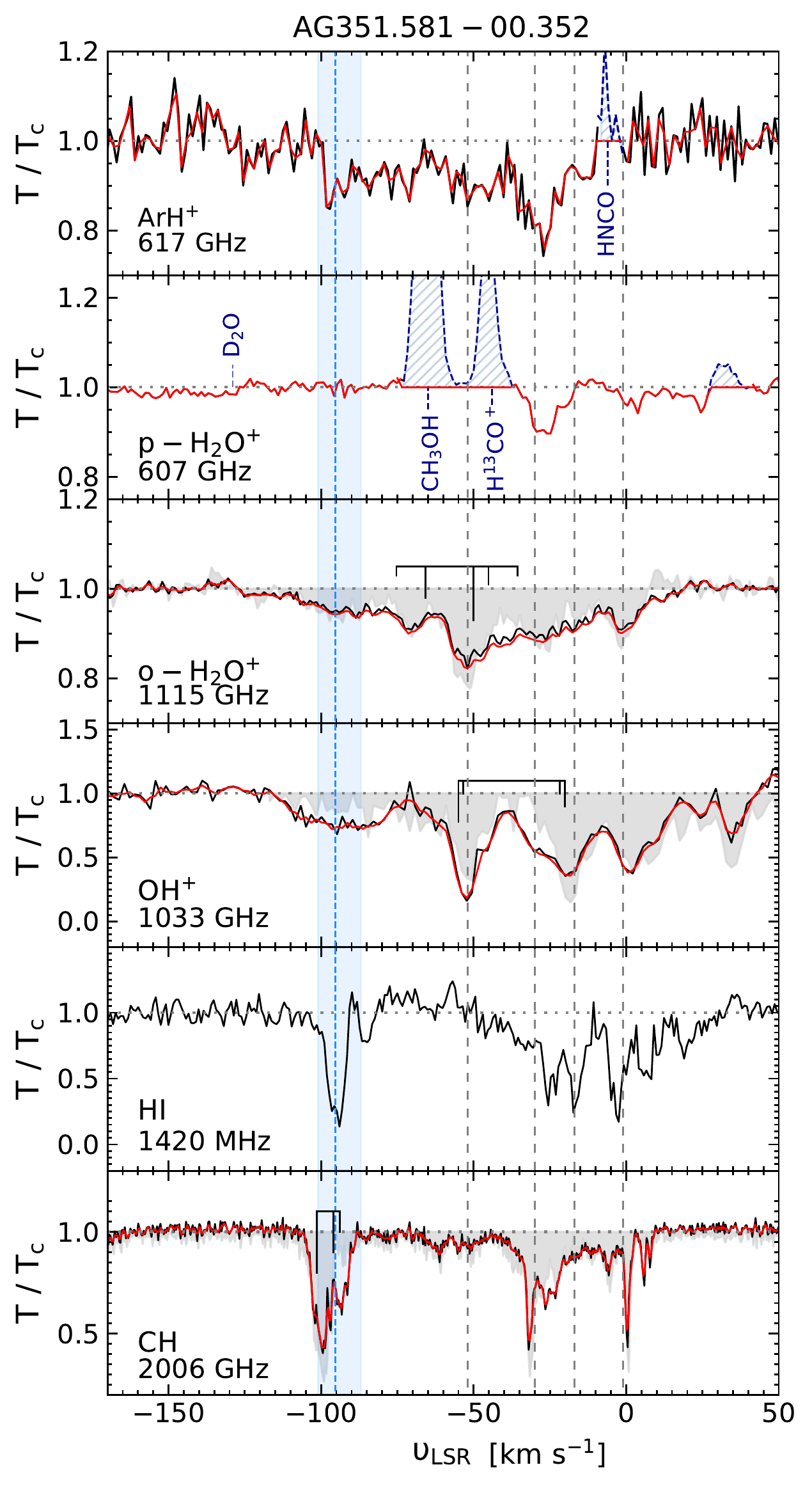}
           \caption{Same as Figure.~\ref{fig:spectra_G10P47} but towards AG351.581$-$00.352. }
    \label{fig:spectra_G351P58}
\end{figure}

%We successfully detected \arhp towards all the sources in the observed sample except towards Arp~220 down to a noise level of 4.9~mK at a velocity resolution of 4.5~km~s$^{-1}$. Therefore, we do not include Arp200 in our analysis. 
Figs.~\ref{fig:spectra_G10P47}-\ref{fig:spectra_G351P58} present the calibrated and baseline subtracted spectra of all the transitions discussed in this work for all the sources observed. In general, spectra observed along the LOS towards hot-cores often show emission from a plethora of molecules, including complex organic species, to the extent that they can form a low level background, `weeds', of emission lines, many of which remain unidentified \citep[see, e.g.][]{Belloche2013}. Features from these species `contaminate' the absorption profiles and make it difficult to gauge the true depth of the absorption features, potentially leading to gross underestimates of the subsequently derived column density values. In the following paragraphs we briefly discuss the main weeds that contaminate those parts of the spectra that are relevant for our absorption studies.\\

towards all the Galactic sources in our sample, except for AG31.412+00.307, the LOS \arhp absorption is blended with emission from the high-lying (${E_{\text{u}} = 473}$~K) HNCO, ${\varv=0,~(28_{1,27}-27_{1,26})}$ transition at 617.345~GHz \citep{Hocking1975} that originates in the hot molecular cores associated with the SFR that provides the continuum background radiation. Additionally, the HNCO emission line contaminant at 617.345~GHz lies very close to the H$_{2}$CS ($18_{2,17}-17_{2,16}$) transition at 617.342~GHz \citep{Johnson1972}. To infer the extent of the contribution from H$_{2}$CS we compared it to another H$_{2}$CS transition covered in the same sideband. We do not detect the H$_{2}$CS ($18_{2,16}-17_{2,15}$) transition at 620.165~GHz which has a comparable upper level-energy, and -degeneracy, and Einstein A coefficient as that of the H$_{2}$CS ($18_{2,17}-17_{2,16}$) line, above a noise level of ${\sim}$83~mK at a spectral resolution of 1.1~km~s$^{-1}$. This leads us to conclude that the H$_{2}$CS transition at 617.342~GHz may not significantly contaminate our spectra.

In the \phtop absorption spectra at 607~GHz, we see blended emission features particularly at the systemic velocity of the SFRs in this study, from H$^{13}$CO$^{+}$ ($J = 7-6$) and CH$_{3}$OH $J_k = 12_2-11_1~\text{E}$ transitions at 607.1747 \citep{Lattanzi2007} and 607.2159~GHz \citep{Belov1995}, respectively. The LOS is also contaminated by the D$_{2}$O ($J_{K_a,K_c} = {1_{1,1}-0_{0,0}}$) transition at 607.349~GHz \citep{matsushima2001frequency}. Detected thus far only towards the solar type protostar IRAS16293-2422 \citep{Butner2007,Vastel2010}, it is unlikely that there is significant contamination from D$_{2}$O absorption in our spectra. Observations of the \ohtop spectra which were performed in double sideband mode, covered the $^{13}$CO ($10-9$) transition at 1101.3~GHz \citep{Zink1990} in the LSB, alongside the \ohtop line in the USB. This $^{13}$CO emission feature blends with the LOS absorption profile of \ohtop towards AG10.472+00.027 and to a lesser extent with that of AG19.609$-$00.234 and AG31.412+00.307.

While, it is imperative to accurately model the degree of contamination in order to determine physical quantities within the velocity intervals concerned, there are several uncertainties in the `residual' absorption obtained from fitting the observed emission features. Therefore, we do not model contributions from the different contaminants present along the different sight lines but rather exclude the corresponding velocity intervals from our modelled fits and analysis. However, by doing so, it is possible that we ignore impeding emission features and emission line-wings, which remain undetected because they are absorbed away below the continuum. This can lead to potential uncertainties in the derived optical depths particularly in the velocity intervals neighbouring the emission. In the following sections, we carry out a qualitative and quantitative comparison between the (uncontaminated) absorption features, for the different species and for each individual source.

\subsection{Line-of-sight properties}\label{subsec:LOS_properties}
The Galactic sources we have selected in this work, which predominantly provide the background radiation for our absorption studies, are luminous (${L > 10^4 L_{\odot}}$) as they are dusty envelopes around young stellar objects with typical signposts of massive star-formation such as class II methanol masers % watermasers 
and/or ultracompact (UC) H{\small II} regions. Much of this work addresses the investigation of trends of various quantities with Galactocentric distance. To allow this, the spectrum towards each sight line is divided into local standard of rest (LSR) velocity intervals that correspond to absorption features arising from different spiral-arm and inter-arm crossings. The velocities of these LOS absorption components are then mapped on a model of Galactic rotation to relate them to the
% This is done by visualising the LOS velocities towards each source as a function of 
Galactocentric distance, $R_{\text{GAL}}$. The values for $R_{\text{GAL}}$ were computed by assuming a flat rotation curve, with the distance between the Sun and the Galactic centre (GC), $R_{0}$, and the Sun's orbital velocity, $\Theta_{0}$, assumed to be 8.15~kpc and 247~km~s$^{-1}$, respectively, as determined by \citet{Reid2019}. We have also compared our distance solutions with those reported in \citet{urquhart2018atlasgal}, who have analysed the kinematic properties of dense clumps present in the ATLASGAL survey, which covers our sources. Furthermore, it is to be noted that our assumption of a flat rotation curve is only valid for $R_{\text{GAL}} \gtrsim 4~$kpc as shown in Fig.~11 of \citet{Reid2019}. For smaller values of $R_{\text{GAL}}$ and for a general check on all distances, we employed the parallax-based distance calculator accessible from the website of the Bar and Spiral Structure Legacy survey (BeSSeL)\footnote{See, \url{http://www.vlbi-astrometry.org/BeSSeL/node/378})}.

In general, over velocity intervals that correspond to the envelopes of the individual molecular cloud cores, the `systemic' velocities, \arhp shows very weak to almost no absorption unlike the other molecules. The absence of \arhp at these velocities is in line with chemical models that predict \arhp to exist exclusively in low-density atomic gas, $f_{\text{H}_{2}} = 10^{-4}$-$10^{-2}$ \citep{schilke2014ubiquitous, neufeld2016chemistry}. As discussed in Sect.~\ref{sec:obs}, the LSB of the SEPIA660 receiver was tuned to cover the \phtop transitions near 604 and 607~GHz. Invariably across our entire sample of sources, we do not detect any clear absorption features from the \phtop line at 604~GHz at an average noise level of 23~mK (at a spectral resolution of 1.1~km~s$^{-1}$), owing to its weaker line strength in comparison to the \phtop line at 607~GHz. Hence, we do not discuss the former \phtop transition in the remainder of our study.%We do not detect any absorption  observed spectra of the 604~GHz \phtop transitions are presented in Appendix~\ref{appendix:phtop_604}. 

\subsubsection*{AG10.472+00.027} 
 The sight line towards AG10.472+0.027, which is at a distance of 8.55~kpc \citep{sanna2014trigonometric}, crosses several spiral arms in the inner Galaxy. This results in a broad absorption spectrum covering an LSR velocity range from $-$35 to 180~km~s$^{-1}$. %large number of velocity components through the inner Galaxy and exhibits a 
 All the molecules show absorption at $\upsilon_{\text{LSR}}$ between $-35$ and $48$~km~s$^{-1}$, which arises partly from inter-arm gas and partly from the near-side crossing of the Sagittarius arm and Scutum-Centaurus arm. Within this velocity range, the absorption profiles form a nearly continuous blend except for p-H$_{2}$O$^{+}$, which has an unidentified emission feature at ${\sim}3$~km~s$^{-1}$ and generally narrower features. In contrast, the \ohp absorption covers a larger velocity range than the other molecules and almost saturates within 0 and 45~km~s$^{-1}$. % almost saturated profile. 
 The features at $\upsilon_{\text{LSR}} > 80~$km~s$^{-1}$ are associated with the 3~kpc arm and the Galactic bar. Both \phtop and \ohtop only weakly absorb at $\upsilon_{\text{LSR}} > 130~$km~s$^{-1}$; the former is also %hindered 
 affected by some unknown emission features. However, \arhp shows absorption features similar to \ohp with strong absorption %peaks
 dips at 80 and 127~km~s$^{-1}$. The high-velocity component likely traces gas that lies beyond the GC belonging to the 135~km~s$^{-1}$ arm \citep{sormani2015recognizing}. CH has a narrow absorption dip %peak 
 at ${\sim}$150~km~s$^{-1}$ unlike the other atomic gas tracers, while \arhp and \ohp have prominent absorption at  170~km~s$^{-1}$. The corresponding H{\small I} spectrum with very little continuum, has a poor signal-to-noise ratio, which results in negative values in regions with large noise levels. However, the Bayesian analysis of the H{\small I} spectrum described in \citet{winkel2017hydrogen}, which is used to derive the column density and spin temperature, correctly accounts for these noise effects amongst other spatial variations along the LOS. 

\subsubsection*{AG19.609-00.234} 
AG19.609$-$00.234 is a massive star-forming clump at a %Galactocentric 
heliocentric distance of 12.7~kpc \citep{urquhart2014rms} with an absorption spectrum extending from $-10$ to 150~km~s$^{-1}$. The absorption at $\upsilon_{\text{LSR}}$ between $-10$ and 20~km~s$^{-1}$ primarily traces inter-arm gas between the Perseus and outer arms and parts of the Aquila Rift, followed by weak absorption near 27~km~s$^{-1}$ from the envelope of the molecular cloud. While the absorption dips between 35 and 74~km~s$^{-1}$ (seen most clearly in the \ohtop spectrum), arising from the near and far side crossings of the Sagittarius spiral-arm. Unsurprisingly, there is no \arhp absorption at 40~km~s$^{-1}$ but \arhp shows red-shifted absorption tracing layers of infalling material associated with the molecular cloud. Infalling signatures towards this hot molecular cloud have previously been studied by \citet{furuya2011infall}, whose observations of the $J=3\rightarrow2$ transitions of $^{13}$CO and $^{18}$CO show inverse P-Cygni profiles or red-shifted absorption alongside a blue-shifted emission component. The \arhp spectrum also shows a narrow absorption feature at 80~km~s$^{-1}$ possibly from the edge of the spiral arm. The strongest absorption feature seen in the \arhp spectrum lies close to 113~km~s$^{-1}$ from the Scutum-Centaurius arm, which is notably a weaker component in the o-H$_{2}$O$^{+}$ and H{\small I} spectra.  

\subsubsection*{AG31.412+00.307} 
Located on the Scutum-Centaurus arm at a distance of 5.55~kpc \citep{moises2011spectrophotometric}, the line of sight towards this young-stellar object, AG31.412+00.307, crosses the Sagittarius arm at velocities roughly between 30 and 55~km~s$^{-1}$ and the Perseus arm at velocities \textless$20$~km~s$^{-1}$. The \arhp absorption profile mimics that of H{\small I} with only minor offsets in the peak velocities ($<1~$km~s$^{-1}$) of the absorption components. Moreover, the \arhp absorption seen close to the source's intrinsic velocity at 98~km~s$^{-1}$ traces infalling material. Unfortunately, because of the poor signal-to-noise ratio of the \ohp spectrum it is very difficult to make an accurate component-wise comparison.

\subsubsection*{AG330.954-00.182} 
Located at a distance of 5.3~kpc \citep{wienen2015atlasgal}, AG330.954$-$00.182 lies in the Norma spiral arm and is a bright ATLASGAL $870~\mu$m source in the fourth quadrant. It has an absorption profile that extends from $-130$ to 20~km~s$^{-1}$. The gas between $-60$ and 20~km~s$^{-1}$ traces the near and far-side crossings of both the Scutum-Centaurius as well as the Sagittarius spiral-arms. The \arhp spectrum shows absorption at $-115$~km~s$^{-1}$, a feature which is not present in the other species. There is weak \arhp absorption close to 90~km~s$^{-1}$, which maybe associated with the molecular cloud itself. The line profile resembles a single absorption component as seen in the H{\small I} spectrum and not two components of comparable depths as seen in the \ohp and o-H$_{2}$O$^{+}$ spectra.

\subsubsection*{AG332.826-00.549} 
The sight line towards AG332.826$-$00.549, which has a near kinematic distance of 3.6~kpc \citep{moises2011spectrophotometric}, primarily traces the Scutum-Centaurius and Sagittarius spiral-arms. Overall, the observed profiles of the different species in terms of their shape, are in concordance with each other. Akin to the ArH$^+$ spectrum, the spectra of the other ionised species in our study, p-, and o-H$_{2}$O$^+$, and OH$^+$, all show their strongest absorption features near $-40$~km~s$^{-1}$ and red-shifted absorption tracing infalling material similar to G19.609$-$00.234. This feature whilst present in the H{\small I} spectrum is much weaker in comparison to the absorption at velocities corresponding to the molecular cloud which is expected, but is weaker, also in comparison to other LOS features and is not prominent in the CH spectrum.

\subsubsection*{AG337.704-00.054} 
At a distance of ${\sim}12.3$~kpc \citep{green2011distances}, AG337.704$-$00.054 is one of the most distant Galactic sources in our sample. Absorption features cover a range of velocities from -$135$ to ${\sim}$ +20~km~s$^{-1}$. The LOS absorption at negative velocities between -140 and -60~km~s$^{-1}$ arise from the 3~kpc and the Norma spiral arms. The \arhp spectrum shows contiguous absorption over the entire velocity range cited, unlike the spectra of the other molecular ions and even H{\small I}.

\subsubsection*{AG351.581-00.352} 
AG351.581$-$00.352 is a bright and compact H{\small II} region at a distance of  6.8~kpc \citep{green2011distances} that is located between the 3~kpc arm and the beginning of the Norma arm. Notably, \arhp has its strongest absorption peak at -27~km~s$^{-1}$, arising from the Scutum arm while \ohp and \ohtop both peak close to 50~km~s$^{-1}$, tracing gas from the Norma arm. At the current rms noise level of the \arhp spectrum, it is difficult to %decipher 
ascertain the presence of absorption components at positive velocities that are clearly observed %similar 
in the spectra of the %to those of the 
other molecular ions, H{\small I} and CH.

    
%\begin{figure*}
%    \centering 
%    \includegraphics[width=7.5cm]{NGC253_alles_spectra.pdf}\quad
%    \includegraphics[width=7.5cm]{NGC4945_alles_spectra.pdf}
%    \caption{From top to bottom: spectra of the \arhp 617~GHz, \phtop 607~GHz, \ohtop 1115~GHz, and \ohp 971~GHz lines towards NGC~253 (left) and NGC~4945 (right). The individual fit components are displayed by the dotted red line and the solid red line is their sum. The dashed blue lines mark the systemic velocity of the sources. In dark blue we indicate the contamination from CH$_{3}$OH, E $\varv t= 0-2$ ($11_{(2,9)}-10_{(1,10)}$) at 959.345~GHz.}
%    \label{fig:extragal_all_spectra}
%\end{figure*}

\subsection{Spectral fitting} \label{subsec:spec_fit}
The optical depth, $\tau$ is determined from the radiative transfer equation for the specific case of absorption spectroscopy ($T_{\text{l}} = T_{\text{c}}\text{e}^{-\tau}$), given knowledge of the continuum background temperature, $T_\text{c}$. The optical depth profile is then analysed using the Wiener filter fitting technique as described in \citet{jacob2019fingerprinting}. This two step fitting procedure first fits the spectrum based on the Wiener filter, which minimises the mean square error between the model and observations and then deconvolves the hfs from the observed spectrum, if any. When fitting lines that do not exhibit hfs, the procedure simply assumes that there is only a single hfs component whose frequency corresponds to that of the fine-structure transition itself. In addition to the observed spectrum and the spectroscopic parameters of the line to be fit, the only other input parameter required by the Wiener filter technique is the spectral noise, which is assumed to be independent of the signal. Furthermore, the technique also assumes that the lines are optically thin and under conditions of local thermodynamic equilibrium (LTE). %As discussed above, where possible we have modelled the LOS contamination features using the CLASS software. Despite modelling the contaminating lines, there is a great deal of uncertainty in the \phtop spectra particularly from the H$^{13}$CO$^{+}$ emission that contaminates the observed absorption. To avoid the described problems, in our analysis we only consider those components which are in velocity intervals that are free from significant contamination.

\subsection{Column density} \label{subsec:column_densities}

From the resultant optical depth profile, we determine the column densities per velocity interval, d$N$/d$\upsilon$, using
\begin{equation}
    \left( \frac{\text{d}N}{\text{d}\upsilon}\right)_{i} = \frac{8\pi\nu^3_{i} }{c^3 } \frac{Q(T_{\text{ex}})}{g_{\text{u}} A_{\text{E}}} \text{e}^{E_{\text{u}}/T_{\text{ex}}} \left[ \text{exp} \left(\frac{h\nu_{i}}{k_{\text{B}}T_{\text{ex}}}\right) - 1 \right]^{-1} (\tau_{i})
    \label{eqn:col_dens}
\end{equation}
for each velocity channel, $i$. For a given hfs transition, all of the above spectroscopic terms remain constant, except for the partition function, $Q$, which itself is a function of the rotation temperature, $T_{\text{rot}}$. Under conditions of LTE $T_{\text{rot}}$ is equal to the excitation temperature, $T_{\text{ex}}$. We further assume that almost all the molecules occupy the ground rotational state where the excitation temperature, $T_{\text{ex}}$, is determined by the cosmic microwave background radiation, $T_{\text{CMB}}$, of  2.73~K. This is a valid assumption for the conditions that prevail in diffuse regions where the gas densities, (${n(\text{H}) \leq  100~\text{cm}^{-3}}$), are sufficiently low enough that collisional excitation becomes unimportant. However, this assumption may no longer be valid for those features that arise from the dense envelopes of the molecular clouds in which collisions can dominate, resulting in higher excitation temperatures. Therefore, the column densities derived by integrating Eq.~\ref{eqn:col_dens} over velocity intervals associated with the molecular cloud components, only represent lower limits to the column density values. Combining all our sources, we are broadly able to identify a total of 38 distinct absorption features; however, we could only assign 33, 15, and 35 of these features to ArH$^+$, p-, and \ohtop absorption, respectively, due to the blending of different species present along the different LOS. The column densities that we have derived for each species and towards each sight line are summarised in Table~\ref{tab:column_densities}. The \arhp column densities are found to vary between ${{\sim} 6\times10^{11}}$ and $10^{13}$~cm$^{-2}$ across inter-arm and spiral-arm clouds along the different sight lines. The derived \phtop column density values also show a similar variation, towards our sample of sources.% and from 5-$12\times 10^{13}$~cm$^{-2}$ towards the extra-Galactic sight lines.  

The accuracy of the reported column density values is mainly limited by uncertainties present in the continuum level and not the absolute temperature calibration. The errors in the WF analysis are computed using  Bayesian methods to sample a posterior distribution of the optical depths. In this approach, we sampled 5000 artificial spectra each generated by iteratively adding a pseudo-random noise contribution to the absorption spectra, of each of the different species prior to applying the WF deconvolution. The standard deviation in the additive noise is fixed to be the same as that of the line free part of the spectrum. The deconvolved optical depths and the subsequently derived column densities per velocity interval, sample a point in the channel-wise distribution of the column densities, across all the spectra. Since, the posterior of the optical depths is not normally distributed, we describe the profiles of the distributions by using the median and the inter-quartile range. This removes any bias introduced on the mean by the asymmetric nature of the distributions. We then determined the sample mean and standard deviation following \citet{Ivezic2019}. The deviations from the newly derived mean are then used to determine the errors.
%\begin{table*}
\begin{sidewaystable*}
\sisetup{table-format=<2.4}
\sisetup{table-format=>2.4}
\sisetup{table-text-alignment=right }
\caption{Synopsis of derived quantities.}
\begin{tabular}{llr rcrcr rrr rl}

\hline 
\hline 
 Source & $\upsilon_{\text{LSR}}$ Range & $R_{\text{GAL}}$\tablefootmark{a} & \multicolumn{1}{c}{$N(\text{ArH}^{+})$} &  \multicolumn{1}{c}{$N(\text{OH}^{+})$} &\multicolumn{1}{c}{ $N(\text{p-H}_{2}{\text{O}}^{+})$}
 & $N(\text{o-H}_{2}{\text{O}}^{+})$
 & \multicolumn{1}{c}{H$_{2}$O$^{+}$}
 & \multicolumn{1}{c}{$N(\text{H{\small I}})$} & %\multicolumn{1}{c}{$N(\text{H}_{2})$} 
  \multicolumn{2}{c}{$f_{\text{H}_{2}}$} 
 & \multicolumn{1}{c}{$\zeta_\text{p}(\text{H})$} 
 \\
 & \multicolumn{1}{c}{[km~s$^{-1}$]} & [kpc] & [10$^{12}$ cm$^{-2}$] & [10$^{14}$ cm$^{-2}$] & \multicolumn{1}{c}{[10$^{13}$ cm$^{-2}$]} & [10$^{13}$ cm$^{-2}$] &  \multicolumn{1}{c}{OPR} & \multicolumn{1}{c}{[10$^{21}$ cm$^{-2}$]} & \multicolumn{1}{c}{ OH$^{+}$\tablefootmark{c}} &
  \multicolumn{1}{c}{CH\tablefootmark{d}}
 & \multicolumn{1}{c}{[$10^{-16}~$s$^{-1}$]} %& \multicolumn{1}{c}{[$10^{-14}~$s$^{-1}$]}
 \\
\hline 
 AG10.472 &  ($-35, 0$) & 10.82 & $1.94^{+0.04}_{-0.06}$ & $2.12^{+0.11}_{-0.16}$ & \multicolumn{1}{c}{---} & $1.71^{+0.22}_{-0.34}$ & \multicolumn{1}{c}{---} & $4.62^{+1.34}_{-1.96}$& \multicolumn{1}{c}{$<0.019$} &   $0.77^{+0.04}_{-0.04}$ & \multicolumn{1}{c}{$<4.52$}\\%& \tablefootmark{c}$7.92^{+0.58}_{-0.81}$\\
 +00.027 & (0, 48) & 5.56 & $10.27^{+0.17}_{-0.21}$ & $3.60^{+0.28}_{-0.33}$ & \multicolumn{1}{c}{$>1.05$} & $3.42^{+0.49}_{-0.62}$ & \multicolumn{1}{c}{$<3.26$} &  $8.60^{+2.09}_{-3.66}$ & \multicolumn{1}{c}{$>0.031$} & $0.87^{+0.04}_{-0.05}$ & \multicolumn{1}{c}{$>4.52$}\\
 %& $0.042^{+0.005}_{-0.005}$ & 
 
 & (48, 77)\tablefootmark{b} & 3.50 & $1.51^{+0.04}_{-0.05}$ & $1.07^{+0.09}_{-0.12}$ & \multicolumn{1}{c}{---} & $0.22^{+0.03}_{-0.05}$ & \multicolumn{1}{c}{---} &   $2.64^{+0.84}_{-1.11}$ & \multicolumn{1}{c}{$>0.005$} &  $0.88^{+0.02}_{-0.03}$ & \multicolumn{1}{c}{$>2.48$}
 %& $0.034^{+0.009}_{-0.004}$
 \\
 & (77, 105) & 3.12 & $6.48^{+0.19}_{-0.23}$ & $1.23^{+0.15}_{-0.19}$ & \multicolumn{1}{c}{---} & \multicolumn{1}{c}{---} & \multicolumn{1}{c}{---} &   $0.80^{+0.26}_{-0.33}$ 
 & \multicolumn{1}{c}{---} &  $0.89^{+0.06}_{-0.06}$ & \multicolumn{1}{c}{---}\\
 & (105, 138) & 2.38 & $4.46^{+0.10}_{-0.14}$ & $1.62^{+0.14}_{-0.18}$ & \multicolumn{1}{c}{---} &\multicolumn{1}{c}{---} & \multicolumn{1}{c}{---} &  $0.64^{+0.22}_{-0.27}$ 
 & \multicolumn{1}{c}{---}  & $0.92^{+0.05}_{-0.06}$ & \multicolumn{1}{c}{---}  \\
 & (138, 190) & 1.48 & \multicolumn{1}{c}{---}%$4.30^{+0.08}_{-0.08}$
 & $1.46^{+0.10}_{-0.13}$ & \multicolumn{1}{c}{---} & $0.34^{+0.03}_{-0.04}$ & \multicolumn{1}{c}{---} &  $0.76^{+0.32}_{-0.27}$ & $0.005^{+0.001}_{-0.001}$&  $0.94^{+0.06}_{-0.06}$ & $12.03^{+1.30}_{-1.33}$ \\
AG19.609 & ($-12, 16$) & 8.04 & $1.19^{+0.03}_{-0.04}$ & \multicolumn{1}{c}{---} & $0.31^{+0.01}_{-0.02}$  & \multicolumn{1}{c}{---} & \multicolumn{1}{c}{---} &  $5.38^{+0.46}_{-0.68}$ & \multicolumn{1}{c}{---} &  \multicolumn{1}{c}{---}& \multicolumn{1}{c}{---}
\\
$-$00.234 & (16, 35) & 6.75 & $0.91^{+0.03}_{-0.05}$ & \multicolumn{1}{c}{---} & $0.38^{+0.07}_{-0.08}$&  $0.42^{+0.10}_{-0.11}$ & $1.10^{+0.33}_{-0.37}$ &  $5.51^{+0.37}_{-0.76}$ & \multicolumn{1}{c}{---} & \multicolumn{1}{c}{---}  &  \multicolumn{1}{c}{---} 
\\
 & (35, 60)\tablefootmark{b} & 4.50 & $4.44^{+0.14}_{-0.18}$ & \multicolumn{1}{c}{---} & \multicolumn{1}{c}{---} & $1.12^{+0.22}_{-0.22}$ & \multicolumn{1}{c}{---} &  $16.99^{+27.4}_{-6.38}$ & \multicolumn{1}{c}{---} &  \multicolumn{1}{c}{---} & \multicolumn{1}{c}{---}
 \\
 & (60, 87) & 4.34 & $7.82^{+0.24}_{-0.28}$ & \multicolumn{1}{c}{---} & \multicolumn{1}{c}{---} & $1.23^{+0.19}_{-0.17}$ & \multicolumn{1}{c}{---} &  $4.09^{+0.34}_{-0.52}$ & \multicolumn{1}{c}{---} & \multicolumn{1}{c}{---}& \multicolumn{1}{c}{---}
 \\
 & (87, 105) & 3.32 & $6.95^{+0.33}_{-0.36}$& \multicolumn{1}{c}{---} & \multicolumn{1}{c}{---} & $0.38^{+0.07}_{-0.09}$  & \multicolumn{1}{c}{---}& $1.36^{+0.13}_{-0.16}$ & \multicolumn{1}{c}{---}& \multicolumn{1}{c}{---} & \multicolumn{1}{c}{---}
 \\
 & (105, 115) & 3.01 & $4.43^{+0.36}_{-0.42}$ & \multicolumn{1}{c}{---} & \multicolumn{1}{c}{---} & $0.13^{+0.08}_{-0.09}$ & \multicolumn{1}{c}{---} &  $0.82^{+0.08}_{-0.09}$ &  \multicolumn{1}{c}{---} & \multicolumn{1}{c}{---} & \multicolumn{1}{c}{---} 
 \\
 & (115, 135) & 2.73 & \multicolumn{1}{c}{---} & \multicolumn{1}{c}{---} & $0.18^{+0.01}_{-0.01}$& $0.29^{+0.07}_{-0.10}$ & $1.61^{+0.40}_{-0.56}$&   $1.33^{+0.13}_{-0.15}$ & \multicolumn{1}{c}{---}  & \multicolumn{1}{c}{---} & \multicolumn{1}{c}{---}
 \\
 %& \\
AG31.412 & ($-8, 31$)\tablefootmark{*} & 7.05 & $2.41^{+0.05}_{-0.06}$ & $0.60^{+0.05}_{-0.07}$ & $0.16^{+0.01}_{-0.02}$ & $0.21^{+0.04}_{-0.04}$ & $1.31^{+0.26}_{-0.30}$ &  $11.98^{+3.23}_{-6.83}$
& $0.013^{+0.003}_{-0.003}$ &  \multicolumn{1}{c}{---} & $0.37^{+0.00}_{-0.03}$ %& $0.12^{+0.03}_{-0.02}$
\\ 
+00.307& (31, 46)\tablefootmark{*} &  6.39 & $0.68^{+0.03}_{-0.05}$ & $0.12^{+0.03}_{-0.05}$ & \multicolumn{1}{c}{---} & \multicolumn{1}{c}{$<0.05$} & \multicolumn{1}{c}{---} & $2.71^{+0.93}_{-1.30}$
& $0.031^{+0.026}_{-0.030}$  &  \multicolumn{1}{c}{---} & $0.18^{+0.07}_{-0.06}$ %& $1.18^{+0.06}_{-0.07}$
\\
& (46, 65)\tablefootmark{*} & 5.49 & $0.86^{+0.03}_{-0.05}$ & $0.10^{+0.04}_{-0.06}$ & \multicolumn{1}{c}{$>0.08$} & $ 0.09^{+0.02}_{-0.03}$ & \multicolumn{1}{c}{$<1.13$} & $2.78^{+0.92}_{-1.35}$
& $0.100^{+0.010}_{-0.008}$  & \multicolumn{1}{c}{---} & $0.29^{+0.08}_{-0.07}$ %& $1.30^{+0.07}_{-0.07}$
\\
& (65, 88)\tablefootmark{*} & 4.71 & $2.27^{+0.07}_{-0.09}$ & $0.19^{+0.01}_{-0.03}$ & $0.24^{+0.02}_{-0.02}$ & $0.25^{+0.04}_{-0.06}$ &$1.04^{+0.17}_{-0.26}$ &  $8.11^{+1.01}_{-4.26}$
& $0.108^{+0.020}_{-0.015}$  &  \multicolumn{1}{c}{---} & $0.29^{+0.03}_{-0.02}$ %& $0.99^{+0.02}_{-0.04}$
\\
& (88, 108)\tablefootmark{b,*} & 4.64 & $0.68^{+0.02}_{-0.03}$
& $0.09^{+0.01}_{-0.02}$ & \multicolumn{1}{c}{---} & $0.63^{+0.13}_{-0.18}$ 
& \multicolumn{1}{c}{---} & $14.69^{+11.7}_{-9.96}$
& \multicolumn{1}{c}{$>0.776$}  & \multicolumn{1}{c}{---} & \multicolumn{1}{c}{$>1.08$} %& $1.11^{+0.06}_{-0.04}$
\\
AG330.954 & ($-130, -74$)\tablefootmark{b} & 4.43 & \multicolumn{1}{c}{$>1.24$} & $1.20^{+0.05}_{-0.07}$ & \multicolumn{1}{c}{---} & $0.73^{+0.06}_{-0.09}$ & \multicolumn{1}{c}{---}&  $7.75^{+1.71}_{-3.48}$ %& \tablefootmark{c}$31.47^{+1.23}_{-2.13}$
& $0.014^{+0.004}_{-0.004}$  & $0.30^{+0.08}_{-0.06}$ & $1.34^{+0.01}_{-0.02}$ %& $0.30^{+0.02}_{-0.01}$
\\ 
$-$00.182& ($-74, -57$) & 4.92 & $1.44^{+0.06}_{-0.08}$ & $0.42^{+0.09}_{-0.13}$ & \multicolumn{1}{c}{---} & $0.45^{+0.12}_{-0.16}$ & \multicolumn{1}{c}{---} & $3.01^{+0.52}_{-1.33}$ %& $19.60^{+2.81}_{-4.10}$
& \multicolumn{1}{c}{$>0.026$}  & $0.43^{
+0.03}_{-0.03}$ & \multicolumn{1}{c}{$>1.62$} %& $0.37^{+0.23}_{-0.12}$
\\
& ($-57, -32$) & 5.78 & $2.85^{+0.09}_{-0.11}$ & $0.70^{+0.10}_{-0.13}$ & $0.16^{+0.00}_{-0.01}$ & $0.51^{+0.07}_{-0.08}$ & $3.18^{+0.44}_{-0.54}$&  $3.66^{+1.02}_{-1.57}$ %& $23.69^{+2.11}_{-3.86}$
& $0.023^{+0.004}_{-0.004}$  &  $0.34^{+0.02}_{-0.02}$ & $1.78^{+0.84}_{-1.01}$ %& $0.85^{+0.18}_{-0.14}$
\\
& ($-32, -10$) & 7.03 & $1.13^{+0.04}_{-0.05}$ & $0.05^{+0.01}_{-0.02}$ & $0.06^{+0.01}_{-0.01}$ & $0.10^{+0.02}_{-0.03}$ & $1.67^{+0.43}_{-0.57}$& $2.43^{+0.91}_{-1.02}$  %& $1.36^{+0.14}_{-0.23}$
& $0.103^{+0.004}_{-0.004}$  & $0.04^{+0.01}_{-0.01}$ & $0.39^{+0.03}_{-0.02}$ %& $1.02^{+0.09}_{-0.10}$
\\
& ($-10, 20$) & 8.54 & \multicolumn{1}{c}{---}%$0.46^{+0.01}_{-0.01}$ 
& $0.06^{+0.01}_{-0.02}$ & \multicolumn{1}{c}{---} & $0.13^{+0.02}_{-0.03}$ & \multicolumn{1}{c}{---} & $2.63^{+0.95}_{-1.11}$ %& $2.13^{+0.16}_{-0.25}$
& \multicolumn{1}{c}{$>0.058$}  & $0.14^{+0.06}_{-0.04}$ & $0.45^{+0.11}_{-0.04}$ %& $0.18^{+0.02}_{-0.02}$
\\
%& \\
AG332.826 & ($-85, -33$)\tablefootmark{b}& 5.30 & \multicolumn{1}{c}{---}%$2.65^{+0.04}_{-0.05}$ 
& $1.18^{+0.04}_{-0.07}$ & \multicolumn{1}{c}{---} & $1.21^{+0.09}_{-0.16}$ & \multicolumn{1}{c}{---}& $9.19^{+3.80}_{-5.06}$ %& \tablefootmark{c}$43.99^{+1.87}_{-3.12}$
& $0.025^{+0.006}_{-0.007}$  & $0.35^{+0.07}_{-0.05}$ & $1.45^{+0.10}_{-0.07}$ %& $1.03^{+0.02}_{-0.02}$
\\
$-$00.549& ($-33, -12$) & 7.02 &  $1.10^{+0.04}_{-0.05}$ & $0.42^{+0.07}_{-0.09}$ & \multicolumn{1}{c}{---} & $0.41^{+0.10}_{-0.15}$ & \multicolumn{1}{c}{---} & $2.26^{+0.30}_{-0.79}$ %& $4.55^{+0.48}_{-0.85}$ 
& \multicolumn{1}{c}{$>0.24$}  & $0.16^{+0.04}_{-0.03}$ & \multicolumn{1}{c}{$>2.04$} %& $0.44^{+0.27}_{-0.12}$
\\
& ($-12, 20$) & 8.67 & $1.26^{+0.03}_{-0.04}$ & $0.15^{+0.01}_{-0.02}$ & $0.13^{+0.00}_{-0.01}$ & $0.25^{+0.05}_{-0.06}$ & $1.92^{+0.38}_{-0.48}$ & $2.81^{+0.15}_{-0.52}$ %& $1.67^{+0.13}_{-0.17}$
& $0.073^{+0.014}_{-0.010}$  & $0.10^{+0.01}_{-0.01}$& $0.88^{+0.07}_{-0.07}$
\\
%& \\
AG337.704 & ($-145, -110$) & 3.12 & $3.53^{+0.08}_{-0.11}$ & $0.10^{+0.01}_{-0.01}$ & $0.25^{+0.09}_{-0.10}$ & $0.60^{+0.08}_{-0.11}$ & $2.41^{+0.92}_{-1.06}$ & $2.43^{+0.87}_{-2.13}$ & $1.080^{+0.070}_{-0.064}$ &   \multicolumn{1}{c}{---} & $4.06^{+0.41}_{-0.22}$ %& $20.38^{+0.19}_{-0.10}$
\\
$-$00.054 & ($-110, -97$) & 3.25 & $2.19^{+0.15}_{-0.18}$ & $0.07^{+0.01}_{-0.03}$ & \multicolumn{1}{c}{---}& $0.32^{+0.13}_{-0.19}$ & \multicolumn{1}{c}{---} & $1.34^{+0.47}_{-0.12}$ &  $0.190^{+0.030}_{-0.031}$ &   \multicolumn{1}{c}{---} & $2.67^{+0.42}_{-0.28}$ %& $10.60^{+0.59}_{-2.78}$
\\
& ($-97, -80$) & 3.73 & $2.78^{+0.13}_{-0.16}$ & $0.07^{+0.02}_{-0.02}$ & $0.14^{+0.01}_{-0.01}$ & $0.63^{+0.21}_{-0.22}$ & $4.50^{+1.53}_{-1.60}$ &  $1.89^{+0.00}_{-0.69}$ & \multicolumn{1}{c}{---\tablefootmark{e}} &  \multicolumn{1}{c}{---} & \multicolumn{1}{c}{---} %& $30.70^{-3.55}_{+6.36}$
\\  
& ($-80, -65$) & 4.43 & $2.60^{+0.13}_{-0.17}$ & $0.08^{+0.01}_{-0.02}$ & $0.11^{+0.01}_{-0.01}$ & $0.33^{+0.13}_{-0.14}$ & $3.00^{+1.21}_{-1.30}$&  $2.42^{+0.55}_{-0.85}$ & $0.255^{+0.030}_{-0.030}$ &  \multicolumn{1}{c}{---} &  $1.56^{+0.21}_{-0.15}$ %& $27.23^{+0.29}_{-0.58}$
\\
& ($-65, -45$)\tablefootmark{b} & 5.17 & $3.03^{+0.12}_{-0.14}$ & $0.18^{+0.02}_{-0.03}$ & \multicolumn{1}{c}{---}& $0.79^{+0.15}_{-0.16}$ & \multicolumn{1}{c}{---} &  $3.88^{+1.31}_{-2.00}$ & $0.162^{+0.030}_{-0.040}$ &   \multicolumn{1}{c}{---} & $2.18^{+0.15}_{-0.41}$ %& $11.52^{+0.17}_{-0.13}$
\\
& ($-45, -25$) & 5.29 & $2.31^{+0.09}_{-0.11}$ & $0.16^{+0.02}_{-0.04}$ & \multicolumn{1}{c}{---} & $0.70^{+0.16}_{-0.18}$ & \multicolumn{1}{c}{---} & $3.12^{+1.04}_{-1.57}$ & $0.089^{+0.032}_{-0.029}$ &   \multicolumn{1}{c}{---} & $2.14^{+0.38}_{-0.32}$ %& $18.86^{+0.16}_{-0.13}$
\\
& ($-25, -13$) & 6.91 & $0.71^{+0.04}_{-0.06}$ & $0.06^{+0.01}_{-0.02}$ & \multicolumn{1}{c}{---} & $0.10^{+0.05}_{-0.08}$ & \multicolumn{1}{c}{---} &  $2.01^{+0.56}_{-1.20}$ & \multicolumn{1}{c}{$>0.048$} &   \multicolumn{1}{c}{---} & $0.50^{+0.08}_{-0.11}$ %& $0.79^{+0.16}_{-0.09}$
\\
& ($-13, 7$) & 8.62 & $1.56^{+0.06}_{-0.08}$ & $0.12^{+0.03}_{-0.04}$ & \multicolumn{1}{c}{---} & $0.12^{+0.02}_{-0.04}$ & \multicolumn{1}{c}{---} & $2.44^{+0.91}_{-1.11}$ & \multicolumn{1}{c}{$>0.026$}&  \multicolumn{1}{c}{---} & $0.57^{+0.02}_{-0.01}$ %& $0.35^{+0.13}_{-0.14}$
\\
%& \\
AG351.581 & ($-120, -70$)\tablefootmark{b} & 1.66 & $5.56^{+0.08}_{-0.11}$ & $0.80^{+0.03}_{-0.04}$ & \multicolumn{1}{c}{---} &   $0.46^{+0.05}_{-0.06}$ & \multicolumn{1}{c}{---} &  $2.20^{+0.20}_{-0.98}$ %& \tablefootmark{c}$24.83^{+1.02}_{-2.00}$
& $0.087^{+0.003}_{-0.004}$  & $0.57^{+0.04}_{-0.04}$&  $1.98^{+0.04}_{-0.06}$ %& $3.28^{+0.84}_{-0.21}$
\\
$-$00.352 & ($-70, -39$) & 3.18 & $4.26^{+0.11}_{-0.13}$ & $0.92^{+0.08}_{-0.13}$ &  \multicolumn{1}{c}{---} & $0.87^{+0.15}_{-0.20}$ &
\multicolumn{1}{c}{---}&
$1.34^{+0.52}_{-0.56}$%& $6.25^{+0.51}_{-0.71}$ 
& \multicolumn{1}{c}{$>0.554$}  & $0.47^{+0.05}_{-0.03}$ &$12.85^{+1.44}_{-1.12}$ %& $8.58^{+0.41}_{-0.46}$
\\
& ($-39, -7$) & 4.50 & $6.40^{+0.15}_{-0.20}$ & $0.80^{+0.08}_{-0.09}$ & $0.43^{+0.10}_{-0.15}$ & $0.61^{+0.11}_{-0.14}$ &
$1.41^{+0.41}_{-0.60}$ &
$4.85^{+1.87}_{-2.03}$ %& $21.40^{+1.56}_{-2.41}$
& $0.314^{+0.005}_{-0.008}$  & $0.45^{+0.05}_{-0.04}$ & $1.23^{+0.22}_{-0.25}$ %& $2.14^{+0.16}_{-0.19}$
\\
& ($-7, 25$) & 9.74 & \multicolumn{1}{c}{---}%$2.38^{+0.06}_{-0.07}$ 
& $0.44^{+0.04}_{-0.07}$ & $0.24^{+0.05}_{-0.09}$ & $0.36^{+0.05}_{-0.08}$ &
$1.50^{+0.37}_{-0.61}$ &
$4.78^{+1.84}_{-2.00}$ %&  $10.15^{+0.55}_{-1.33}$
& $0.027^{+0.010}_{-0.004}$ &  $0.28^{+0.03}_{-0.03}$& $1.02^{+0.12}_{-0.13}$ %& $0.73^{+0.06}_{-0.07}$
\\
%&~~~~~~~~~~~~~~~~~~~~~~~~~~~~ \\
%NGC~253 & (25, 275) & --- & $10.37^{+0.06}_{-0.07}$ & $9.69^{+0.13}_{-0.17}$ & $5.14^{+0.02}_{-0.02}$ & $8.16^{+0.15}_{-0.17}$\\
%& \\
%NGC~4945 & (425, 585) & ---& 10.37$^{+0.06}_{-0.86}$ & $2.03^{+0.08}_{-0.11}$ & $7.80^{+0.04}_{-0.05}$ & $20.75^{+1.10}_{-1.40}$ &  \\
%& (585, 735) & --- & 13.75$^{+0.08}_{-0.10}$ & $3.34^{+0.16}_{-0.19}$ &  $11.99^{+0.07}_{-0.09}$ & $43.21^{2.72}_{-3.11}$\\
%& \\
         \hline 
    \end{tabular}
\tablefoot{\tablefoottext{a}{The galactocentric distance for each velocity interval is computed using, $R_{\text{GAL}}$ = $R_{0}\frac{\Theta(R_{\text{GAL}})\text{sin}(l)\text{cos}(b)}{\upsilon_{\text{lsr}} + \Theta_{0}\text{sin}(l)\text{cos}(b)}$, with $R_{0} = 8.15$~kpc, $\Theta_{0}$ = 247~km~s$^{-1}$ \citep{Reid2019} and assuming a flat Galactic rotation curve, i.e, $\Theta(R_{\text{GAL}}) = \Theta_{0}$ for $R_{\text{GAL}}>5~$kpc. We used the BeSSeL distance calculator, to both cross-check and determine lower $R_{\text{GAL}}$ values. The distances are computed at the mean velocity within each velocity interval.} \tablefoottext{b}{Indicates the velocity intervals corresponding to the molecular cloud.}\tablefoottext{c,d}{represent $f_{\text{H}_{2}}$ values derived using Eqs.~\ref{eqn:molecular_fraction_oh} and ~\ref{eqn:mole_frac}, respectively. Where $\text{H}_{2}$ column densities for the latter are determined from those of CH \citep{jacob2019fingerprinting} using [CH]/[H$_{2}$] = $3.5^{+2.1}_{-1.4}\times10^{-8}$ \citep{sheffer2008ultraviolet}.\tablefoottext{e}{Faces a singularity as the denominator approaches 0.}\tablefoottext{*}{Represents velocity intervals with low signal-to-noise ratios, for which molecular hydrogen fractions are calculated using the upper limits of the derived $N(\text{OH}^+)$ values.}}
}
    \label{tab:column_densities}
\end{sidewaystable*}
%\end{table*}

\subsection{Cosmic-ray ionisation rate} \label{subsec:cosmic_ray_ionisation}
Ionisation by cosmic-rays is the primary ionisation mechanism that drives ion-molecular reactions within diffuse atomic clouds. As discussed in Sect.~\ref{sec:intro}, the formation of \arhp is initiated by the ionisation of argon atoms by cosmic-rays. Therefore, knowledge of the primary ionisation rate per hydrogen atom, $\zeta_\text{p}(\text{H})$, is vital for our understanding of the ensuing chemistry. In this section, we investigate the cosmic-ray ionisation rates towards our sample of sources, which can be determined by analysing the steady-state ion-molecular chemistry of \ohp and \ohtop. 

In the diffuse ISM, both \ohp and \ohtop are mainly destroyed either via reactions with H$_{2}$ or recombination with an $e^{-}$. The relevant reactions are summarised below, alongside their reaction rates which are taken from \citet{Tran2018}, \citet{Mitchell1990}, and the KIDA database\footnote{ \url{http://kida.obs.u-bordeaux1.fr}} \citep{wakelam2012kinetic}:
\begin{align*}
    \text{OH}^+  &+  \text{H}_{2} \rightarrow \text{H}_{2}\text{O}^{+} + \text{H}; \quad k_{1} = 1.0\times 10^{-9}~\text{cm}^3\text{s}^{-1}\\
    \text{OH}^+  &+ e^{-} \rightarrow \text{H} + \text{O}; \quad \quad \,\,\,\,\, k_{2} = 6.6\times 10^{-8}~\text{cm}^3\text{s}^{-1}\\
    \text{H}_{2}\text{O}^{+} &+ \text{H}_{2}  \rightarrow \text{H}_{3}\text{O}^{+} + \text{H}; \quad k_{3} = 9.7\times 10^{-10}~\text{cm}^{3}\text{s}^{-1} 
\end{align*}
\vspace{-0.7cm}
\begin{align*}
\begin{rightcases}
   \text{H}_{2}\text{O}^{+} + {e}^{-} & \rightarrow \text{O} + \text{H}_{2};  \\
    & \rightarrow \text{H} + \text{OH}; \\
    &\rightarrow \text{H} + \text{O} + \text{H}; 
\end{rightcases}
    \,\, k_{4} = 7.4\times 10^{-7}~\text{cm}^3\text{s}^{-1} \, .
    \end{align*}
Since our analysis does not require the exact branching information between the products formed from the recombination of H$_{2}$O$^{+}$ with an $e^-$, $k_4$ cites the total reaction rate. The %reaction 
rates for reactions with H$_{2}$ are independent of the gas temperature, $T$, while, the recombination reactions have a weak dependence on it with $ k \propto T^{-0.49}$. The reaction rates for the latter were computed at a temperature of 100~K corresponding to the typical H{\small I} spin temperature along the LOS towards the sources in our sample. Moreover, the transitions that we study mainly probe warm diffuse gas with typical values of $T$ between 30 and 100~K \citep{Snow2006}. Decreasing the temperature from 100~K to 30~K would result in an 80\%~ increase in the reaction rates. %Since this increase in reaction rates does not lead to %reflect
%large variations in our results, we do not consider the uncertainties in the temperature to be significant.

Following the steady-state analysis presented in \citet{indriolo2012chemical}, the cosmic-ray ionisation rate may be written as
\begin{equation}
\epsilon \zeta_\text{p}(\text{H}) = \frac{N(\text{OH}^+)}{N(\text{H})}n_{\text{H}}\left[ \frac{f_{\text{H}_{2}}}{2}k_{1} + x_\text{e}k_{2} \right] \, ,   
\label{eqn:CRIR}
\end{equation}
where, $k_{1}$ and $k_{2}$ are the rates of the reactions between \ohp and H$_{2}$, and \ohp and a free electron. As introduced in \citet{neufeld2010herschel}, $\epsilon$ is an efficiency factor that represents the efficiency with which ionised H forms OH$^{+}$. The efficiency parameter was predicted by chemical models \citep{hollenbach2012chemistry} to vary between 5\% and 20\% as it is dependent on the physical properties of the cloud. Observationally, this efficiency has been determined only towards the W51 region by \citet{indriolo2012chemical} to be $7\pm4$\%, using the spectroscopy of OH$^{+}$, H$_{2}$O$^+$ and H$_{3}^+$. For our calculations we use a value of $\epsilon=7\%$.

While, $N(\text{OH}^{+})$ and $N(\text{H})$ are determined from observations, $n_{\text{H}}$ refers to the gas density, for which we adopt a value of $n_{\text{H}}\approx 35$~cm$^{-3}$. This value corresponds to the mean density of the cold neutral medium at the solar circle, $R_{\text{GAL}} \approx 8.3~$kpc \citep{indriolo2012chemical, indriolo2015herschel}. The electron abundance, $x_\text{e}$ in diffuse clouds is often considered to be equal to the fractional abundance of ionised carbon, $x_\text{e} = 1.5\times10^{-4}$ \citep{sofia2004interstellar}. Under the conditions that prevail over the diffuse regions of the ISM, this is a valid assumption because C$^+$ is responsible for most of the free electrons and therefore the electron abundance, with negligible contributions from other ionised species. This assumption however, breaks down in regions of high $\zeta_\text{p}(\text{H})/n(\text{H})$ where ionisation by H becomes significant, like in the Galactic centre \citep{LePetit2016}.\\ 

\noindent
The molecular fraction, $f_{\text{H}_{2}}$ is typically defined as
\begin{equation}
    f_{\text{H}_{2}} = \frac{2n(\text{H}_{2})}{n(\text{H}) + 2n(\text{H}_{2})} 
    \label{eqn:mole_frac}
\end{equation}
and can be re-written in terms of the $N(\text{OH}^{+})/N(\text{H}_{2}\text{O}^{+})$ ratio by
\begin{equation}
    f_{\text{H}_{2}} = \frac{2x_\text{e}k_4/k_1}{N(\text{OH}^{+})/N(\text{H}_{2}\text{O}^{+}) - k_3/k_1} \, ,
    \label{eqn:molecular_fraction_oh}
\end{equation}
where, $N(\text{H}_{2}\text{O}^{+})$ is the total column density of  H$_{2}$O$^{+}$, ${N(\text{H}_{2}\text{O}^+) = N(\text{p-H}_{2}\text{O}^{+}) + N(\text{o-H}_{2}\text{O}^{+})}$. For those velocity intervals for which absorption is detected only in one of either the o- or p-H$_{2}$O$^{+}$ lines we present a lower limit on the molecular fraction.
%For those velocity intervals for which absorption is observed neither in the o- nor p-H$_{2}$O$^{+}$ line %symmetry states %in which either of the symmetric states are not present we report a lower limit on the molecular fraction. 
The molecular hydrogen fractions hence derived using Eq.~\ref{eqn:molecular_fraction_oh} has a median value of $f_{\text{H}_{2}} = (5.8\pm3.0)\times10^{-2}$. Alternatively, we can also compute the column-averaged molecular hydrogen fraction ${f^{N}_{\text{H}_{2}} = 2N(\text{H}_{2})/\left( N(\text{H}) + 2N(\text{H}_{2}) \right)}$ by using $N(\text{CH})$ as a tracer for $N(\text{H}_{2})$, following the relationship between the two molecules estimated by \citet{sheffer2008ultraviolet}, [CH]/[H$_{2}$] = $3.5^{+2.1}_{-1.4}\times10^{-8}$. However, the column-averaged molecular fractions derived towards the different velocity intervals using CH column densities are higher (by almost an order of magnitude) than those derived using Eq.~\ref{eqn:molecular_fraction_oh}. This difference arises from the fact that CH resides in, and therefore traces molecular gas unlike both OH$^{+}$ and H$_{2}$O$^{+}$.%We assume the rate coefficient for the forward reaction that produces Ar$^+$ ions via cosmic-rays (Ar$\rightarrow$ Ar$^{+}$ + $e^{-}$) or the total ionisation rate for Ar ($\zeta$)(Ar) to be (10 + 3.85$\phi$)$\zeta$(H) following \citet{jenkins2013fractional}. The quantity $\phi$ is the number of secondary ionisation products of H, per primary ionisation by cosmic-rays. Its value has been determined by \citet{schilke2014ubiquitous} to lie between 0.26 and 0.48 by adopting the fit given by \citet{dalgarno1999electron}. In this analysis we use $\phi = 0.37$ (mid-way within the range of values) which implies $\zeta(\text{Ar}) = 11.42\zeta_\text{p}(\text{H})$.

The cosmic-ray ionisation rates we infer from the steady-state chemistry of \ohp (and H$_{2}$O$^+$) using Eq.~\ref{eqn:CRIR} lie between $1.8\times10^{-17}$ and $1.3\times10^{-15}~$s$^{-1}$. Resulting in an average value of ${\zeta_\text{p}(\text{H}) = (2.28 \pm 0.34) \times10^{-16}~\text{s}^{-1}}$ towards the sight lines studied here. The uncertainties in $\zeta_\text{p}(\text{H})$ quoted in this work reflect the uncertainties in the derived column density values. The impact of the uncertainties associated with the assumptions made in this analysis, on the derived cosmic-ray ionisation rates are discussed in more detail in \citet{schilke2014ubiquitous} and \citet{indriolo2015herschel}. Within the statistical errors, this result is consistent with the cosmic-ray ionisation rates that were previously determined using H$_{3}^{+}$ by \citet{indriolo2012cosmic} and OH$^+$ and H$_{2}$O$^{+}$ by \citet{indriolo2015herschel, neufeld2017cosmic}. 

\begin{figure*}
\sidecaption
    \includegraphics[width=11cm]{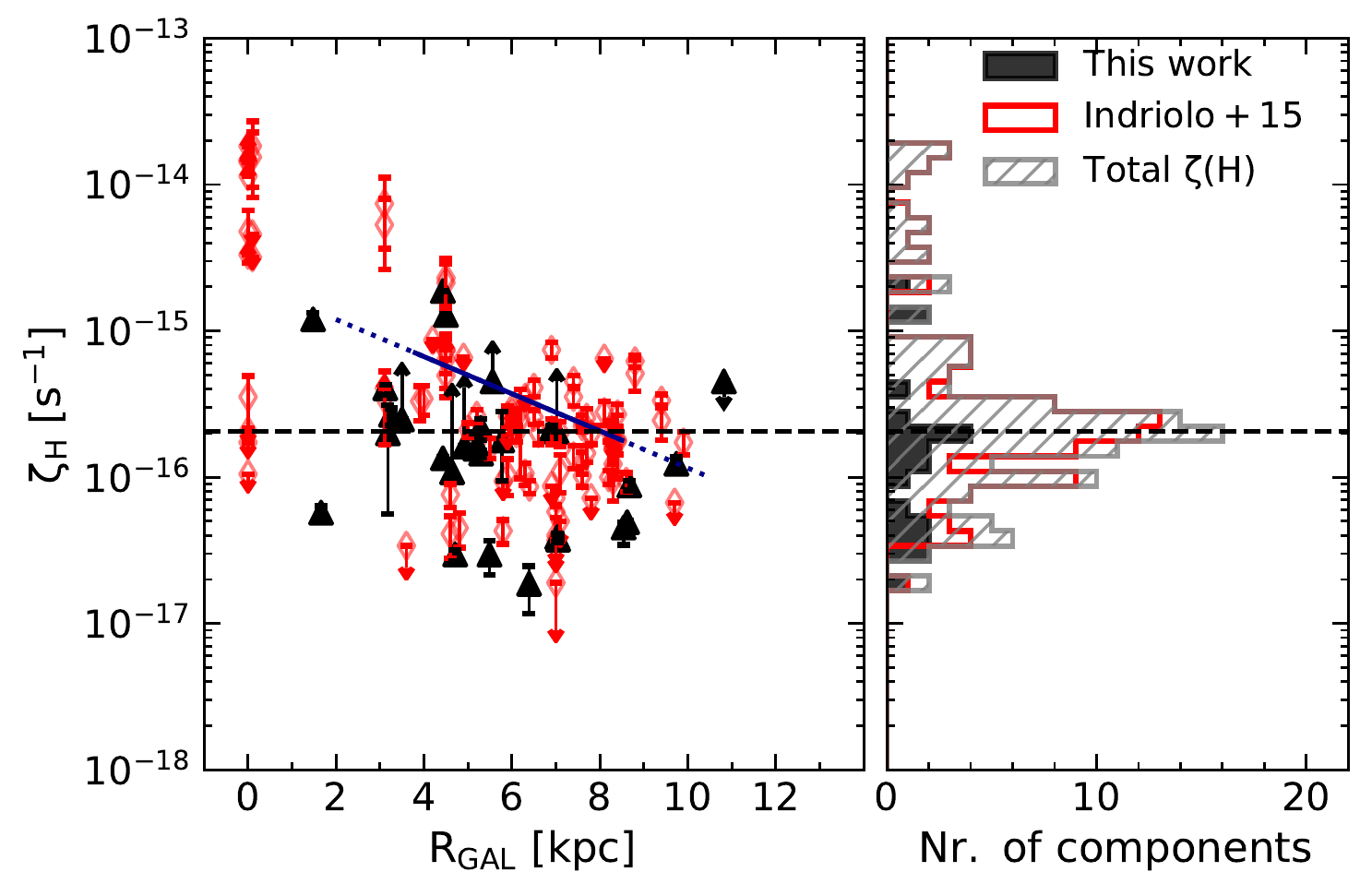}
    \caption{Left: Cosmic-ray ionisation rates derived from \ohp vs. galactocentric distance. The filled black triangles and unfilled red diamonds represent $\zeta_\text{p}(\text{H})$ values derived from the LOS absorption features observed in this study and in \citet{indriolo2015herschel}, respectively. The dashed black line marks the median value of the total $\zeta_\text{p}(\text{H})$ distributions. The solid-dotted line represents the derived cosmic-ray ionisation rate gradient. Right: Histogram distributions of $\zeta_\text{p}(\text{H})$ derived in this work displayed in black, that from \citet{indriolo2012chemical} in red, and the combined distribution by the hatched grey region.}
    \label{fig:CRIR_distribution}
\end{figure*}

The derived values for the cosmic-ray ionisation rate are summarised in Tab.~\ref{tab:column_densities} and in Fig.~\ref{fig:CRIR_distribution} we display their variation with galactocentric distance. The cosmic-ray ionisation rates in the extreme environment of the GC, are generally found to be higher than in the general Galactic environment and cover a range of values over two orders of magnitude between ${\sim}10^{-16}$ and $1.83\times10^{-14}~\text{s}^{-1}$ as discussed in \citet{indriolo2015herschel} and references therein.
%The higher rates derived in these regions reflect the widespread energetics present in the inner Galaxy, which decreases as we move away from the GC. 
At galactocentric distances \textless 5~kpc the ionisation rates that we derive in this study are comparable to those derived by \citet{indriolo2015herschel} at larger distances. 
This might hint to the fact that the gradient seen by these authors at ${3 < R_{\text{GAL}} < 5~}$kpc from the LOS components towards the GC sources (like M-0.13-0.08, M-0.02-0.07, Sgr~B2(M) and Sgr~B2(N)) are higher than the average cosmic-ray ionisation rate present in these spiral arms because these velocity components may contain contributions from the GC. At ${R_{\text{GAL}}>5~}$kpc, the cosmic-ray ionisation rates derived in this work are systematically lower than those derived in \citet{indriolo2015herschel} however, several of the ionisation rates derived by these authors, that lie below ${{\sim}5\times10^{-17}~\text{s}^{-1}}$ only represent upper limits to the cosmic-ray ionisation rates.

We derive a slope, $m$, of $-0.012$ from $\zeta_\text{p}(\text{H})$ versus $R_{\text{GAL}}$, for galactocentric distances that lie within ${4 < R_{\text{GAL}} < 8.5~}$kpc. Our analysis thus far assumes a constant value for $n(\text{H})$ equal to the mean gas density of cold gas near the solar circle, which itself varies with galactocentric distances as discussed in \citet{Wolfire2003}. \citet{neufeld2017cosmic} took this, as well as variations in the UV radiation field ($\chi_{\text{UV}}$) over Galactic scales into account and described the galactocentric gradient in the cosmic-ray ionisation rate as follows,
\begin{equation}
\frac{d\text{log}_{10}\zeta_\text{p}(\text{H})}{dR_{\text{GAL}}} = m + 1.7\frac{d\text{log}_{10}n(\text{H})}{dR_{\text{GAL}}} - 0.7 \frac{d\text{log}_{10} \chi_{\text{UV}}}{dR_{\text{GAL}}} \, ,
\end{equation}
where, $\zeta_\text{p}(\text{H})/n(\text{H}) \propto \left[ \chi_{\text{UV}}/n(\text{H})\right]^{-0.7}$. From the models presented in \citet{Wolfire2003}, which characterise the nature of neutral gas present within the Galactic disk, ${d\text{log}_{10}n(\text{H})}/{dR_{\text{GAL}}}$ and ${d\text{log}_{10} \chi_{\text{UV}}}/{dR_{\text{GAL}}}$ have values of $-0.110~\text{kpc}^{-1}$ and $-0.106~\text{kpc}^{-1}$, respectively. Using, the scale length, ${\left[ \text{dln}\zeta_\text{p}(\text{H})/\text{d}R_{\text{GAL}}\right]^{-1}}$, hence derived and the mean value of the cosmic-ray ionisation rate of their data-set, these authors derive the Galactic gradient to be $\zeta_\text{p}(\text{H}) = \left(2.2\pm0.3\right)\text{exp}\left[ \left(R_{\text{o}} - R_{\text{GAL}}\right]/{4.7~\text{kpc}}\right)\times10^{-16}~\text{s}^{-1}$, where ${R_{\text{o}} = 8.5}$~kpc and $R_{\text{GAL}}$ between 3-8.5~kpc. Following the analysis presented by \citet{neufeld2017cosmic}, briefly discussed above, we derive a revised cosmic-ray ionisation rate gradient 
as, $\zeta_\text{p}(\text{H}) = \left(1.80\pm0.70\right)\text{exp}\left[ \left(R_{\text{o}} - R_{\text{GAL}}\right)/{3.46~\text{kpc}}\right]\times10^{-16}~\text{s}^{-1}$ using a scale length of 3.46~kpc and a mean cosmic-ray ionisation rate of $(1.80\pm0.70)\times10^{-16}~\text{s}^{-1}$ (where the value in parenthesis is the standard error), across the combined sample of sight lines studied in \citet{indriolo2015herschel} and this work for galactocentric distances that lie within ${4 < R_{\text{GAL}} < 8.5~}$kpc and a gas density of 35~cm$^{-3}$. We have excluded the $\zeta_\text{p}(\text{H})$ values derived for the Galactic centre sources.

%\begin{equation}
%      \left(  \frac{\zeta_\text{p}(\text{H})}{ \text{s}^{-1}} \right) = \left( 9.33 \pm 0.70 \right) \times 10^{-17} \text{exp}\left[\frac{\left( R_0 - R_{\text{g}}\right)}{\left(8.19 \pm 0.02 \right)~\text{kpc}} \right]  \, ,
%      \label{eqn:CRIR_gradient}
%\end{equation}
%where, $R_{0} = 8.3~$kpc and $R_{\text{g}} > 2~$kpc.
%, $\zeta_\text{p}(\text{H})_{\text{ArH}^+}$ is found to be almost two orders of magnitude larger than $\zeta_\text{p}(\text{H})$. Our results greatly suggest that \arhp and OH$^+$ probe different layers of a given diffuse cloud. This is supported by previous chemical models presented by \citet{neufeld2016chemistry}, who have demonstrated that the difference in ionisation rates between these ionised species results from absorption against different cloud populations characterised by their size and visual extinction. In their follow-up paper, \citet{neufeld2017cosmic} investigate another likely explanation for this discrepancy, which is that the cosmic-ray ionisation rate decreases as a function of the visual extinction, $A_{\text{v}}$. This is conceivable since, \arhp observed on the surface of clouds is exposed to higher rates of cosmic-ray ionisation than \ohp and H$_{2}$O$^+$ formed deeper within the cloud. They find the ionisation rate to be marginally decreasing with $A_{\text{v}}$, with a slope of $-1$ for $A_{\text{v}} \geq 0.5$. 

\subsection{\texorpdfstring{$\text{H}_{2}\text{O}^{+}$}{HtOp} analysis}
For H$_{2}$O$^+$, the lower energy spin variant is \ohtop owing to its $C_{2\varv}$ symmetry and ${}^{2}B_1$ ground state, unlike the case of H$_{2}$O, for which the lowest energy level is the $0_{00}$ level of %the absolute ground state of the molecule corresponds to the ground state of its 
the p-H$_{2}$O spin isomer. The lowest ortho- and para-rotational levels of H$_{2}$O$^+$ have an energy difference of ${\approx} 30.1~$K and %where 
the fine structure levels of the ortho ground state ($I = 1$) further undergo hfs splitting, while those of the para ($I = 0$) state do not. Under the typical conditions present in the diffuse ISM, the rotational temperature (as discussed in Sect.~\ref{subsec:column_densities}) is close to $T_{\text{CMB}}$, which makes it valid to assume that most of the molecules will exist in the lowest ortho and para rotational states. %Similar to H$_{2}$ and H$_{2}$O, H$_{2}$O$^+$ with two protons should show an ortho-to-para ratio (OPR) of at least 3:1 \citep{townes1975microwave}. 

The conversion between the ortho- and para-states of a species, typically occurs via proton exchange reactions with either atomic H or H$_{2}$. Since, atomic H is the most abundant species along our diffuse LOS, we expect the proton exchange in H$_{2}$O$^+$ to primarily take place via gas phase reactions with H. Moreover, in the presence of H$_{2}$, H$_{2}$O$^+$ ions may energetically react to form H$_{3}$O$^+$ and a hydrogen atom:
\begin{equation*}
    \text{p-H}_{2}\text{O}^+ + \text{H} \rightleftharpoons \text{o-H}_{2}\text{O}^+ + \text{H} \, .
\end{equation*}
Since, the two spin states can be moderately coupled via collisions, the OPR is given by
\begin{equation}
    \text{OPR} = \frac{Q_{\text{ortho}}}{Q_{\text{para}}}\text{exp}\left( \Delta E / T_{\text{ns}}\right) \, ,
    \label{eqn:OPR}
\end{equation}
where $Q_{\text{ortho}}$ and $Q_{\text{para}}$ represent the partition functions of the respective states, $\Delta E$ is the energy difference between the two states, $\Delta E = -30.1~$K (it is negative because the lowest ortho state lies at a lower energy level than the lowest para state, unlike in H$_{2}$O) and $T_{\text{ns}}$ is the nuclear spin temperature. At low temperatures, the partition function of the para- and ortho-states are governed by the degeneracy of their lowest fine-structure and hfs levels, respectively. Both states have the same quantum numbers and degeneracy, which implies that, as the rotational temperature approaches 0~K, the ratio of the partition functions approaches unity \citep[see, Appendix A of][]{schilke2010}. Therefore, in the analysis that follows we assume that, $Q_{\text{ortho}}$ = $Q_{\text{para}}$. Since, the largest uncertainty in our derived OPR arises from the uncertainties in the absorption features of the \phtop profile, we only carry out this analysis in those velocity intervals that are the least contaminated. From our column density calculations, we derive OPR values between ${\sim 1:1}$ and 4.5:1, which is compatible with the equilibrium value of three, within the error bars. We find a mean OPR for H$_{2}$O$^+$ of 2.1:1 which corresponds to a mean nuclear spin temperature of 41~K.

\section{Discussion}\label{sec:discussion}
\subsection{Properties of \texorpdfstring{ArH$^{+}$}{ArHp} as a tracer of atomic gas}
\begin{figure*}
    \centering 
    \includegraphics[width=0.9\textwidth]{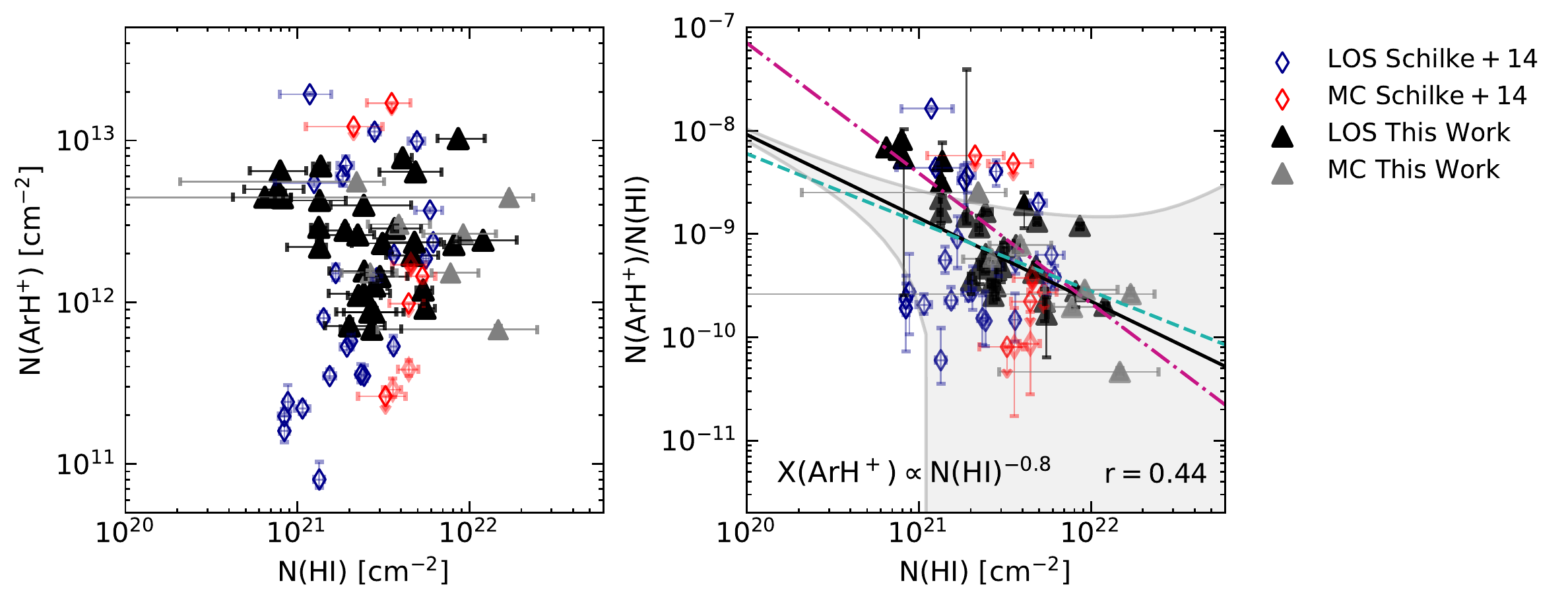}
    \caption{Left: $N(\text{ArH}^+)$ vs. $N(\text{H})$. Right: $X(\text{ArH}^+) = N(\text{ArH}^+
    )/N(\text{H{\small{I}}})$ vs. $N(\text{H})$. The data points include column density values derived towards the MC and LOS features from our sample of sources indicated by filled grey and black triangles, as well as those from \citet{schilke2014ubiquitous} marked by unfilled red and blue diamonds, respectively. The solid black line displays the power-law relation to the combined data-set, $X(\text{ArH}^+) \propto N(\text{H\text{\small I}})^{-0.78}$, with the grey shaded region representing the 1$\sigma$-level of the weighted fit. The dashed cyan and dashed-dotted pink curves represent the fits to only the LOS and only the MC components, respectively.}
    \label{fig:ArHp_abundance}
\end{figure*}

We derive \arhp abundances relative to atomic H column densities, $X({\text{ArH}^+}) = N(\text{ArH}^+)/N(\text{H{\small{I}}})$, that span over roughly two orders of magnitude varying between $4.6\times10^{-11}$ and $1.6\times10^{-8}$, with an average value of $(1.6\pm 1.3)\times10^{-9}$. In Fig.~\ref{fig:ArHp_abundance}, by combining our \arhp data points with values from the sight lines studied in \citet{schilke2014ubiquitous}, we compare the derived column densities and abundances of \arhp with atomic hydrogen column densities. Over the entire range, the ArH$^+$ column densities cluster at ${N(\text{H}) = (2.7\pm1.6)\times10^{21}~\text{cm}^{-2}}$. We observe a large spread in the \arhp abundances derived, even for the molecular cloud (MC) components. This is mainly because of the high optical depths of the H{\small I} data and to a lesser extent because, towards some of the sources in our sample, we see that \arhp traces layers of infalling material associated with the molecular cloud. Figure~\ref{fig:ArHp_abundance} does not reveal a strong correlation between \arhp and H{\small I}, as we had expected because H{\small I} traces different phases of the ISM with varying degrees of molecular fraction. Moreover, the H{\small I} emission at certain velocity intervals may be a superposition of both the near- and far-side components of the Milky Way, often leading to blending effects which can potentially bias the
inferred spin temperatures and column densities. %Unsurprisingly, those data points that %correspond to represent absorption in velocity intervals typical for material %near the 
%from the envelopes of the molecular clouds in our sample, (displayed by red and grey markers in Fig.~\ref{fig:ArHp_abundance}) show lower ArH$^+$ abundances with a median value of $X(\text{ArH}^+) = 2.37\times10^{-10}$. This is because these components arise from dense clouds with a higher molecular gas content, relative to the diffuse sight lines. However, even amongst these MC components, there exists a spread in the abundance values derived as, towards some of the sources in our sample we observe high \arhp abundances tracing layers of infalling material associated with the molecular cloud. 
%KMM: I don't understand the logic in the last half sentence (starting with: as,)

%Over 
We subjected the entire ArH$^+$/H{\small I} data-set to a regression analysis and find $\text{log}(X(\text{ArH}^+)) = (-0.80\pm0.20)~\text{log}(N(\text{H})) + (7.84\pm 4.28)$  with a correlation coefficient of 38\% at a 99.99\% confidence level. This translates to a power-law relation between \arhp and H{\small I} column densities of $N(\text{ArH}^+) \propto N(\text{H})^{0.2}$. This can be viewed as a global relation as the behaviour of the power-law index (or slope in logarithmic-scales) does not appreciably vary between the diffuse LOS and dense molecular cloud components and there is no clear power-law break.
Observations in the past have shown that considerable amounts of H{\small I} gas exist within molecular clouds, whose population is maintained via the destruction of H$_{2}$ by cosmic-rays \citep[e.g.][]{Wannier1991, Kuchar1993}. Hence, the H{\small I} gas along any given LOS in our study, also contains gas that resides inside molecular clouds with a fractional abundance, [H]/[H$_{2}$], ${\sim 0.1\%}$. From Fig.~\ref{fig:ArHp_abundance}, we find that as $N$(H{\small I}) increases, $X$(ArH$^+$) decreases due to the increasing amounts of dense gas traced by H{\small I} gas. It is therefore clear that only a small fraction of the H{\small I} gas traces the same cloud population as that traced by ArH$^+$. Based on the astrochemical models presented by \citet{schilke2014ubiquitous}, if we assume the abundance of ArH$^+$ to be a constant ($X(\text{ArH}^+) = 2\times10^{-10}$) for cloud depths $\leq 0.01~$mag which corresponds to an average molecular fraction of 10$^{-3}$, then we find that only 17.3\% of the LOS H{\small I} gas traces these conditions. \\

\begin{figure}
    
    \includegraphics[width=0.47\textwidth]{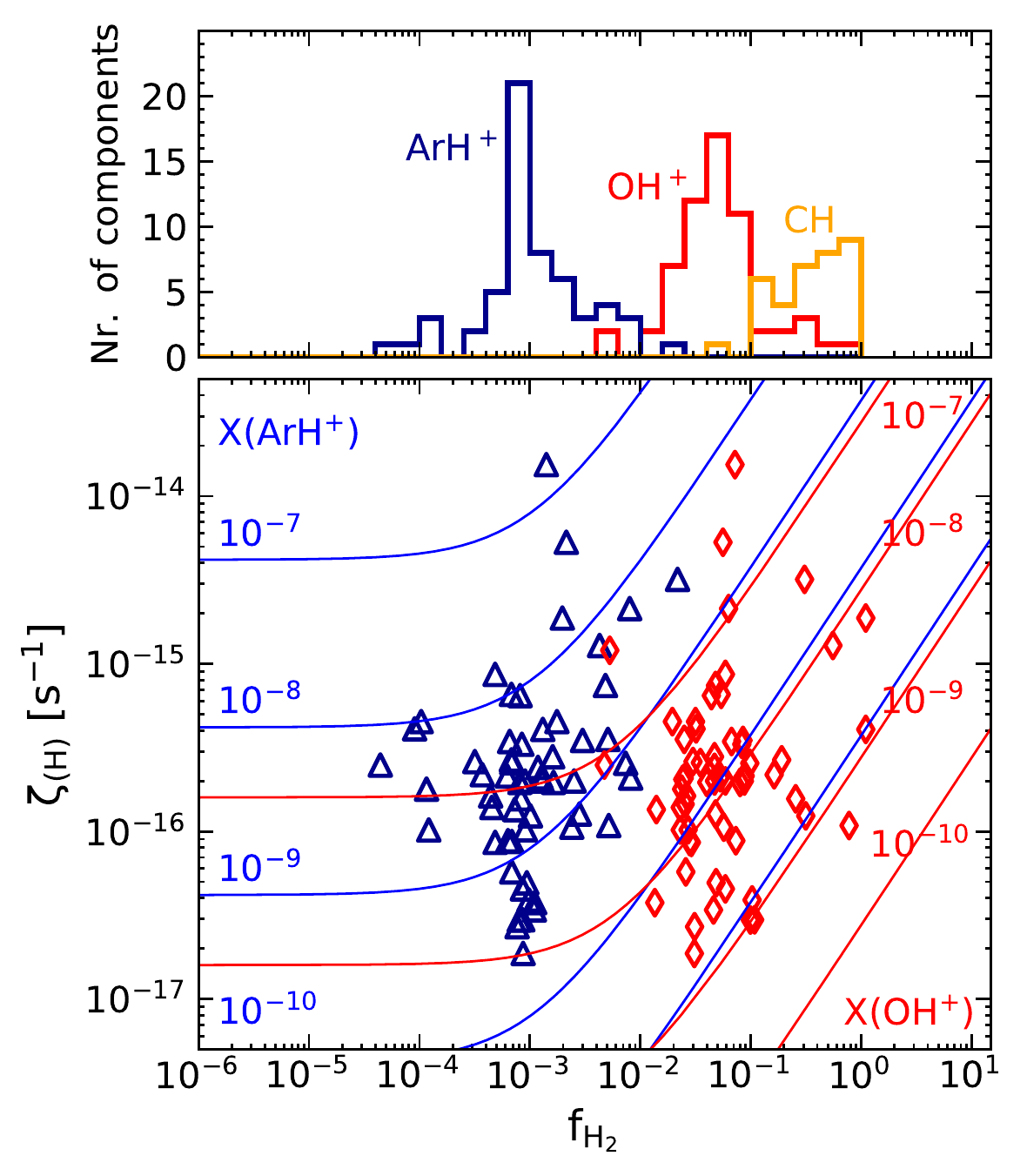}
    \caption{Top: Distribution of molecular gas fraction ($f_{\text{H}_{2}}$) traced by ArH$^+$ (blue), OH$^+$ (red) and CH (orange). Bottom: Contours of $X(\text{ArH}^+)$ (blue) and $X(\text{OH}^+)$ (red) abundances with respect to $N(\text{H})$ in the $f_{\text{H}_{2}}$--$\zeta_{\text{(H)}}$ plane. Blue triangles and red diamonds represent the corresponding values derived from the LOS observations presented in \citet{schilke2014ubiquitous} and this work.}
    \label{fig:arhp_properties}
\end{figure}

As detailed in \citet{schilke2014ubiquitous}, \arhp is destroyed primarily via proton transfer reactions with H$_{2}$ or atomic oxygen, and photodissociation:  
\begin{align*}
    \text{ArH}^{+} &+ \text{H}_{2} \rightarrow \text{Ar} + \text{H}_{3}^{+} \, ; \quad \quad \quad k_{5} = 8\times 10^{-10}~\text{cm}^{3}\text{s}^{-1} \\
    \text{ArH}^{+} &+ \text{O} 
    \rightarrow \text{Ar} + \text{OH}^{+} \, ; \quad  \quad \,\,\, \, k_{6} = 8\times 10^{-10}~\text{cm}^{3}\text{s}^{-1} \\ 
    \text{ArH}^{+} &+ \text{h}\nu \rightarrow \text{Ar}^+ + \text{H} \, ; \, \quad \quad \quad k_{7} = 1.0\times 10^{-11}\chi_\text{UV}f_{\text{A}}~\text{s}^{-1} \, .
\end{align*}
The photodissociation rate of \arhp was estimated by \citet{alekseyev2007theoretical} to be ${\sim 1.0\times10^{-11}f_{\text{A}}~\text{s}^{-1}}$ for an unshielded
cloud model uniformly surrounded by the standard Draine
UV interstellar radiation field. However, for the particular environment of the Crab nebula, the photodissociation rate has been calculated to be much higher at $1.9 \times 10^{-9}~\text{s}^{-1}$ \citep{roueff2014photodissociation}. The attenuation factor, $f_{\text{A}}$, is given by an exponential integral and is a function of visual extinction, $A_\text{v}$. \citet{schilke2014ubiquitous} derive values of $f_{\text{A}}$ between 0.30 and 0.56, increasing as you move outwards from the centre of the cloud, for a cloud model with $A_{\text{v}} = 0.3$. In our analysis, we use a value of 0.43 for $f_{\text{A}}$, mid-way through the computed range of values. We further assume a gas density ${n(\text{H}) = 35~\text{cm}^{-3}}$, an atomic oxygen abundance (relative to H nuclei) of ${3.9 \times 10^{-4}}$ and an argon abundance close to its solar abundance of ${3.2\times 10^{-6}}$ \citep{lodders2008solar}. Using these values the cosmic-ray ionisation rate is approximated as follows,
\begin{align}
    \zeta_\text{p}(\text{H}) &= \frac{N(\text{ArH}^{+})}{N(\text{H})}\left(\frac{k_{5}n(\text{H}_{2}) + k_{6}n(\text{O}) + k_{7}}{11.42}\right) \, ,\\ &=\frac{N(\text{ArH}^{+})}{N(\text{H})}\left( \frac{0.5005 + 448f_{\text{H}_{2}}}{1.2 \times 10^{6} }\right) \, .
    \label{eqn:CRIR_arhp}
\end{align}
Re-arranging Eq.~\ref{eqn:CRIR_arhp}, we express it in terms of $f_{\text{H}_{2}}$, as follows:
\begin{equation}
  f_{\text{H}_{2}}  = 2.68\times10^{3} \left[ \frac{\zeta_\text{p}(\text{H})}{X(\text{ArH}^{+})} - 4.17\times10^{-7} \right] \, 
  \label{eqn:hfrac_arhp}
\end{equation}
Substituting the \arhp abundances derived from observations into Eq.~\ref{eqn:hfrac_arhp} and by assuming that the \arhp ions are exposed to the same cosmic-ray flux as %that of 
OH$^+$ and H$_{2}$O$^+$, we derive the molecular fraction of the gas probed by ArH$^{+}$. Of course, there is the caveat that ArH$^+$ does not necessarily trace the same gas as that traced by both OH$^+$, and H$_{2}$O$^+$ and therefore, need not be exposed to the same cosmic-ray ionisation flux. Our results find \arhp to probe a range of molecular fractions, with a median value of 8.8$\times10^{-4}$. This is in agreement with the results obtained from chemical models presented in \citet{neufeld2017cosmic}, %whose models 
which suggest that the observed \arhp abundance resides in regions with molecular fractions that are at most 10$^{-2}$. The molecular fraction analysis displayed in Fig.~\ref{fig:arhp_properties}, derived from the observed abundances of ArH$^+$, OH$^+$, and CH using Eqs.~\ref{eqn:hfrac_arhp}, \ref{eqn:molecular_fraction_oh}, and \ref{eqn:mole_frac}, respectively, clearly shows the transition between different phases of the ISM, from the diffuse atomic to the diffuse/translucent molecular gas.   

\begin{figure*}
    
    \includegraphics[width=0.9\textwidth]{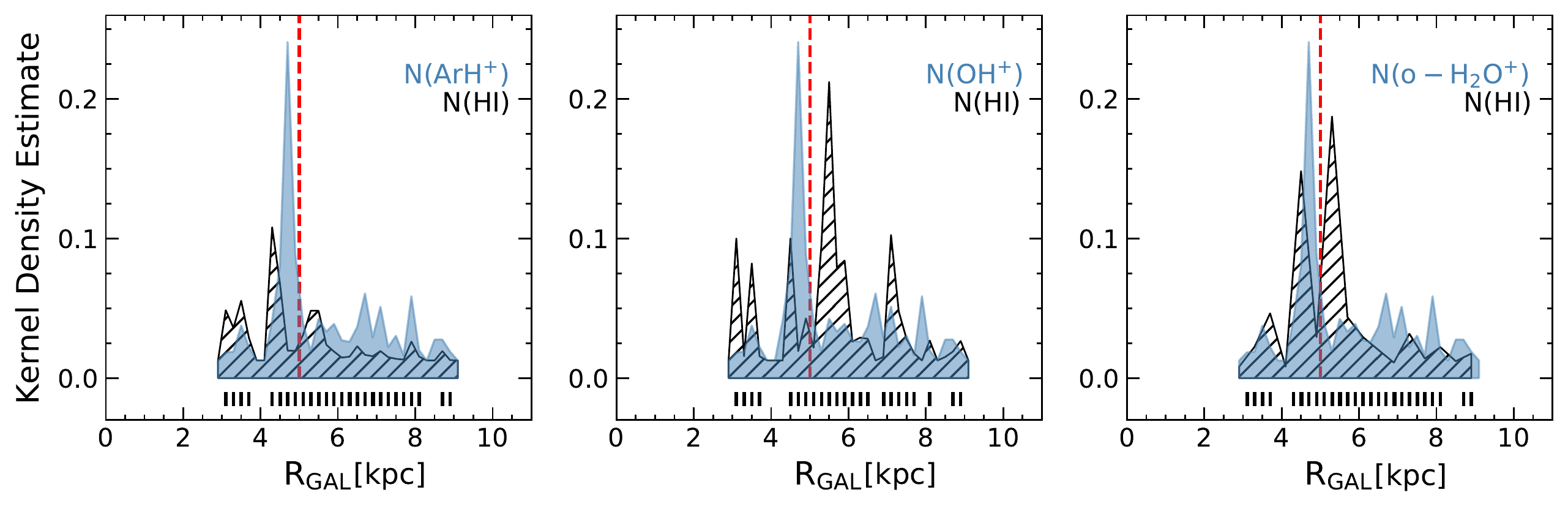}
    \caption{Kernel density distributions of the of \arhp (top), \ohp (centre), and \ohtop (bottom) column densities as a function of galactocentric distances (black hatched regions). The sampling of the distribution is displayed as a rug plot and is indicated by the solid black lines below the density curves. The dashed red line marks the 5~kpc ring. }
    \label{fig:column_density_gradient}
\end{figure*}

We further compare the Galactic radial distribution of the azimuthally averaged ArH$^+$, \ohp and \ohtop column densities with those of H{\small I} towards a combined sample set that contains sight lines presented in both \citet{schilke2014ubiquitous} and this work. The probability distributions of the different column densities were computed using the kernel density estimation method, for a Gaussian kernel. This analysis was carried out for galactocentric distances between 4 and 9~kpc, since we have only a limited number of data points at $R_\text{GAL}$\textless 4~kpc and $R_\text{GAL}$\textgreater 9~kpc making it difficult to comment on the nature of the distribution at these distances. Although  we only cover a small number of background continuum sources in the Galaxy (but sample many LOS), the distribution of $N$(H{\small{I}}) versus galactocentric radii presented in Fig.~\ref{fig:column_density_gradient}, for the common $R_{\text{GAL}}$ coverage, resembles that of the H{\small{I}} emissivity presented by \citet{pineda2013herschel} for 500 LOS across the Galaxy. The H{\small{I}} column densities we derived peak at $R_{\text{GAL}} =5.1$~kpc corresponding to the intersection of the Perseus and Norma spiral-arms with a possible narrower peak between 7.5 and 8~kpc arising from the Scutum-Crux arm.  

Atomic gas predominantly exists in thermal equilibrium between two phases of the ISM namely, the cold neutral medium (CNM) and the warm neutral medium (WNM). \citet{pineda2013herschel} used the H{\small I} emission line to trace the total gas column but were able to separate the relative contributions from gas in the CNM, and WNM by using constraints from H{\small I} absorption studies. Their analysis revealed that the CNM is the dominant component within the inner Galaxy peaking close to the 5~kpc arm while it is the WNM that contributes towards the peak in the outer Galaxy at 8~kpc. The radial distribution of the \arhp column densities peaks between 4--4.5~kpc slightly offset from the H{\small I} peak with almost an anti-correlation at 4.7~kpc. At $R_{\text{GAL}}$\textgreater 5~kpc, the $N(\text{ArH}^+)$ profile decreases mimicking the radial trend followed by the CNM component. However, there maybe a potential increase in the column densities near 8~kpc but we cannot as yet confirm the presence of this peak as we require more sight line information towards the outer Galaxy ($R_{\text{GAL}} > 8~$kpc). If \arhp does have a WNM component our results are in agreement with the idea that \arhp is formed in the outermost layers of the cloud. The column density distributions of \ohp and \ohtop on the other hand, peak behind the 5~kpc ring, with a significant contribution at 4--4.5~kpc similar to the ArH$^+$, which we do not see in the H{\small I} data. Since, the column densities at 7 and 7.5~kpc are much greater than that at 8~kpc for \ohp and \ohtop distributions, the presence of a secondary peak in the outer Galaxy, seems unlikely.

\subsection{Comparison of \texorpdfstring{ArH$^{+}$}{ArHp} with other atomic gas tracers}

\begin{figure*} 
     \includegraphics[width=0.29\textwidth]{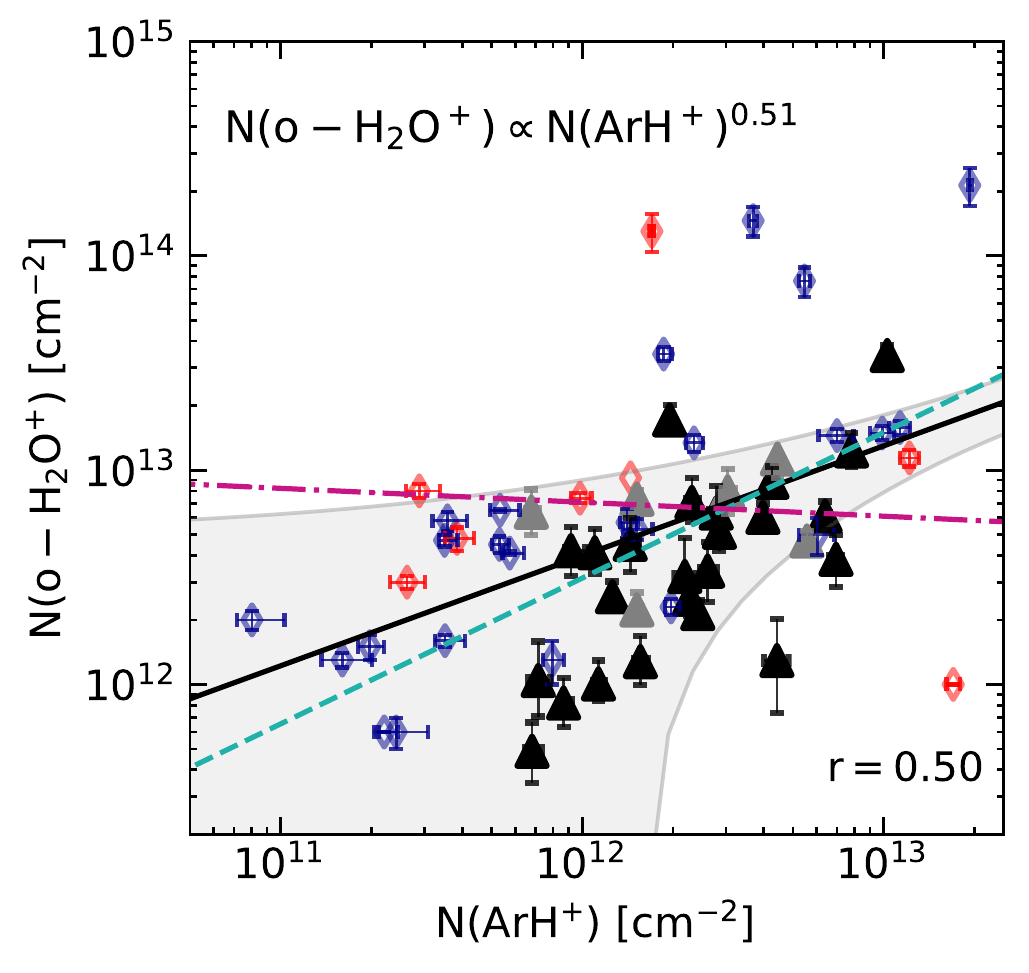}
    \includegraphics[width=0.29\textwidth]{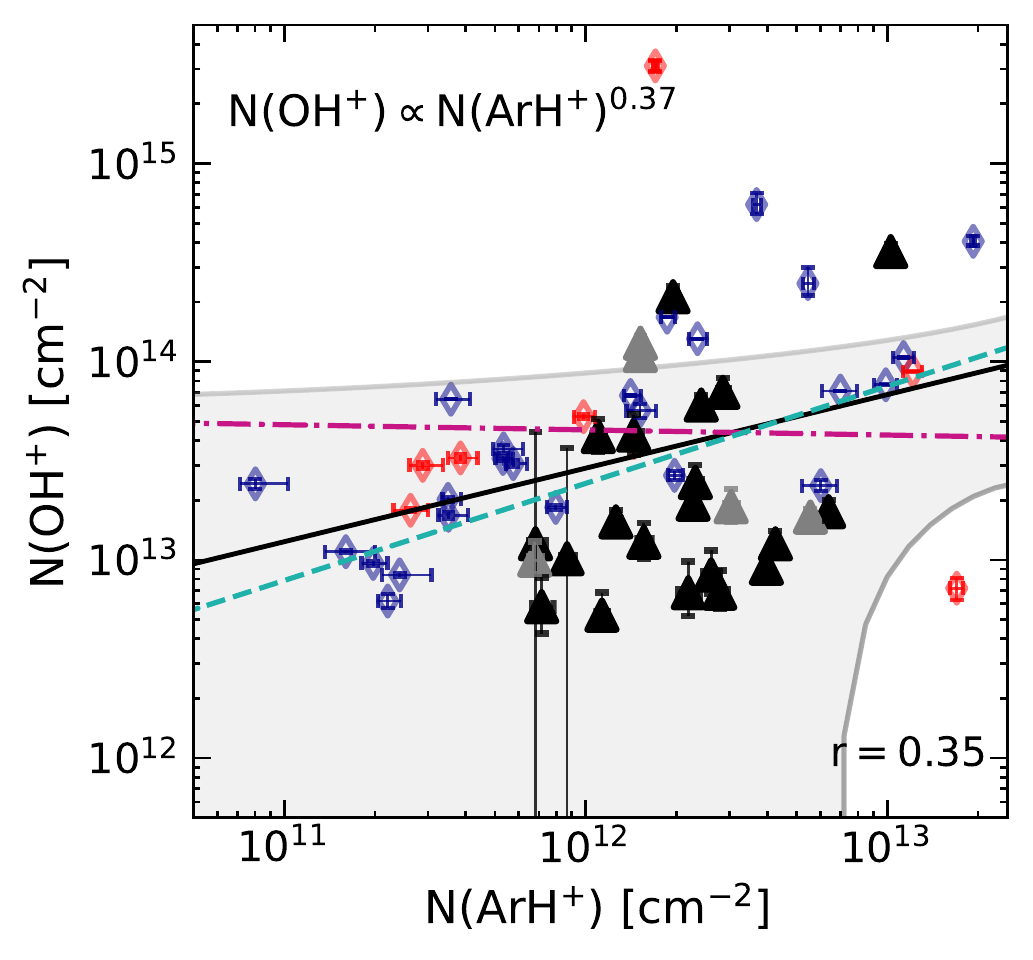}
    \includegraphics[width=0.47\textwidth]{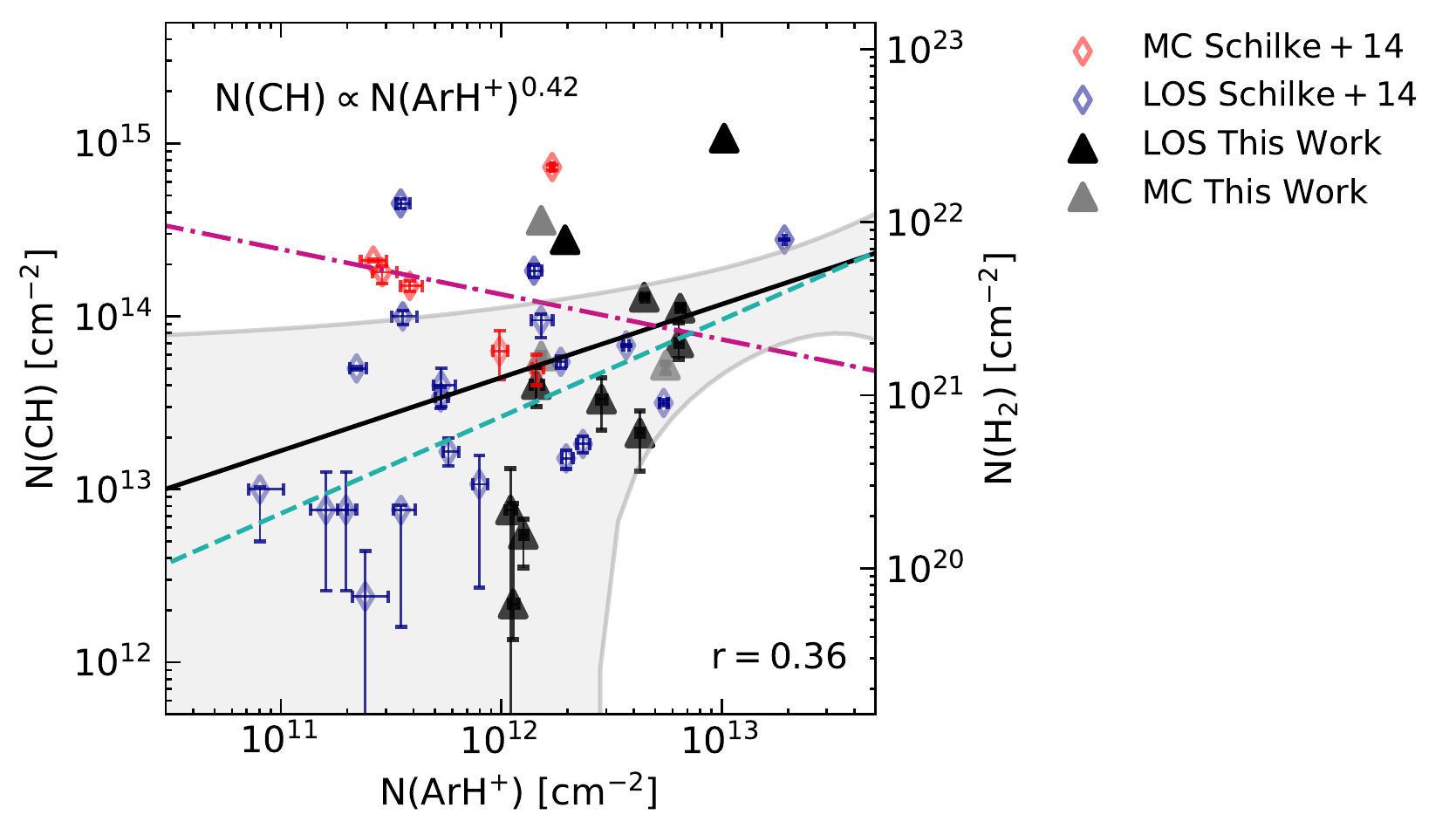}
    \caption{Comparison of the \arhp column density 
to that of \ohtop (left), \ohp (centre), and CH (right). Black and grey filled triangles indicate the column density values derived within velocity intervals corresponding to LOS absorption and molecular cloud (MC) towards the Galactic sources presented in this study. Additionally, the blue and red unfilled diamonds mark the LOS and MC column densities derived towards the sources discussed in \citet{schilke2014ubiquitous}. The solid black curve represents the best fit to the combined data set including both the LOS and MC components with the grey shaded region displaying the 1~$\sigma$ interval of the weighted regression. The dashed cyan and dashed-dotted pink curves represent the fits to only the LOS and only the MC components, respectively.}
    \label{fig:correlation_plots}
\end{figure*}

In this section, we investigate the correlation between the column density of \arhp with that of other tracers of atomic gas such as \ohp and \ohtop and with molecular gas traced by CH. Combining our sample with data taken by \citet{schilke2014ubiquitous, indriolo2015herschel, wiesemeyer2016far, Wiesemeyer2018} and \citet{jacob2019fingerprinting}, in Fig.~\ref{fig:correlation_plots} we show the correlation between the 62, 57, and 40 sight lines towards which both \arhp and o-H$_{2}$O$^+$ (left), OH$^+$ (centre), and CH (right), respectively, have been detected. The distribution of the OH$^+$ and o-H$_{2}$O$^+$ column density components associated with the molecular cloud environment of these SFRs displays a large dispersion in comparison to the LOS components with no clear correlations over the range of ArH$^+$ column densities probed, reflected by weakly negative correlation coefficients of $-0.02$ and $-0.08$, respectively. In addition, as expected the ArH$^+$ and CH column densities derived over molecular cloud velocities are anti-correlated with a correlation coefficient of $-$0.30. This emphasises that \arhp probes cloud layers with slightly different properties from those traced by not only CH but also \ohtop and OH$^+$. 

The global correlation of the LOS components are described using a power-law of the form $N_{\text{Y}} = cN_{\text{X}}^{p}$, where c is the scaling factor and p is the power-law index, and $N_{\text{X}}, N_{\text{Y}}$ represent the column densities of the different species. The fit parameters and correlation coefficients (at a 99.99$\%$ confidence interval), corresponding to the derived correlations, amongst the different species across all the data points (both LOS and MC) are summarised in Tab.~\ref{tab:fit_parameters}. We chose to describe the correlation using a power-law trend after comparing the residuals and correlation coefficients obtained from this fit with those obtained from fitting a linear regression of the form, $N_{\text{Y}} = mN_{\text{X}} + c$. Moreover, a positive y-axis intercept would suggest the presence of either o-H$_{2}$O$^+$, OH$^+$ or H$_{2}$ in the absence of \arhp or vice versa for a negative intercept and interpreting this would require the aid of additional chemical models. 

\begin{figure}
    \centering
    \includegraphics[width=0.48\textwidth]{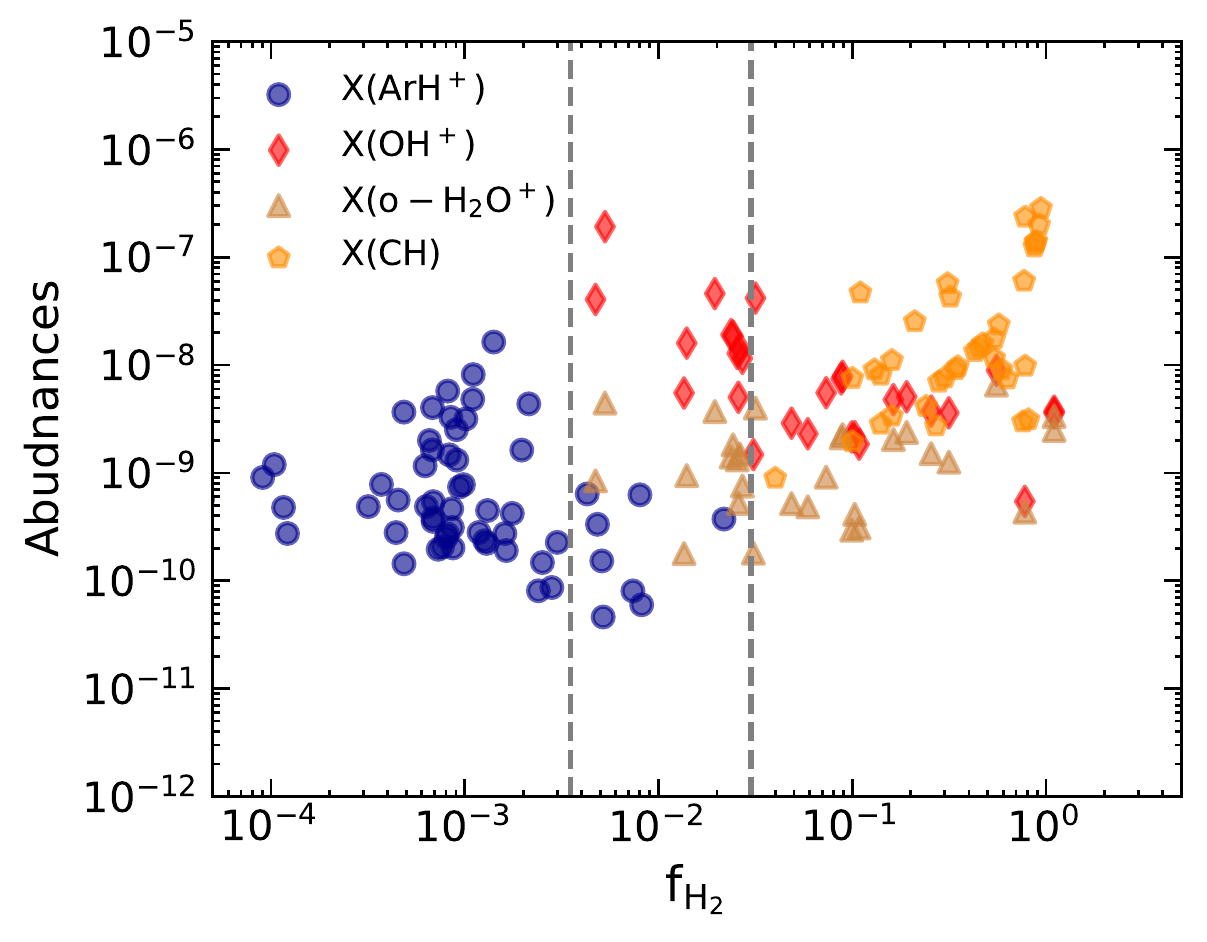}
    \caption{Molecular abundances with respect to atomic hydrogen as a function of molecular gas fraction ($f_{\text{H}_{2}}$). The blue circles, red diamonds, tan triangles, and yellow pentagons represent the abundances of ArH$^+$, OH$^+$, H$_{2}$O$^+$, and CH, respectively. The dashed grey lines enclose within them the cloud layers traced by ArH$^+$, OH$^+$, and H$_{2}$O$^+$.}
    \label{fig:abundance_vs_hfrac}
\end{figure}

As discussed earlier, the different species may not be spatially co-existent and the observed correlations may result from the fact that an increase in the total column density of each sight line subsequently, increases the column density of each phase. Therefore, one would expect there to be a weak correlation between the column densities in each phase, along any given LOS. We investigate this by comparing the abundances of these species as a function of the molecular fraction of the gas they trace. From Fig.~\ref{fig:abundance_vs_hfrac} it is clear that ArH$^+$ and OH$^+$ and H$_{2}$O$^+$ all trace the same cloud layers only in a small range of molecular gas fractions between $1.5\times10^{-3}$--$3\times10^{-2}$, while CH traces the denser molecular gas. We further notice that the distribution of CH abundances shows a positive correlation with $f_{\text{H}_{2}}$ while, that of OH$^+$ is anti-correlated and both, ArH$^+$, and H$_{2}$O$^+$ remain almost constant at $X(\text{ArH}^{+}) = (4.7\pm0.2)\times10^{-10}$ and $X(\text{H}_{2}\text{O}^{+}) = (1.3\pm1.0)\times10^{-9}$.

\begin{table}
  \caption{Summary of correlation power-law fit parameters as a function of $N(\text{ArH}^+)$ ($N_{\text{X}}$).}
    \begin{tabular}{lrlc}
        
        \hline \hline
         $N_{\text{Y}}$ &  Power-law & \multicolumn{1}{c}{Coefficient} & Pearson's\\
         & index ($p$) & \multicolumn{1}{c}{($c$)} & \multicolumn{1}{c}{$r$-value} \\
         \hline
          $N(\text{o-H}_{2}\text{O}^+)$ & $0.51 \pm 0.11$ & $(2.77 \pm 0.60) \times10^{6}$ & 0.50 \\
         $N(\text{OH}^+)$ & $0.37 \pm 0.13$& $(1.02 \pm 0.18)\times10^{9}$ & 0.35 \\
         $N(\text{CH})$ & $0.42 \pm 0.18$ & $(3.65 \pm 0.91)\times10^{8}$ & 0.36\\
         $N(\text{H}_{2})$\tablefootmark{a} & $0.42 \pm 0.18$ & $(1.04^{+0.67}_{-0.50})\times10^{15}$ & 0.36 \\
         \hline
    \end{tabular}
    \tablefoot{\tablefoottext{a}{$N(\text{H}_{2})$ values were derived using $N(\text{CH})$ following the relationship determined by \citet{sheffer2008ultraviolet}, [CH]/[H$_{2}$]= 3.5$^{+2.1}_{-1.4}\times10^{-8}$.}}
    \label{tab:fit_parameters}
\end{table}

\subsection{\texorpdfstring{$\text{H}_{2}\text{O}^{+}$}{HtOp} ortho-to-para ratio}
Of the 14 absorption components for which we were able to compute the OPR of H$_{2}$O$^+$, 9 show values significantly lower than the value of 3:1 with 6 of the derived OPRs lying close to unity. Such low values of the OPR are conceivable as the corresponding spin temperatures (derived in the absence of a rotational excitation term as in Eq.~\ref{eqn:OPR}) reflect the typical kinetic temperatures of the diffuse clouds. However, amongst some sight line components the OPR is as high as $4.50^{+1.53}_{-1.60}$ which point to cold environments with spin temperatures as low as 20~K. Perhaps similar to studies of the OPR of H$_{2}$ by \citet{flower2006importance}, the OPR of H$_{2}$O$^+$ at these low temperatures maybe governed by the kinematic rates of its formation and destruction processes. 

Moreover, the OPR of H$_{2}$ greatly impacts the observed OPR of H$_{2}$O$^+$ because the latter is synthesised in diffuse clouds via an exothermic reaction between \ohp and H$_{2}$. To assess this, requires the determination of the fraction of ortho- and para-H$_{2}$O$^+$ formed from the reacting fractions of both ortho- and para-H$_{2}$ states whilst abiding by the spin selection rules. After estimating the fractions of ortho- and para-H$_2$O$^+$, \citet{herbst2015interstellar} derived an OPR for H$_{2}$O$^+$ of 2:1 and concludes that this approach fails as it requires all the reacting, parent H$_{2}$ molecules to exist in the ortho-state in order to reproduce the higher observed OPR values. 

\begin{figure}
    \includegraphics[width=0.4\textwidth]{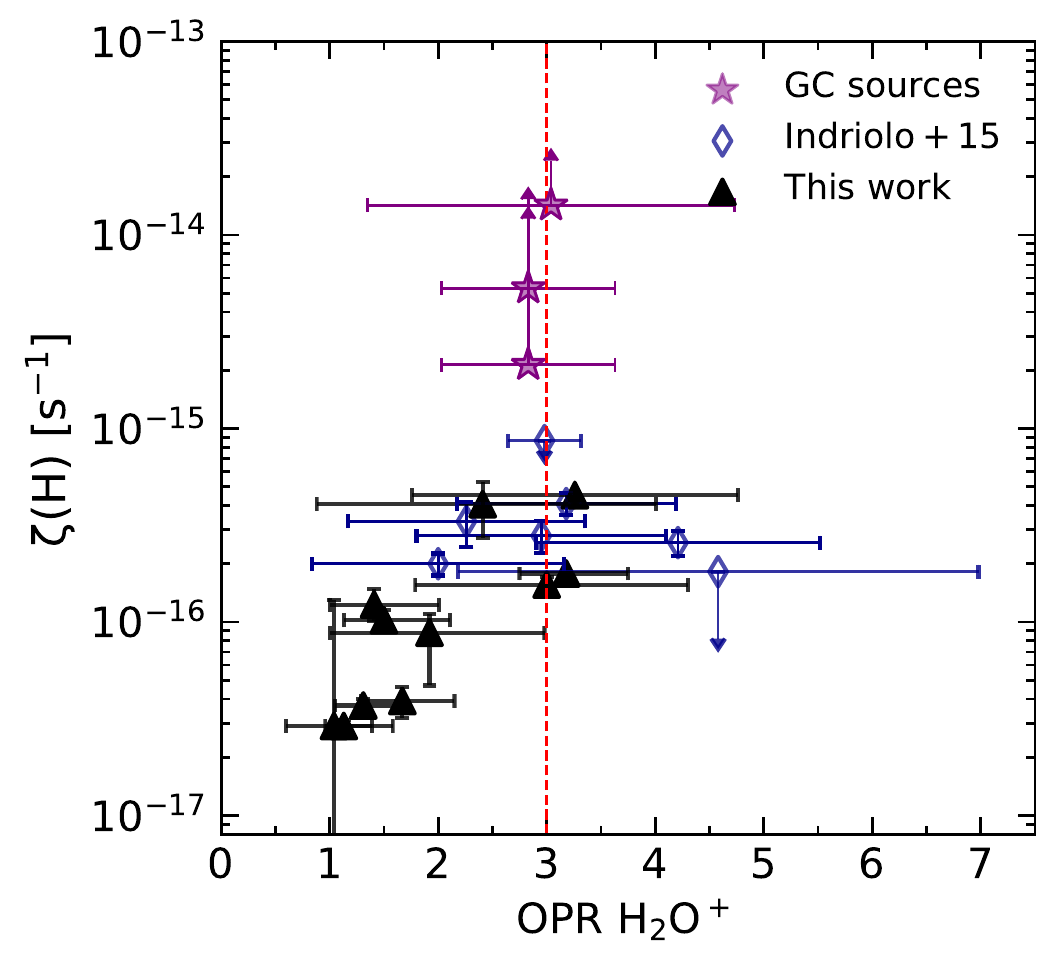}
    \caption{Observed OPR of H$_{2}$O$^+$ vs. the cosmic-ray ionisation rates derived using the steady state analysis of the OH$^+$ and H$_{2}$O$^+$. The results from this work are marked by black triangles while those from \citet{indriolo2015herschel} are displayed by blue diamonds. For comparison we display the OPR of H$_{2}$O$^+$ derived by these authors towards GC sight lines using purple stars. The dashed red line marks the equilibrium OPR value of three.}
    \label{fig:CRIR_OPR}
\end{figure}

In Fig.~\ref{fig:CRIR_OPR}, we investigate the impact of cosmic-ray ionisation rates on the OPR ratio of H$_{2}$O$^+$. In the discussion that follows we do not include those data points corresponding to GC sources studied in \citet{indriolo2015herschel}, as the unique nature of this region results in very high cosmic-ray ionisation rates ($\zeta_\text{p}(\text{H}) > 10^{-15}~\text{s}^{-1}$). As can be seen, the OPR of H$_{2}$O$^+$ thermalises to a value of three when increasing $\zeta_\text{p}(\text{H})$ from ${\sim} 2.8\times10^{-17}$ up to ${\sim4.5\times10^{-16}~}$s$^{-1}$. This is because an increase in the cosmic-ray flux results in an increase in the abundance of atomic H, thus efficiently driving the (H$_{2}$O$^+$ + H) proton-exchange reaction. Within the cited error bars the OPR saturates to a value of three at a median $\zeta_\text{p}(\text{H}) = (1.8\pm 0.2)\times 10^{-16}~\text{s}^{-1}$. The lack of a correlation beyond this value may suggest that the destruction of H$_{2}$O$^+$ via reactions with H$_{2}$ and free electrons dominates over the proton exchange reaction. If this is true, then the increase in the OPR is controlled by the kinematics of the destruction pathways. Such variations in the destruction reactions of both the ortho and the para forms have been experimentally detected for the dissociative recombination reaction of the H$_{3}^+$ ion by \citet{glosik2010binary}. Such a preferential recombination of the para-state of H$_{2}$O$^+$ over that of its ortho-state might explain the higher observed values of the OPR. However, laboratory measurements of the recombination cross-sections of rotational excitation of the ortho- and para-states of H$_{2}$O$^+$  are required to confirm whether the lowest para state undergoes recombination reactions faster than its ortho counterpart. Alternatively, in regions with high values of $\zeta_\text{p}(\text{H})$ where protons are largely abundant, reactions between H$_{2}$O and protons can be a competing formation pathway for the production of H$_{2}$O$^+$ ions (${k \sim 1-4\times 10^{-8}~\text{cm}^{3}\text{s}^{-1}}$ for $T =  10$--100~K).
This further complicates the analysis as we now have to take into account the OPR of the reacting H$_{2}$O molecules and its corresponding efficiency in producing either ortho- or para-H$_{2}$O$^+$ states. However, this requires detailed modelling of the dust-grain and gas-phase processes that govern the OPR of H$_{2}$O, which is beyond the scope of this paper. The main source of uncertainty in this analysis is in the column density estimates of p-H$_{2}$O$^+$ which may be under-estimated in velocity intervals that are affected by contamination from emission lines or over-estimated in regions with D$_{2}$O absorption.

\section{Conclusions}\label{sec:conclusions}
In this paper we present our observations searching for \arhp absorption along the LOS towards seven SFRs in the Galaxy. Our successful search towards all our targets has doubled the sample of known \arhp detections, which allows us to constrain the physical properties of the absorbing medium. We determined \arhp column densities over 33 distinct velocity intervals corresponding to spiral-arm and inter-arm crossings and find a mean \arhp abundance, $X(\text{ArH}^+) = N(\text{ArH}^+)/N(\text{H})$ of $1.5\pm0.9 \times10^{-9}$ with a power-law relationship between \arhp and atomic hydrogen  of $N(\text{ArH}^+) \propto N(\text{H})^{0.2}$. We do not detect any ArH$^+$ absorption at the systemic velocities of these sources except for signatures that trace infalling material.

By analysing the steady state chemistry of OH$^+$, and o-H$_{2}$O$^+$ along with the abundances determined from observations, we derived molecular hydrogen fractions, $f_{\text{H}_{2}}$, of the order of $10^{-2}$, consistent with previous determinations of $f_{\text{H}_{2}}$ by \citet{ neufeld2016chemistry}. We subsequently derived cosmic-ray ionisation rates with a mean $\zeta_\text{p}(\text{H})$ value of ($2.28\pm0.34$)$\times10^{-16}~$s$^{-1}$. Assuming that the observed population of \arhp ions are also exposed to the same amount of cosmic-ray irradiation, we derived the molecular fraction of gas traced by \arhp to be on average $8.8\times10^{-4}$. Moreover, the distribution of the column densities of ArH$^+$--OH$^+$, ArH$^+$--\ohtop and ArH$^+$--CH, while being moderately correlated for material along the LOS, shows almost no correlation or even a negative one for the material associated with the molecular cloud environments. These results observationally highlight that \arhp traces a different cloud population than its diffuse atomic gas counterparts OH$^+$, and o-H$_{2}$O$^+$ and the diffuse molecular gas tracer CH. Additionally, the Galactic distribution profile of the \arhp column densities resembles that of H{\small I} within the inner Galaxy (${R_{\text{GAL}} \lesssim7}$~kpc). A pronounced peak at 5~kpc is dominated by contributions from the CNM component of atomic gas \citep{pineda2013herschel}, while the WNM component primarily contributes in the outer Galaxy. While we find a larger dispersion of $N(\text{ArH}^+)$ values at ${R_{\text{GAL}} > 7~}$kpc, our analysis does not clearly find evidence for a secondary peak at 8~kpc as it is limited by the number of sight lines. Proving or disproving the presence of this feature, to ultimately distinguish between the different components of the neutral gas probed by the ArH$+$, would require further observations covering sight lines present in the outer Galaxy. While our results confirm the value of \arhp as a tracer of purely atomic gas, there are several sources of uncertainties that can bias them. These include the validity of the assumptions we made in computing the molecular fraction and the systematic effects that the blending of the H{\small I} emission particularly at and close to the sources' systemic velocities has on the derived column densities. 

Through our H$_{2}$O$^+$ analysis, we have derived OPRs that are both above and below the expected equilibrium value of three with a mean value of 2.10$\pm 1.0$. This corresponds to a nuclear spin temperature of 41~K which is within the typical range of gas temperatures observed for diffuse clouds. However, a majority of the velocity components show OPRs close to unity, increasing steeply to the standard value of three with increasing cosmic-ray ionisation rates up to ${\zeta_\text{p}(\text{H}) = (1.8 \pm0.2)\times10^{-16}~\text{s}^{-1}}$. Beyond this value, the OPR shows no correlation with $\zeta_\text{p}(\text{H})$, indicating that from that point on reactions with H$_{2}$ take over as the dominant destruction pathway in its chemistry. The general deviations of the OPR of H$_{2}$O$^+$ may also be inherited from the OPR of its parent species, namely H$_{2}$ or H$_{2}$O and the efficiency in converting between the two spin states. 

%Over the last decade, the study of cosmic-rays in the ISM has garnered widespread attention as it is a crucial source of ionisation, responsible of driving varied chemical reactions and plays an important role in shaping the dynamical evolution of different gas environments through interactions with magnetic fields.
%Combining observations of multiple species, examining their correlations and measuring abundance ratios, we have been able to derive the molecular hydrogen fractions traced by these different species and also determine the cosmic-ray ionisation rates.

\begin{acknowledgements}
This publication is based on data acquired with the Atacama Pathfinder Experiment (APEX) under the project id  M-0103.F-9519C-2019. APEX is a collaboration between the Max-Planck-Institut fur Radioastronomie, the European Southern Observatory, and the Onsala Space Observatory. We would like to express our gratitude to the APEX staff and science team for their continued assistance in carrying out the observations presented in this work. The authors would like to thank the anonymous referee for their careful review of the article and valuable input and Marijke Haverkorn for providing us with the SGPS continuum data for AG10.472+00.027. We are thankful to the developers of the C++ and Python libraries and for  making them available as open-source software. In particular, this research has made use of the NumPy \citep{numpy}, SciPy \citep{scipy} and matplotlib \citep{matplotlib} packages.
\end{acknowledgements}

%By combining observations of multiple species, examining correlations and measuring abundance ratios, it has been possible to determine the distribution function for the H 2 fraction within the diffuse ISM; this provides an important constraint on global models for the formation and destruction of molecular hydrogen in a turbulent medium (e.g. Bialy et al. 2019).

\bibliographystyle{aa} % style aa.bst
\bibliography{ref}

\begin{appendix}
%\section{\texorpdfstring{604~GHz p-H$_{2}$O$^{+}$}{pHtOp} spectra}\label{appendix:phtop_604}
%In this appendix we present the spectra of the 604~GHz \phtop lines towards individual sources.  

\section{H{\small I} analysis} \label{appendix:hi_analysis}
In this appendix we present the H{\small{I}} absorption ($T_\text{on}$) and emission spectra ($T_\text{off}$), along with derived quantities such as the optical depth, spin temperature and column density. The solid black curve in the continuum normalised on-source spectrum, $T^{\text{on,obs}}/ T^{\text{sou,obs}_{\text{cont}}}$ (top panel) represents the absorption profile smoothed to the resolution of the emission data. %In the off-source spectrum, we plot the interpolated brightness temperature (in black).
Classically the off-source spectrum is taken from a position next to the source of interested (well outside of the beam width). However, in the Milky Way disk, HI is subject to strong spatial fluctuations and using a single off position would be error prone. Therefore, in \citet{winkel2017hydrogen} a spatial filtering technique using a ring (or doughnut-shaped) kernel was applied to obtain an interpolated brightness temperature, shown in the second panel (black curve). For comparison, in the panel displaying spin temperatures, we also plot the brightness temperature in blue and lastly, the column density panel also displays the uncorrected column density profile, $N^{*}_{\text{HI}}$. The blue and grey shaded regions indicate the $1\sigma$ (68\% percentile) and $3\sigma$ (99.7\% percentile) confidence intervals, respectively. Additionally in each panel, we mark the systemic velocity and velocity dispersion of each source by the dashed pink line and shaded pink regions, respectively. A detailed description is presented in \citet{winkel2017hydrogen}. 

\begin{figure}
    \centering
    \includegraphics[width=0.48\textwidth]{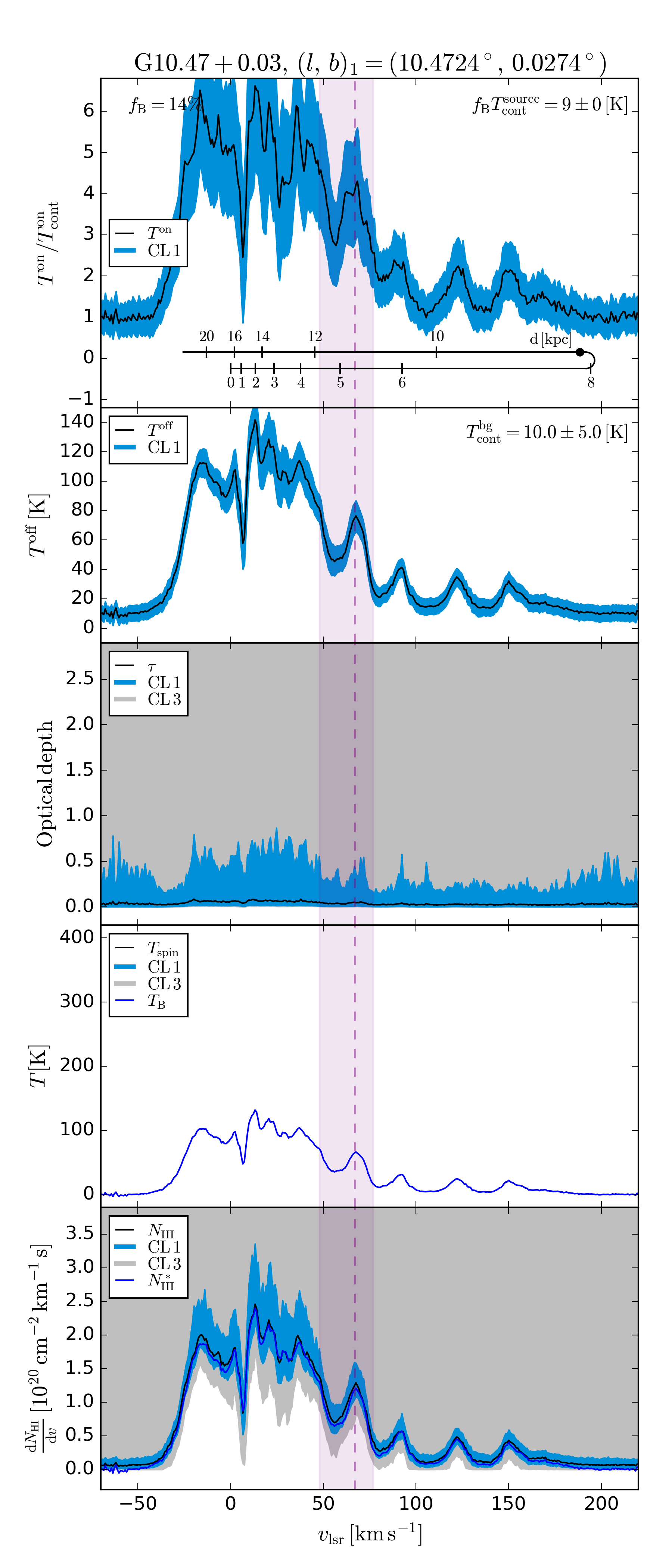}
    \caption{From top to bottom: H{\small I} absorption and emission spectra, optical depth, spin temperature, and H{\small I} column density towards AG10.472$+$00.027. The pink dashed line and shaded region mark the systemic velocity of the source and highlight the velocity dispersion of the source. }
    \label{fig:G10P47_HI}
\end{figure}

\begin{figure}
    \centering
    \includegraphics[width=0.48\textwidth]{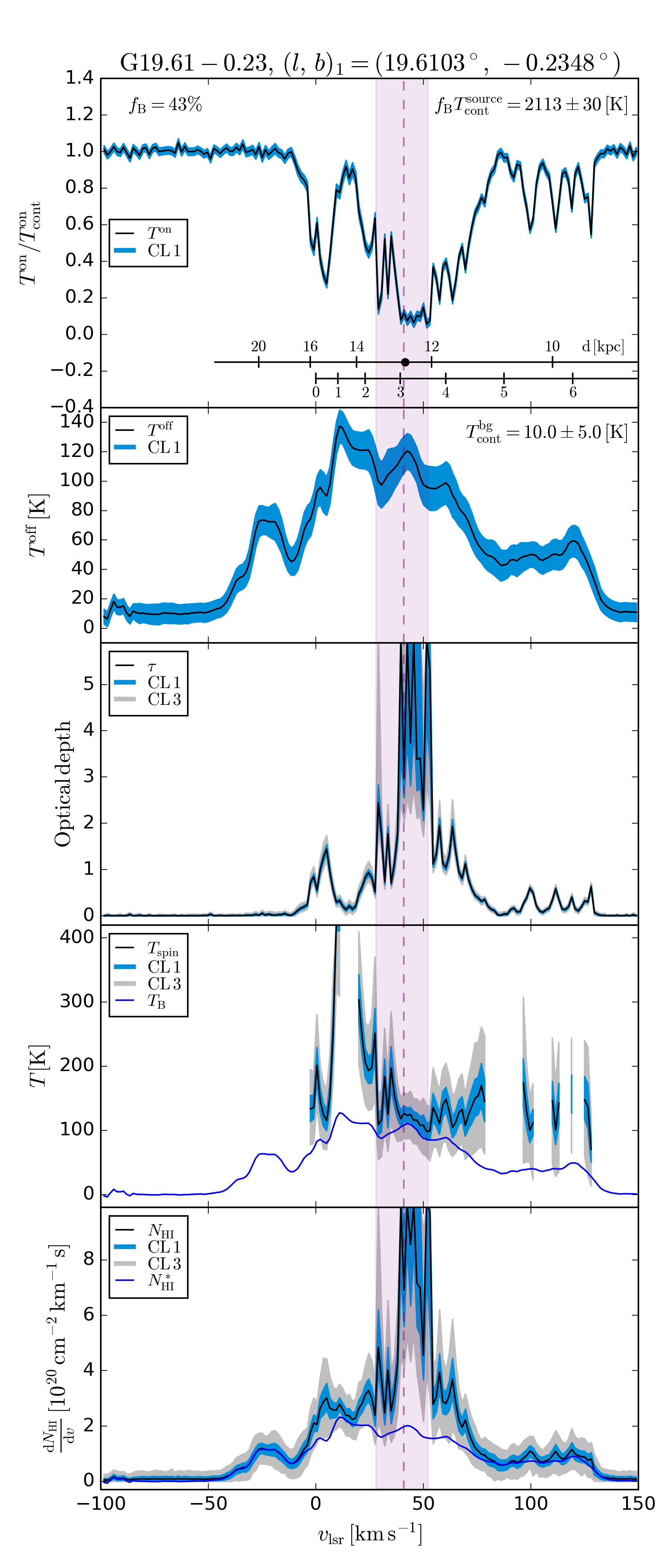}
    \caption{Same as Fig.~\ref{fig:G10P47_HI} but towards AG19.609$-$00.234.}
    \label{fig:G19P61_HI}
\end{figure}

\begin{figure}
    \centering
    \includegraphics[width=0.48\textwidth]{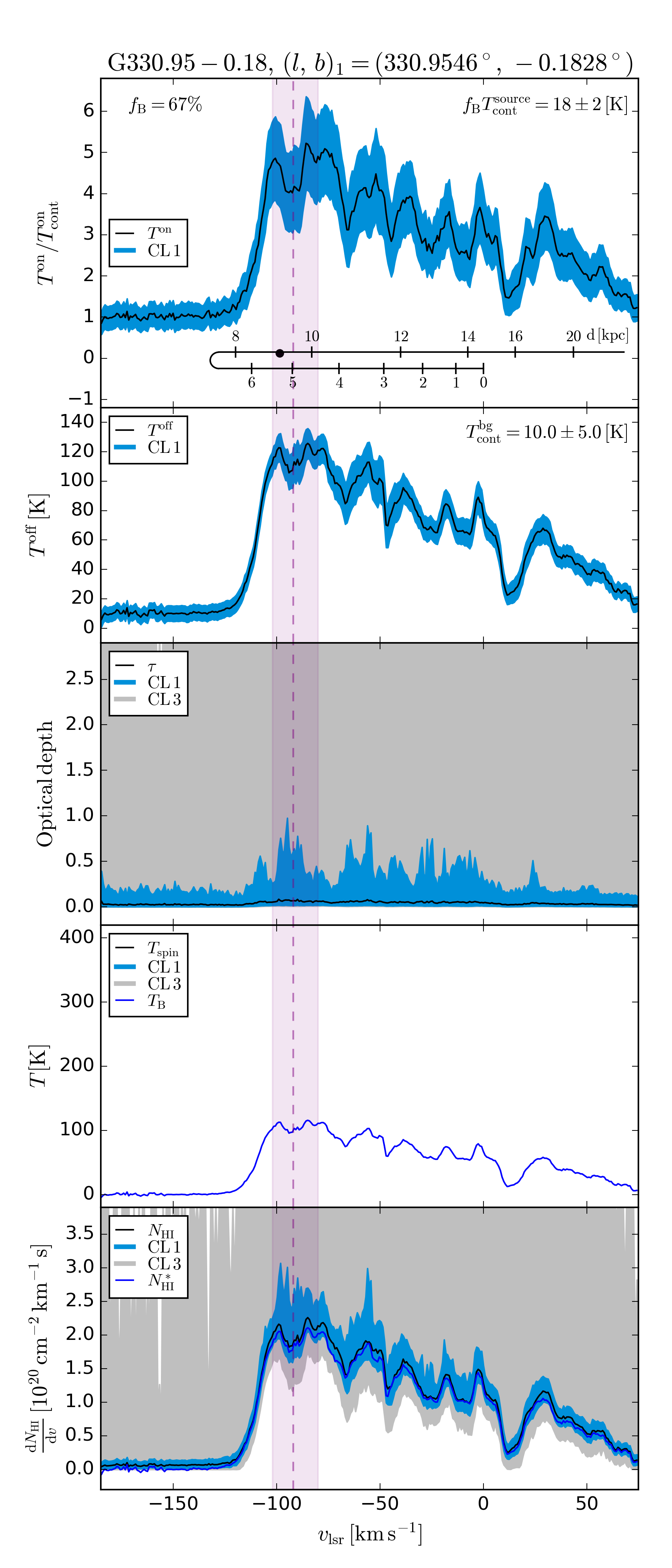}
    \caption{Same as Fig.~\ref{fig:G10P47_HI} but towards AG330.954$-$00.182.}
    \label{fig:G330P95_HI}
\end{figure}

\begin{figure}
    \centering
    \includegraphics[width=0.48\textwidth]{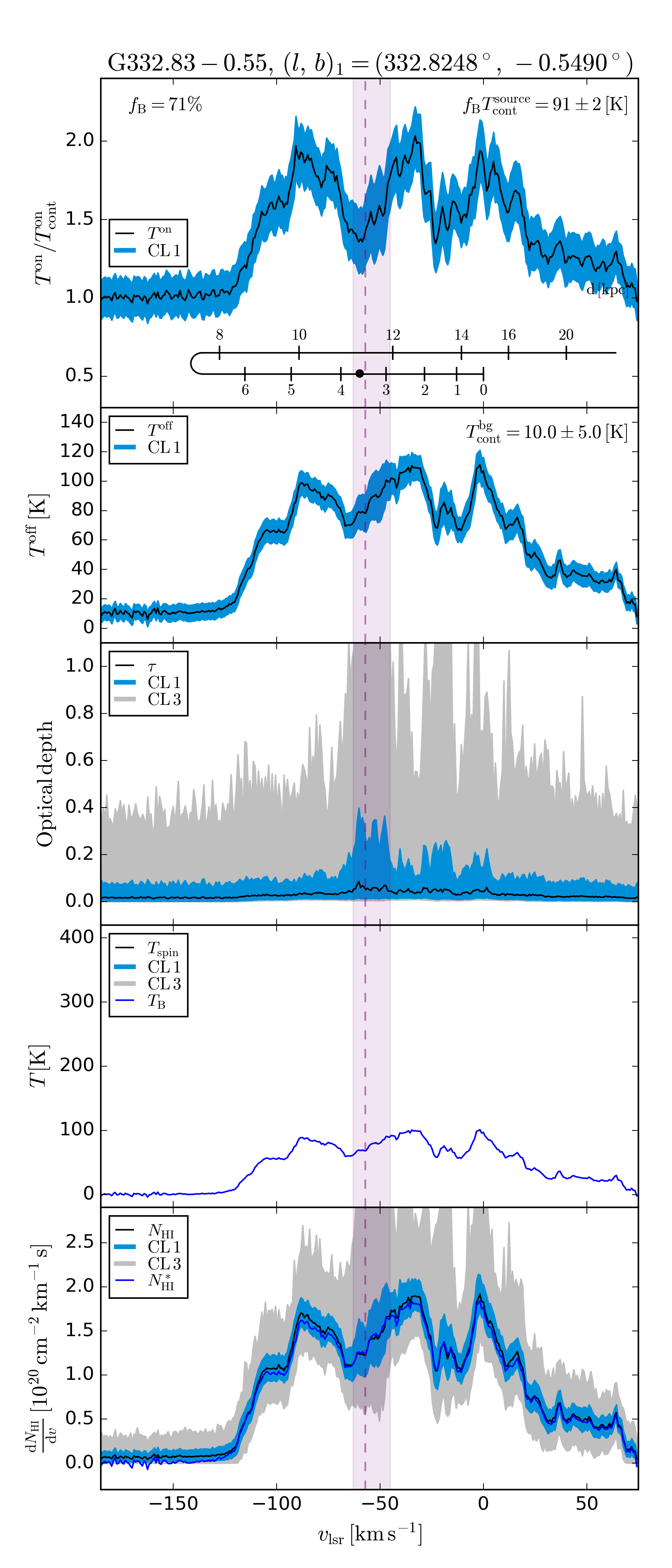}
  \caption{Same as Fig.~\ref{fig:G10P47_HI} but towards AG332.826$-$00.549.}
    \label{fig:G332P83_HI}
\end{figure}

\begin{figure}
    \centering
    \includegraphics[width=0.48\textwidth]{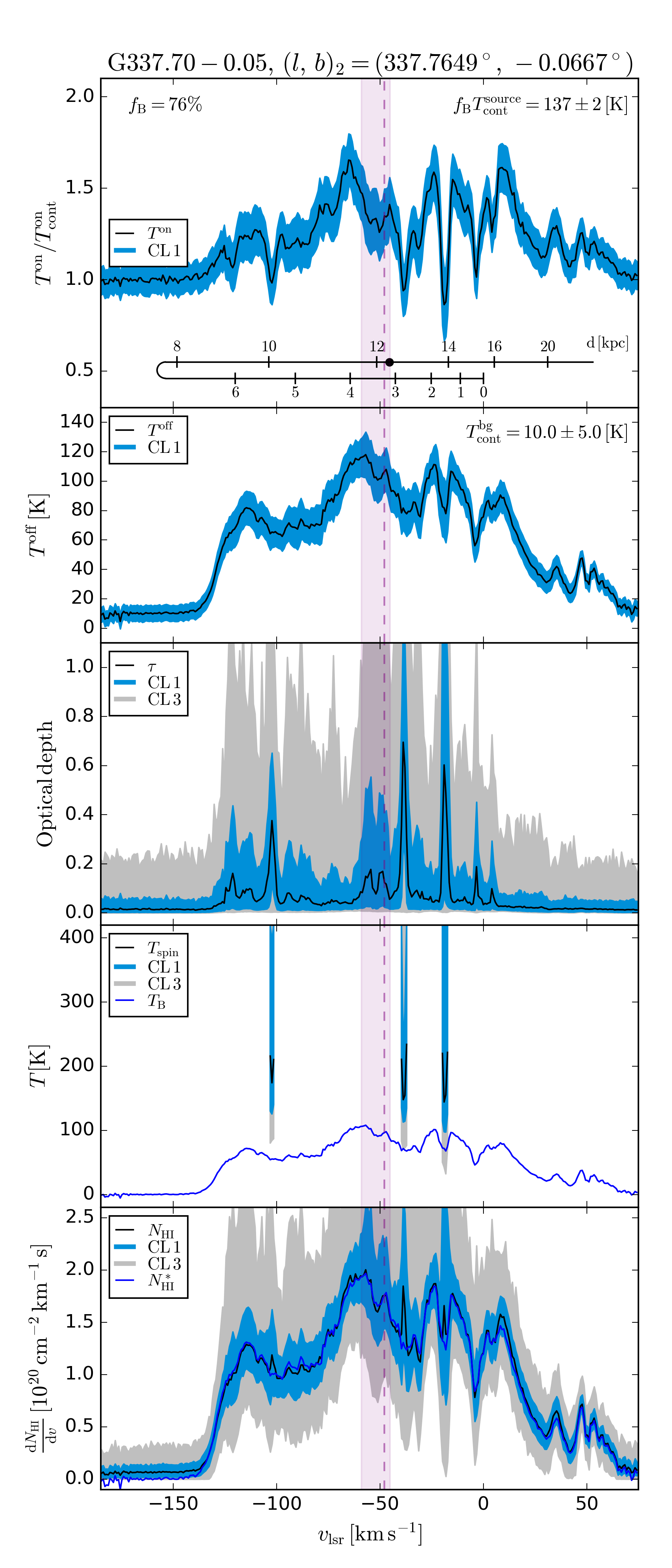}
  \caption{Same as Fig.~\ref{fig:G10P47_HI} but towards AG337.704$-$00.054.}
    \label{fig:G337P70_HI}
\end{figure}

\begin{figure}
    \centering
    \includegraphics[width=0.48\textwidth]{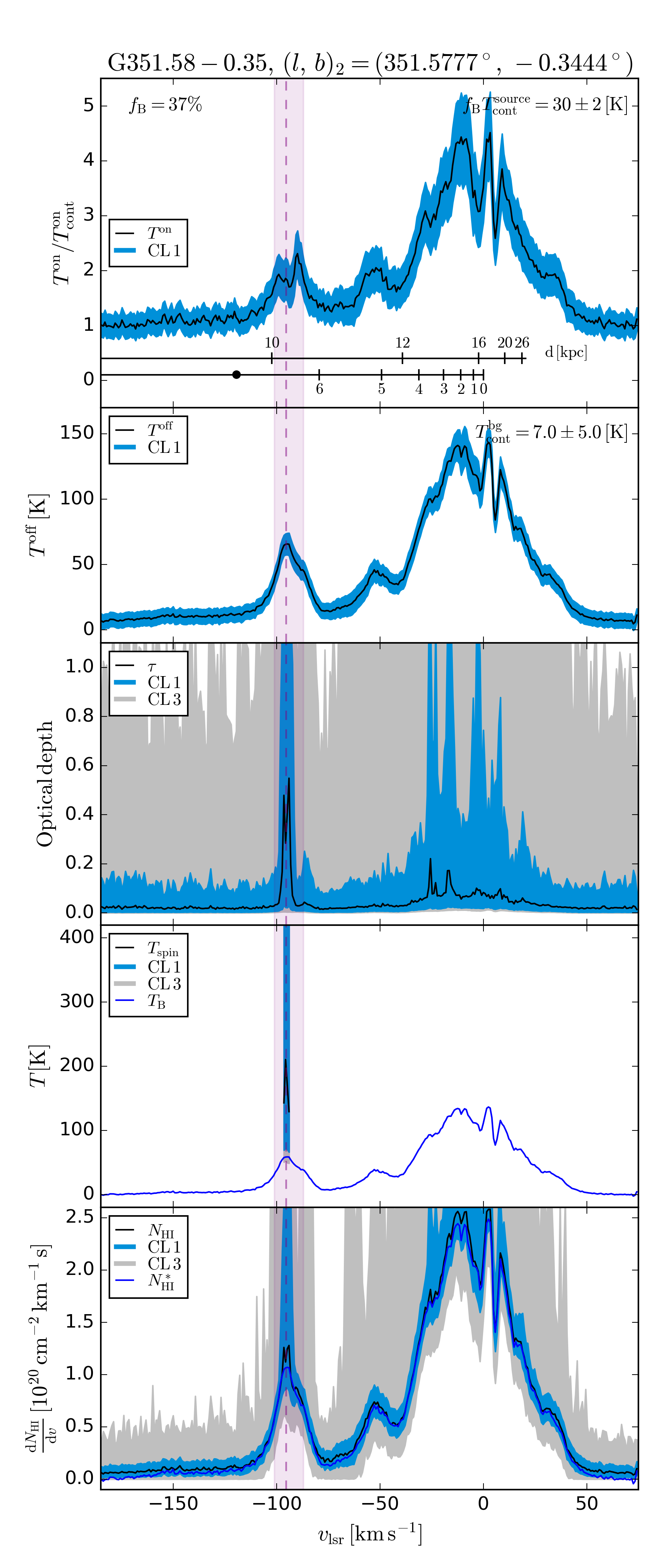}
     \caption{Same as Fig.~\ref{fig:G10P47_HI} but towards AG351.581$-$00.352.}
    \label{fig:G351P58_HI}
\end{figure}

\section{Continuum level uncertainties} \label{appendix:continuum_calib}
\begin{figure*}
    
    \includegraphics[width=11cm]{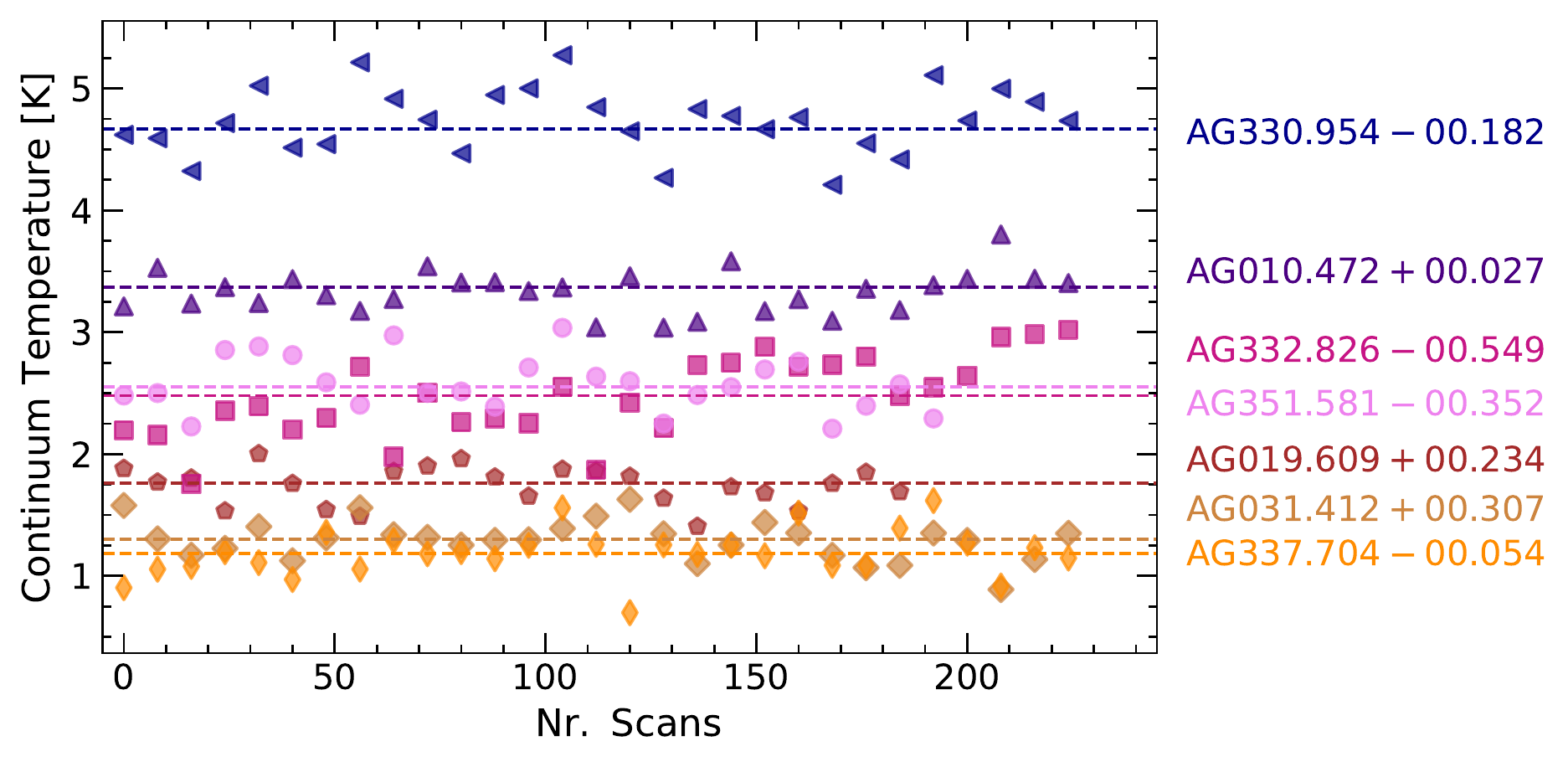}\quad
    \includegraphics[width=5.5cm]{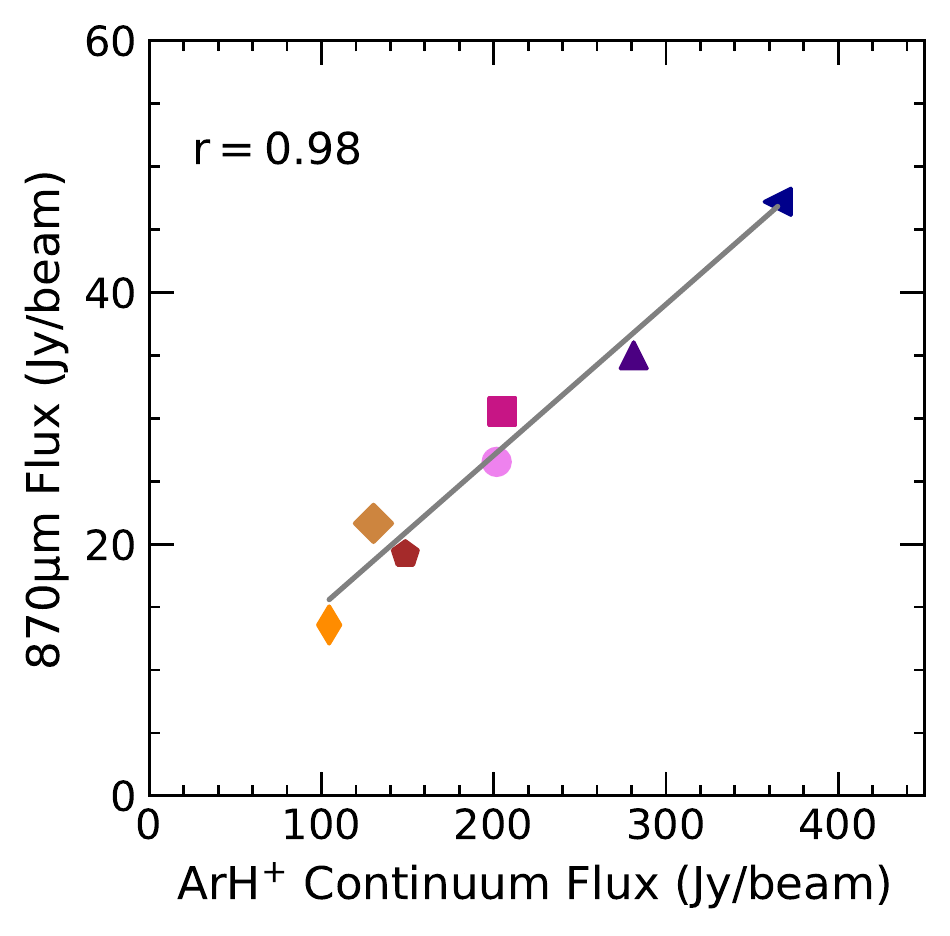}
    \caption{Left: Continuum fluctuations across scans. The dashed lines represent the median continuum level for each source. Right: Correlation between the observed ArH$^+$ continuum flux and the 870~$\mu$m continuum flux. }
    \label{fig:continuum_scans}
\end{figure*}

In this Appendix we briefly discuss the reliability of the absolute calibration of the continuum brightness temperatures used in our analysis. As a first check, we binned our observations across scans (time), towards each individual source over velocity intervals/channels previously used to estimate the continuum and baseline. Assessing the relative fluctuations in the continuum level will reveal the stability of its measurement. From Fig.~\ref{fig:continuum_scans}, it can be seen that the scatter in the continuum level, over different scans, is less than 10\% for all the sources in our study except AG330.954$-$00.182 and AG332.826$-$00.549 which shows a slightly large scatter of 18\%. Not showing any major variations, leads us to conclude that the continuum levels are fairly reliable. It is important to note here that the observations were carried out using the wobbling secondary with a fast switching rate which was essential in removing any drifts due to atmospheric instabilities.

As discussed earlier in the text, we have also carried out a comparison between the continuum flux levels of our ArH$^+$ observations with the peak fluxes obtained from 870~$\mu$m continuum data from the ATLASGAL survey. We see that our data is well correlated with 870~$\mu$m continuum emission with a relative scatter $<5\%$.

\section{Column density profiles} \label{appendix:colun_density_profiles}
In this Appendix we present the column density profiles for all the species studied in this work except H{\small I} (which is already presented in Appendix~\ref{appendix:hi_analysis}) towards all the entire source sample in this study.

\begin{figure}
    \centering
    \includegraphics[width=0.45\textwidth]{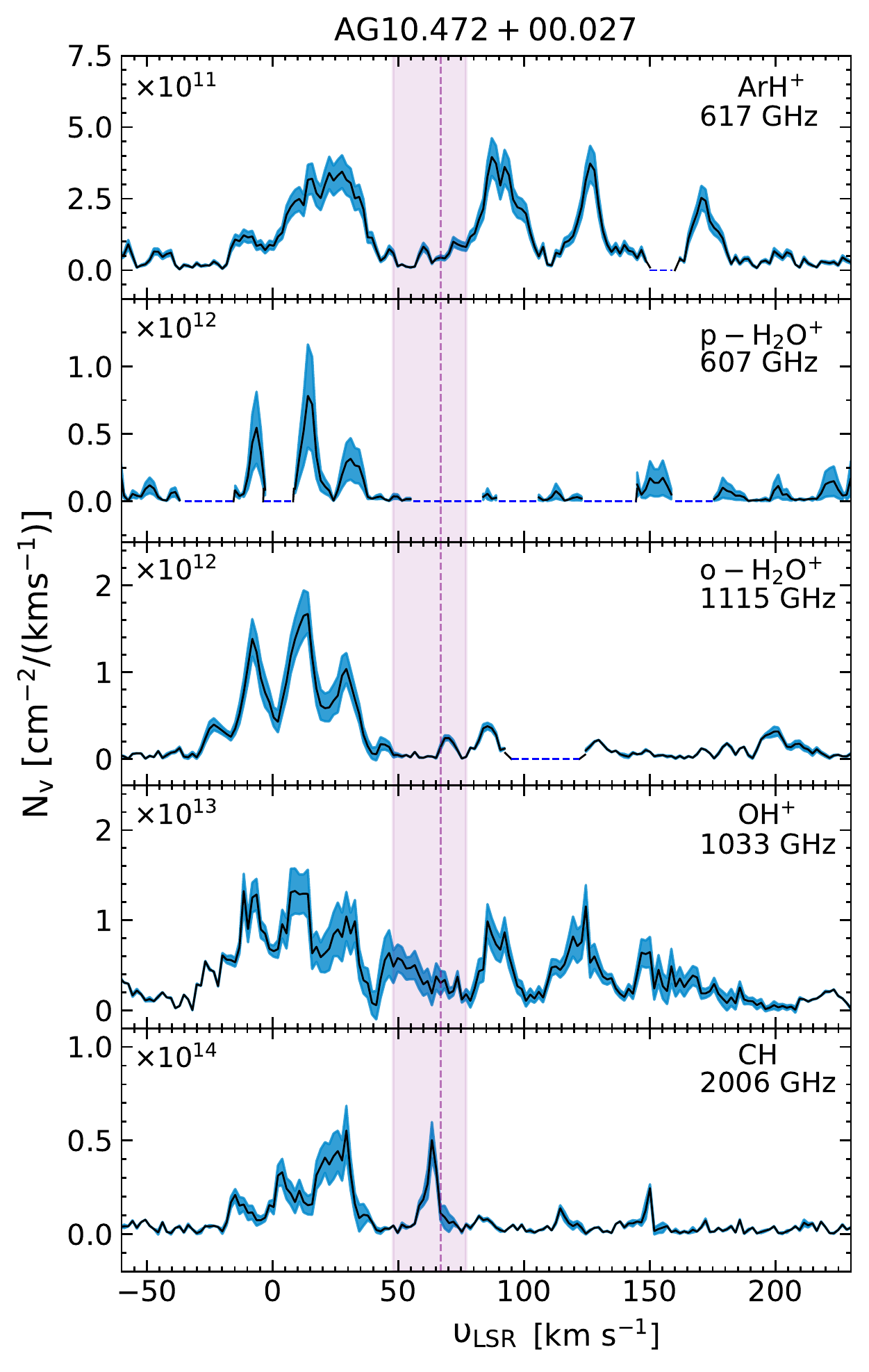}
    \caption{Top to bottom: Column density per velocity channel of ArH$^+$, p-H$_{2}$O$^+$, o-H$_{2}$O$^+$, OH$^+$, and CH towards G10.472+00.027. The blue shaded region represents the uncertainties while the pink dashed line, and shaded region mark the systemic velocity of the source and highlight the velocity dispersion of the source. The dashed blue lines indicate those velocity intervals that were omitted from the fitting routine due to contamination.}
    \label{fig:G10P47_coldens}
\end{figure}
\begin{figure}
    \centering
    \includegraphics[width=0.45\textwidth]{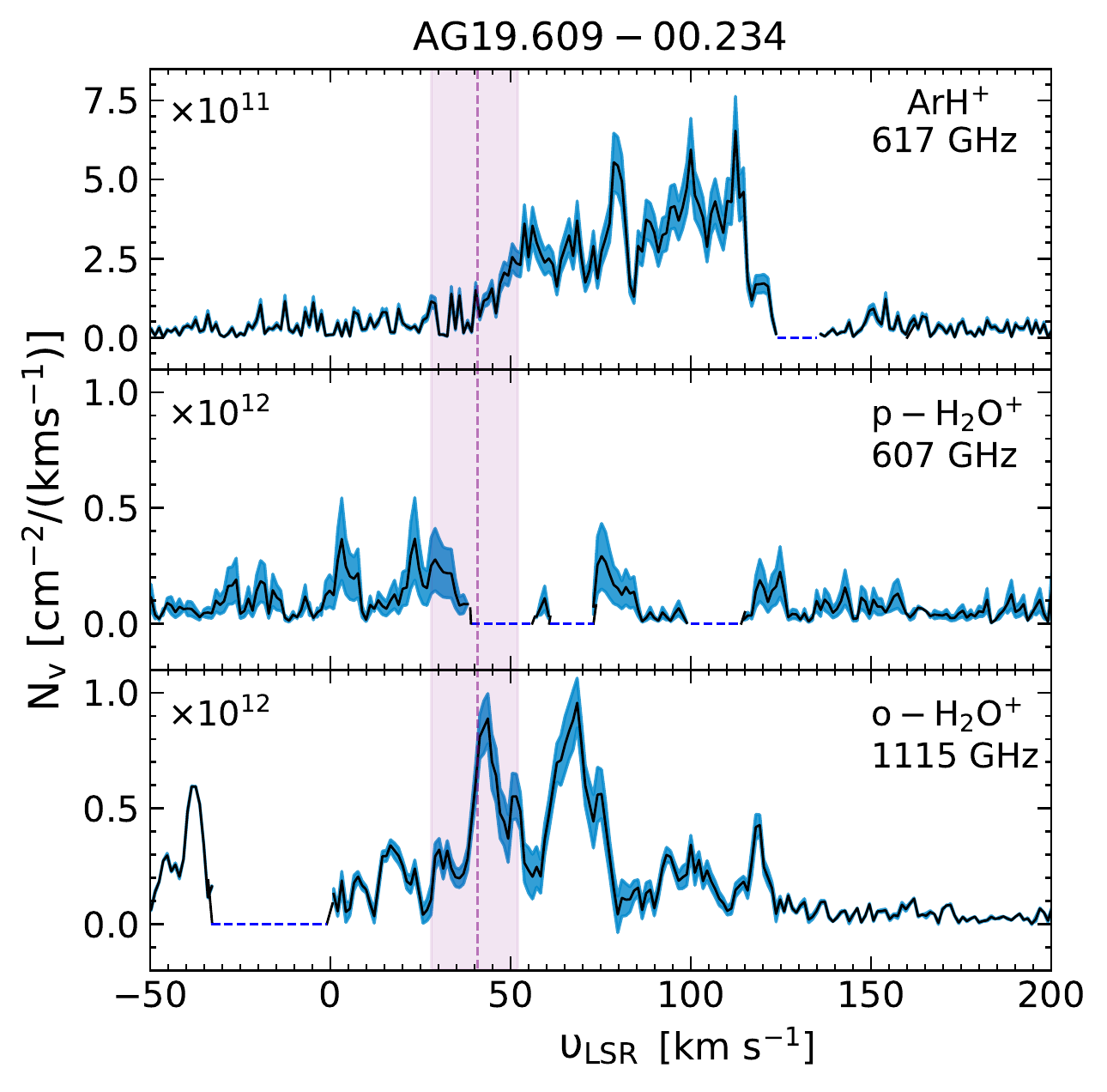}
    \caption{Same as Fig.~\ref{fig:G10P47_coldens} but towards G19.609$-$00.234. There is no OH$^+$ and CH spectra of the transitions studied here, available for this source.}
    \label{fig:G19P61_coldens}
\end{figure}
\begin{figure*}
    \centering
    \includegraphics[width=0.45\textwidth]{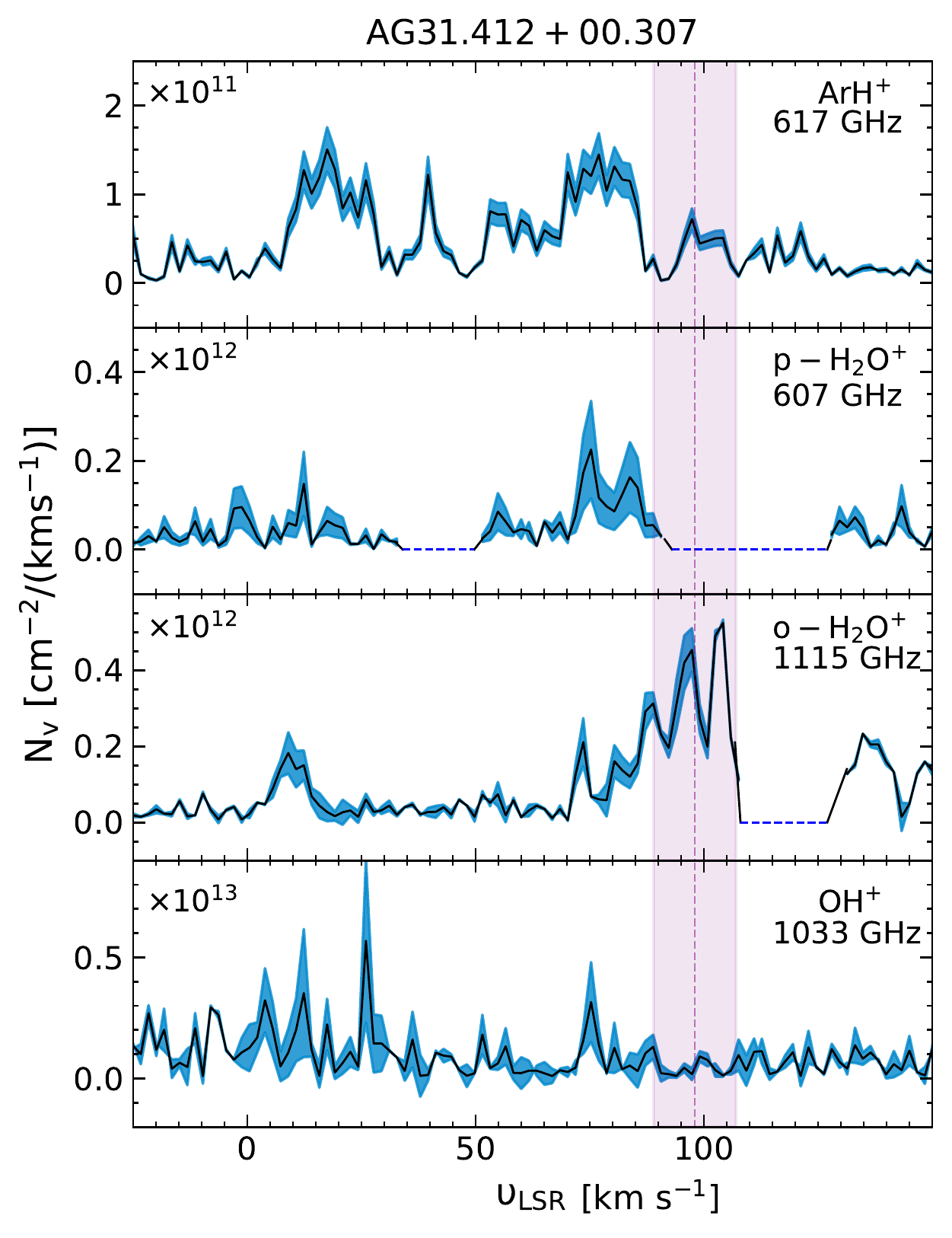}\quad
    \includegraphics[width=0.463\textwidth]{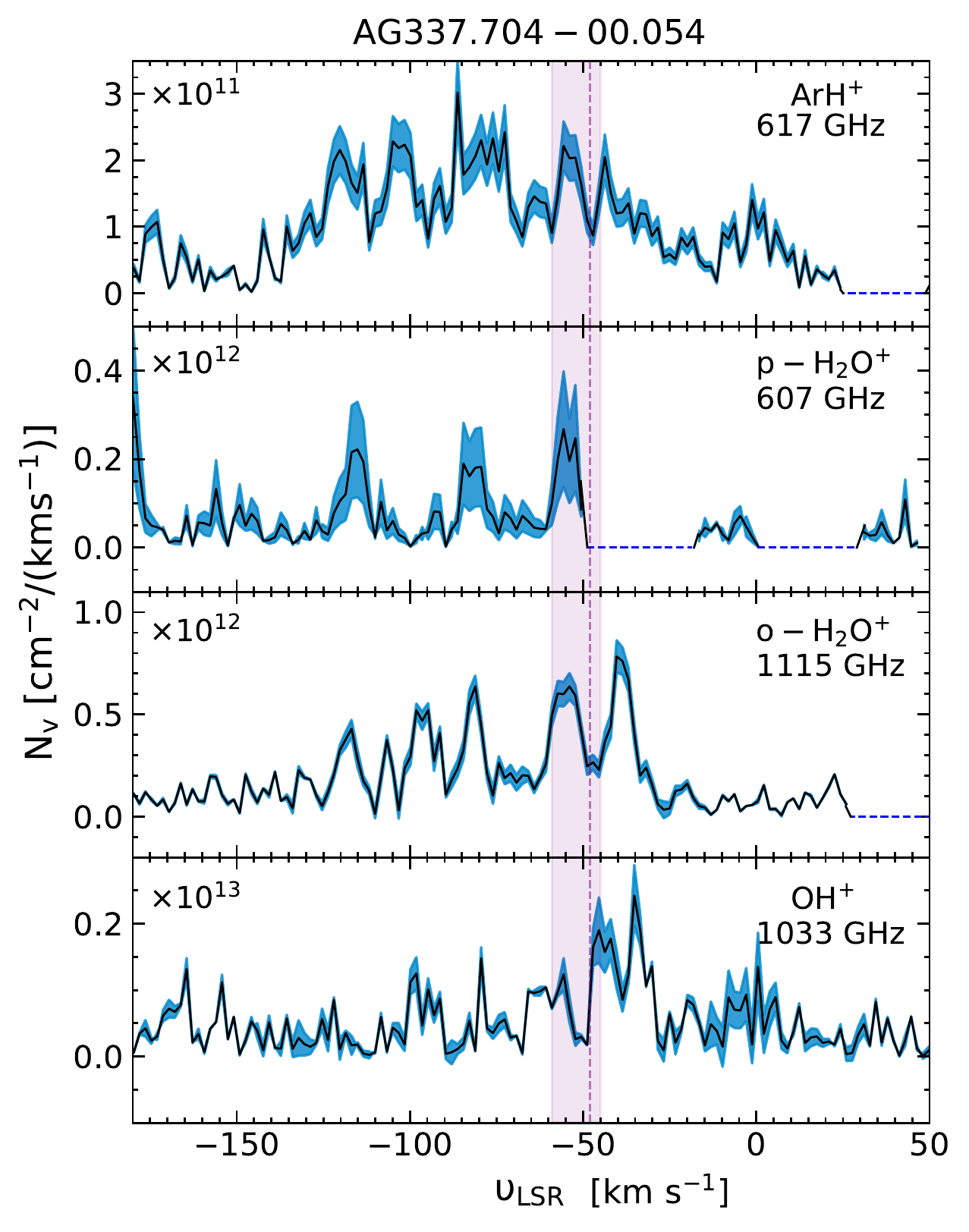}\\
    \includegraphics[width=0.45\textwidth]{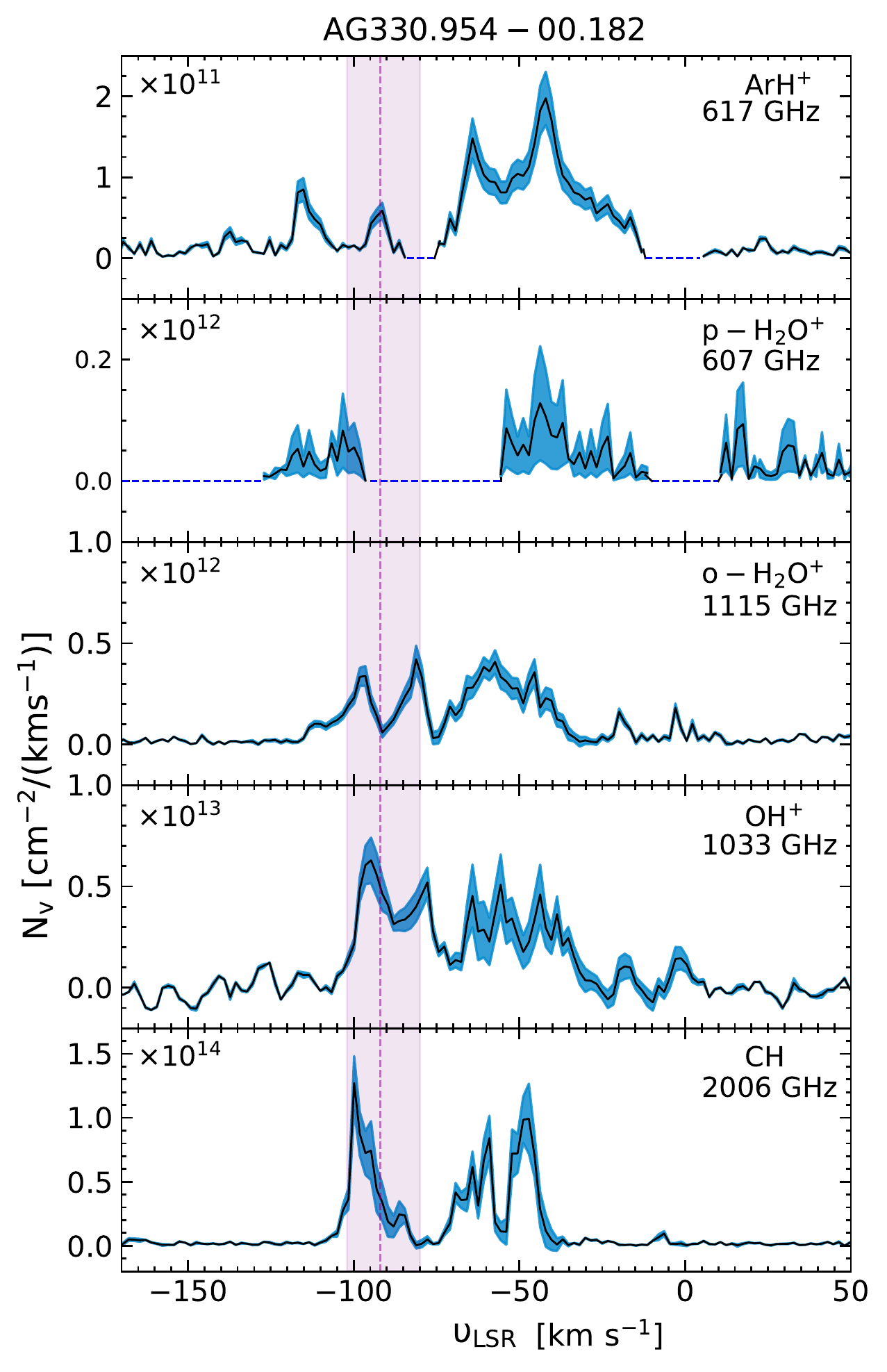}\quad
    \includegraphics[width=0.455\textwidth]{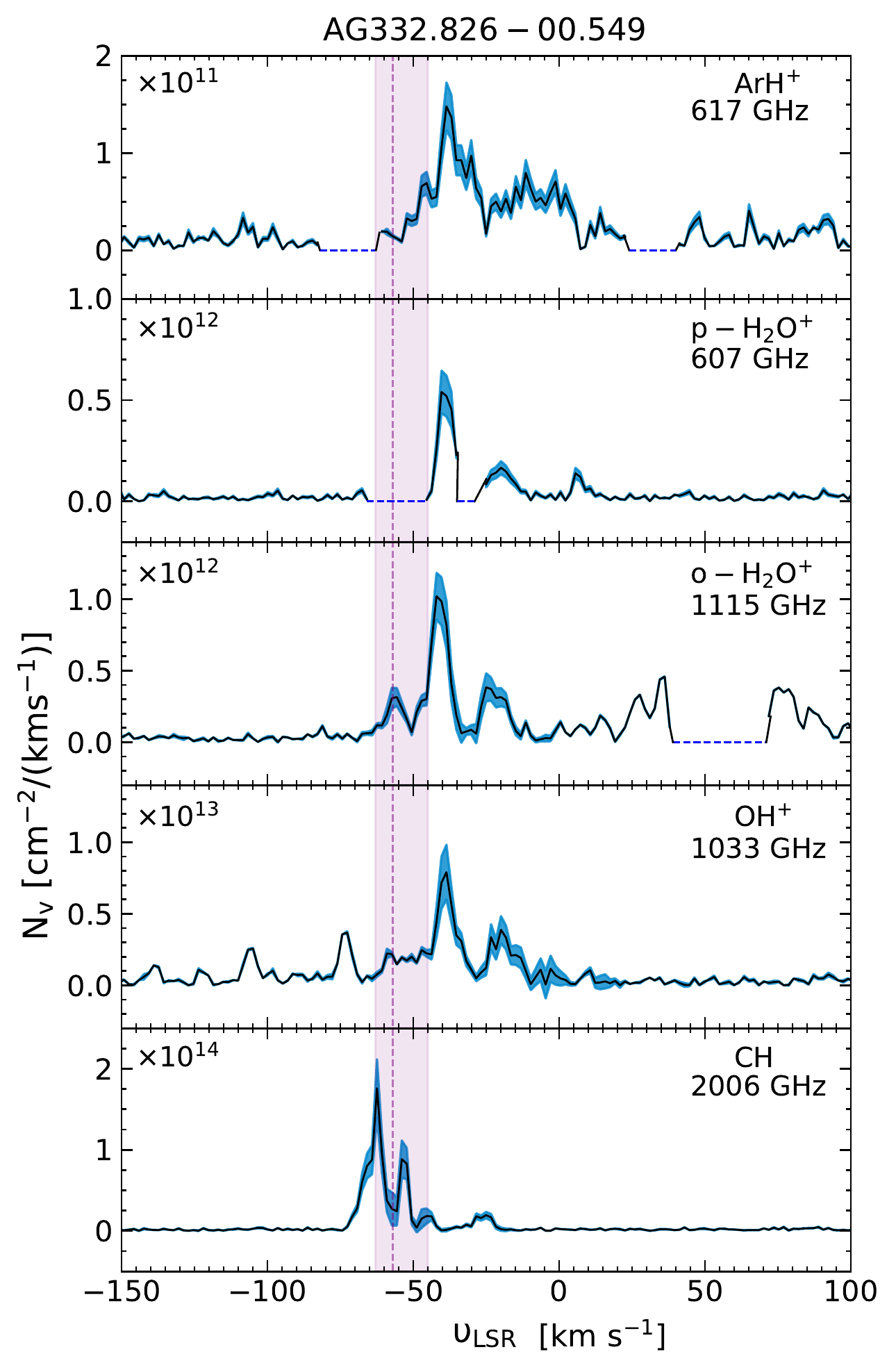}\quad
    
    \caption{Same as Fig.~\ref{fig:G10P47_coldens} but towards (clockwise from top-left) G31.412$+$00.307, G337.704$-$00.054, G332.826$-$00.549, and G330.954$-$00.182. There is no CH spectrum of the transitions studied here, available for G31.412$+$00.307 and G337.704$-$00.054.}
    \label{fig:G337P70_coldens}
\end{figure*}
\begin{figure}
    \centering
    \includegraphics[width=0.45\textwidth]{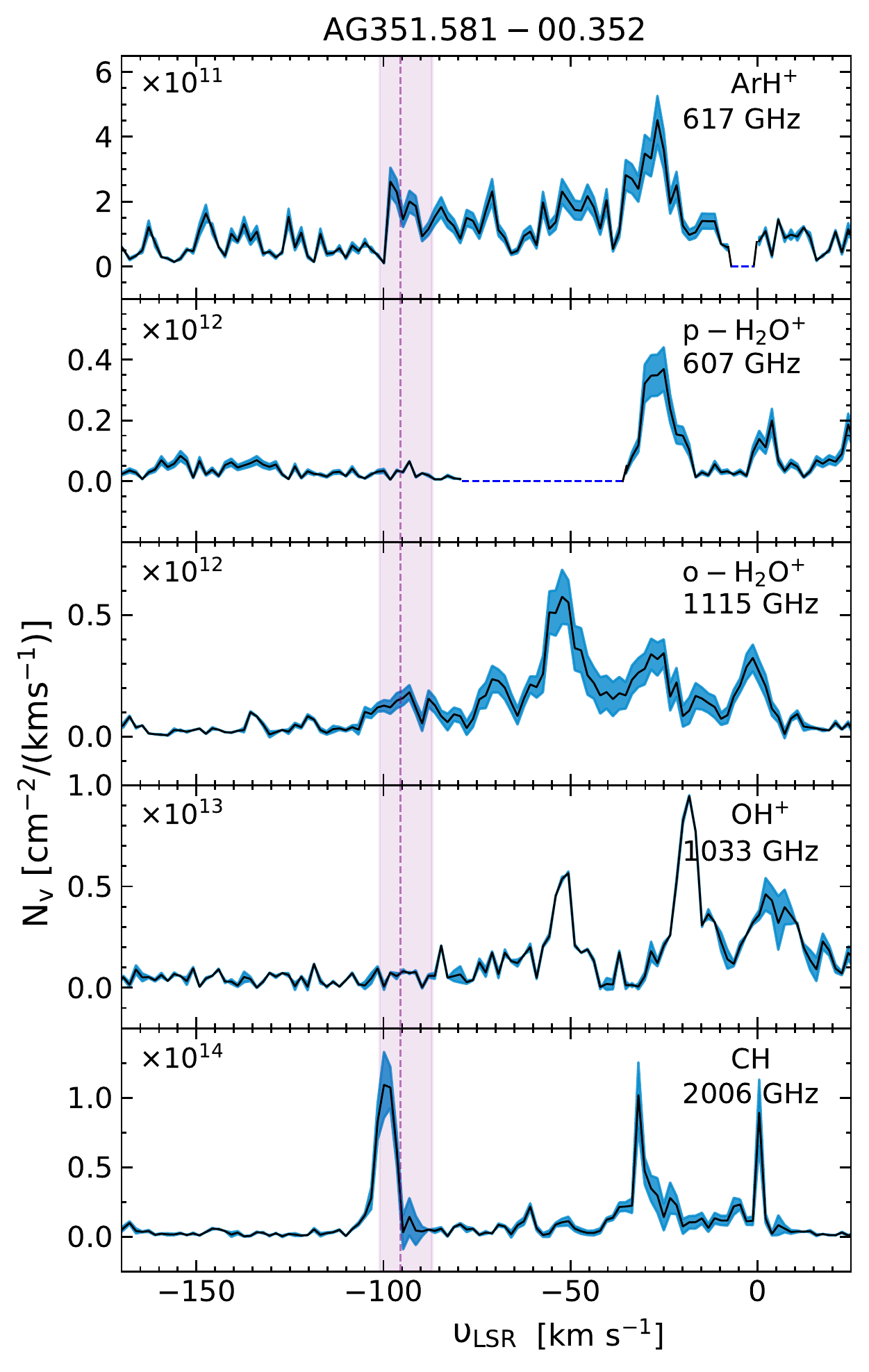}
    \caption{Same as Fig.~\ref{fig:G10P47_coldens} but towards G351.581$+$00.352.}
    \label{fig:G351P58_coldens}
\end{figure}
\end{appendix}
 \end{document}